%% file: ANA-SUSY-2018-30-PAPER.tex
\documentclass[cernpreprint, atlasdraft=false, texlive=2023, UKenglish, texmf, orcidlogo]{atlasdoc}
 
\usepackage[subfigure,block=none]{atlaspackage}
\usepackage{float}
\usepackage{atlasbiblatex}
 
\usepackage{atlasphysics}
\usepackage{atlasbsm}
\usepackage{multirow}
\usepackage{pdflscape}
\usepackage{pifont}

\usepackage{tabulary}
\newcolumntype{K}[1]{>{\centering\arraybackslash}p{#1}}

\graphicspath{{logos/}{figures/}}
 
\usepackage{ANA-SUSY-2018-30-PAPER-defs}

\AtlasTitle{Search for supersymmetry in final states with missing
transverse momentum and three or more $b$-jets in 139 fb$^{-1}$ of proton--proton collisions
at $\sqrt{s} = 13$ TeV with the ATLAS detector}

\AtlasAbstract{
A search for supersymmetry involving the pair production of gluinos decaying via off-shell third-generation squarks into the lightest neutralino ($\displaystyle\tilde\chi^0_1$) is reported. It exploits LHC proton--proton collision data at a centre-of-mass energy $\sqrt{s} = 13$ TeV with an integrated luminosity of 139~fb$^{-1}$ collected with the ATLAS detector from 2015 to 2018. The search uses events containing large missing transverse momentum, up to one electron or muon, and several energetic jets, at least three of which must be identified as containing $b$-hadrons. Both a simple kinematic event selection and an event selection based upon a deep neural-network are used. No significant excess above the predicted background is found.  In simplified models involving the pair production of gluinos that decay via off-shell top (bottom) squarks, gluino masses less than 2.44~TeV (2.35~TeV) are excluded at 95\% CL for a massless $\displaystyle\tilde\chi^0_1$. Limits are also set on the gluino mass in models with variable branching ratios for gluino decays to $b\bar{b}\tilde\chi^0_1$, $t\bar{t}\tilde\chi^0_1$ and $t\bar{b}\tilde\chi^-_1$ / $\bar{t}b\tilde\chi^+_1$.
}

\AtlasRefCode{SUSY-2018-30}
 
\PreprintIdNumber{CERN-EP-2022-213}
 
\AtlasDate{\today}

\arXivId{2211.08028}
 
\HepDataRecord{https://www.hepdata.net/record/95928}
 
\AtlasJournal{Eur. Phys. J. C}
\AtlasJournalRef{\EPJC 83, 561 (2023)}
\AtlasDOI{10.1140/epjc/s10052-023-11543-6}

\hypersetup{pdftitle={ATLAS document},pdfauthor={The ATLAS Collaboration}}
 
\begin{document}
 
\maketitle

\section{Introduction}

Supersymmetry (SUSY)~\cite{Golfand:1971iw,Volkov:1973ix,Wess:1974tw,Wess:1974jb,Ferrara:1974pu,Salam:1974ig} is a generalisation of space-time symmetries that predicts new bosonic partners for the fermions of the Standard Model (SM) and new fermionic partners for its bosons. In SUSY models, if $R$-parity is conserved~\cite{Farrar:1978xj}, SUSY particles are produced in pairs and the lightest supersymmetric particle (LSP) is stable. The scalar partners of the left- and right-handed quarks, the squarks $\squarkL$ and $\squarkR$, can mix to form two mass eigenstates $\tilde{q}_1$ and $\tilde{q}_2$, ordered by increasing mass. SUSY can reduce unnatural tuning in the Higgs sector by orders of magnitude~\cite{Sakai:1981gr,Dimopoulos:1981yj,Ibanez:1981yh,Dimopoulos:1981zb} provided that the superpartners of the top quark (the top squarks, $\stop_1$ and $\stop_2$) have masses not too far above the weak scale~\cite{Papucci:2011wy}. Because of the  SM weak-isospin symmetry, the mass of the lighter bottom squark $\sbottomone$ is also expected to be close to the weak scale. The fermionic partners of the gluons, the gluinos ($\gluino$), are also motivated by naturalness~\cite{Barbieri:1987fn} to have a mass around the \tev\ scale in order to limit their contributions to the radiative corrections to the top squark masses. For these reasons, and because the gluinos are expected to be pair-produced with a high cross-section at the Large Hadron Collider (LHC~\cite{Evans:2008zzb}), the search for gluino production with decays via top and bottom squarks is highly motivated at the LHC.
 
This paper presents a search for pair-produced gluinos decaying via top or bottom squarks in events with multiple jets containing $b$-hadrons ($b$-jets in the following), high missing transverse momentum of magnitude \met, and potentially additional jets and/or an isolated electron or muon (referred to as `leptons' hereafter). The results constitute an update of those obtained using 36.1~\ifb\ of proton--proton ($pp$) collision data~\cite{SUSY-2016-10} from the ATLAS detector~\cite{PERF-2007-01}. They exploit an expanded dataset of 139~fb$^{-1}$ of $pp$ collision data acquired at a centre-of-mass energy of 13~\tev. To make best use of the expanded dataset, the simple kinematic selections used in the earlier analysis have been re-optimised, and a new event selection based upon a deep neural network optimised to discern the gluino signatures from background is employed. The latter optimally combines selections requiring zero leptons or one lepton.
 
Interpretations are provided in the context of several simplified models \cite{Alwall:2008ve,Alwall:2008ag,Alves:2011wf} probing gluino decays into the LSP via off-shell top or bottom squarks. In these models, the LSP is assumed to be the lightest neutralino $\ninoone$, a linear superposition of the superpartners of the neutral electroweak and Higgs bosons. One model also features the lighter charginos $\chinoonepm$,  which are linear superpositions of the superpartners of the charged Higgs and SM electroweak bosons. Several benchmark simplified model scenarios studied in the earlier instances of the analysis~\cite{SUSY-2016-10,Aad:2016eki} are considered: two models, referred to as `Gtt' and `Gbb' respectively, which feature exclusively gluino decays to the LSP via off-shell top or bottom squarks (Figure~\ref{fig:feynman-Gluino-topbottom-offshell}), and a third model, referred to as `Gtb', with variable branching ratios for $\gluino \to t\bar{t}\ninoone$, $\gluino \to b\bar{b}\ninoone$ and $\gluino \to t \bar{b} \chinoonem$ / $\gluino \to \bar{t} b \chinoonep$ (Figure~\ref{fig:feynman-Gluino-topbottom-branchingratio} shows the additional decay processes that this model permits). In the Gtb models the mass difference between the $\chinoonepm$ and the $\ninoone$ is fixed to a small value (2~\GeV), motivated by natural SUSY models in which the $\ninoone$ is an almost pure higgsino (see e.g. Ref.~\cite{Papucci:2011wy}).
 
Pair-produced gluinos with top-squark-mediated decays have also been sought in events containing either pairs of same-sign leptons or three leptons~\cite{Aad:2016tuk,Aaboud:2017dmy}. The same-sign/three-leptons search is comparable in sensitivity to the search presented in this paper only when the masses of the gluino and the LSP are of similar magnitude. Sensitivity to such scenarios is also obtained by searching for events with large jet multiplicity, $\njet\geq \text{7--12}$~\cite{SUSY-2018-17}. Similar searches for pair-produced gluinos by the CMS experiment with 36--137~fb$^{-1}$ of 13~\TeV\ collisions~\cite{CMS-SUS-19-005,CMS-SUS-19-006,CMS-SUS-19-007,CMS-SUS-19-008,CMS-SUS-19-010,CMS-SUS-16-036,CMS-SUS-16-050} produced results comparable to the previous ATLAS results.
 
This paper is organised as follows. Section~\ref{sec:detector} describes the ATLAS detector, and Section~\ref{sec:samples} describes the data and simulated event samples used in the analysis. Section~\ref{sec:object} introduces the event reconstruction methodology, and Section~\ref{sec:strategy} introduces the analysis strategy. The event selection is discussed in Section~\ref{sec:evsel}, and systematic uncertainties in Section~\ref{sec:syst}. The results of the analysis are presented and interpreted in Section~\ref{sec:results}. Section~\ref{sec:conclusion} gives the conclusions.
 
\begin{figure*}[h]
\centering
\subfigure[]{\includegraphics[width=0.27\textwidth]{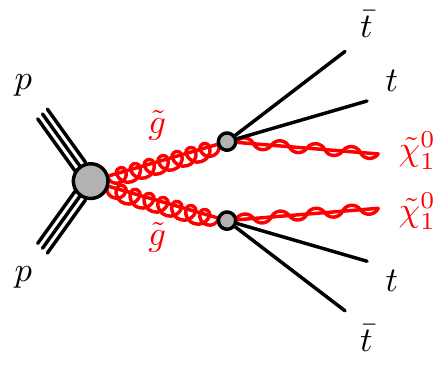}}
\subfigure[]{\includegraphics[width=0.27\textwidth]{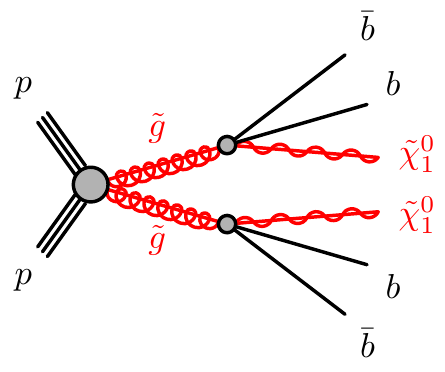}}
\caption{The decay processes in the (a) Gtt and (b) Gbb simplified models.
}\label{fig:feynman-Gluino-topbottom-offshell}
\end{figure*}

\begin{figure*}[h]
\centering
\subfigure[]{\includegraphics[width=0.27\textwidth]{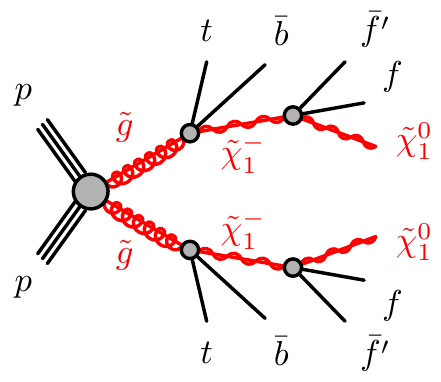}}
\subfigure[]{\includegraphics[width=0.27\textwidth]{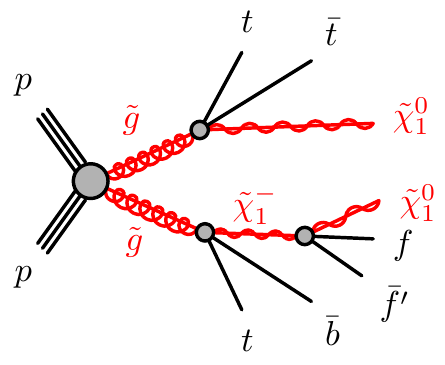}}
\subfigure[]{\includegraphics[width=0.27\textwidth]{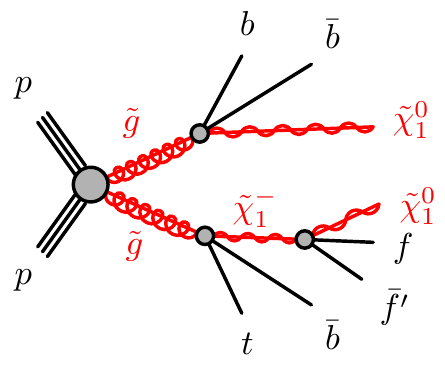}}
\subfigure[]{\includegraphics[width=0.27\textwidth]{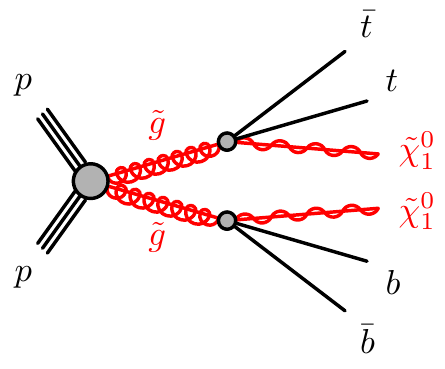}}
\caption{The additional decay processes permitted by the variable gluino branching ratio (Gtb) model, in addition to those shown in Figure~1. In diagram~(a), both gluinos decay via $\tilde{g} \to t \bar{b} \chinoonem$ with $\chinoonem \to f\bar{f}' \ninoone$. In diagrams~(b) and (c), only one gluino decays via the $\chinoonem$ while the other decays via $\tilde{g} \to t\bar{t}\ninoone$ or $\tilde{g} \to b\bar{b}\ninoone$, respectively. In diagram~(d), one gluino decays via $\tilde{g} \to t\bar{t}\ninoone$ and the other via $\tilde{g} \to b\bar{b}\ninoone$.  In each case the charge conjugate processes are implied. The fermions originating from the $\chinoonepm$ decay have low momentum and are not detected because the mass difference  between the $\chinoonepm$ and the $\ninoone$ is fixed to 2~\GeV.
}\label{fig:feynman-Gluino-topbottom-branchingratio}
\end{figure*}


\clearpage
 
\section{ATLAS detector}
\label{sec:detector}

The ATLAS detector is a multipurpose particle physics detector with a forward--backward
symmetric cylindrical geometry and nearly 4$\pi$ coverage in solid angle.\footnote{ATLAS
uses a right-handed coordinate system with its origin at the nominal interaction point in
the centre of the detector. The positive $x$-axis is defined by the direction from the
interaction point to the centre of the LHC ring, with the positive $y$-axis pointing
upwards, while the beam direction defines the $z$-axis. Cylindrical coordinates $(r,\phi)$
are used in the transverse plane, $\phi$ being the azimuthal angle around the $z$-axis.
The pseudorapidity $\eta$ is defined in terms of the polar angle $\theta$ by $\eta=-\ln\tan(\theta/2)$.
Rapidity is defined as $y = 0.5 \ln[(E + p_z )/(E - p_z ) ]$  where $E$ denotes the energy and $p_z$ is the component of the momentum along the beam direction.}
The inner tracking detector (ID) consists of silicon pixel and microstrip detectors
covering the pseudorapidity region $|\eta|<2.5$, surrounded by a transition radiation tracker,
which enhances electron identification in the region $|\eta|<2.0$.
The ID is surrounded by a thin superconducting solenoid providing an axial 2 T magnetic field and by
a fine-granularity lead/liquid-argon (LAr) electromagnetic calorimeter covering $|\eta|<3.2$.
A stainless-steel/scintillator tile calorimeter provides coverage for hadronic showers in
the central pseudorapidity range ($|\eta|<1.7$).
The endcaps ($1.5<|\eta|<3.2$) of the hadronic calorimeter are made of LAr active layers with copper as the absorber material.
The forward region ($3.1 < |\eta| < 4.9$) is instrumented with a LAr calorimeter for both the EM and hadronic measurements.
A muon spectrometer with an air-core toroidal magnet system surrounds the calorimeters.
Three layers of high-precision tracking chambers
provide coverage in the range $|\eta|<2.7$, while dedicated fast chambers allow triggering in the region $|\eta|<2.4$.
The ATLAS trigger system \cite{TRIG-2016-01} consists of a hardware-based level-1 trigger followed by a software-based high-level trigger (HLT).
In terms of software, an extensive suite~\cite{ATL-SOFT-PUB-2021-001} is used in data simulation,  in the reconstruction and analysis of real and simulated data, in detector operations, and in the trigger and data acquisition systems of the experiment.


\section{Data and simulated event samples}
\label{sec:samples}

The data analysed in this paper were collected between 2015 and 2018 at a centre-of-mass energy of 13~\TeV\ with a 25~ns proton bunch crossing interval. The average number of $pp$ interactions per bunch crossing (pile-up) ranged from 13 in 2015 to around 38 in 2017--2018. Application of beam, detector and data-quality criteria \cite{Aad:2700249} results in a total integrated luminosity of 139~fb$^{-1}$. The uncertainty in the combined 2015--2018 integrated luminosity is 1.7\% \cite{ATLAS-CONF-2019-021}, obtained using the LUCID-2 detector \cite{LUCID2} for the primary luminosity measurements and cross-checked by a suite of other systems.
 
Events are required to pass an $\met$ trigger~\cite{TRIG-2016-01,TRIG-2019-01} with thresholds of 70~\gev, 100~\gev\ and 110~\gev\ in the HLT for the 2015, early 2016 and late 2016\,/\,2017\,/\,2018 datasets, respectively. These triggers are fully efficient for events passing the preselection defined in Section~\ref{sec:presel}, which requires the offline reconstructed $\met$ to exceed 200~\GeV~\cite{TRIG-2019-01}.
 
Samples of Monte Carlo (MC) simulated events are used to model the signal and background processes in this analysis, except multijet processes, which are estimated using a data-driven method (Section~\ref{sec:strategy}). The MC simulation strategy is largely the same as that described in Ref.~\cite{SUSY-2016-10}, with one exception, which is described below. A summary of the generators used can be found in Table~\ref{t-MCsamples}.
 
\begin{table}[htb]
\caption{List of MC generators used to simulate different signal and background processes. The third column (`Tune') describes the tuned set of underlying event and hadronisation parameters, the fourth column the PDF set used, and the fifth column the perturbative accuracy in the strong coupling constant used for the calculation of the cross-section used to normalise the sample.}
\label{t-MCsamples}
\vspace{2mm}
\centering{\footnotesize
\begin{tabular}{ccccc}
\toprule
Process &    Generator                   & Tune & PDF set  &  Cross-section  \\
& + fragmentation/hadronisation  &      &          &  order \\
\midrule{\normalfont \bfseries Gbb/Gtb/Gtt} &  \MGNLO[2.2.2]~\cite{Alwall:2014hca}
& {\textsc A14}~\cite{A14tune} & {\NNPDF[2.3]}~\cite{Ball:2012cx} & NLO+NLL~\cite{Beenakker:1996ch,Kulesza:2008jb,Kulesza:2009kq,Beenakker:2009ha,Beenakker:2011fu,Borschensky:2014epjc} \\
& + \PYTHIA[8.186] & & & \\
\\
{\normalfont \bfseries \ttbar} & \POWHEGBOX[v2]
& {\textsc A14} & {\NNPDF[3.0]} & NNLO+NNLL~\cite{Czakon:2011xx} \\
& + \PYTHIA[8.230] & & & \\
\\
{\normalfont \bfseries Single top} & \POWHEGBOX[v2] & {\textsc A14} & {\NNPDF[3.0]} & NLO~\cite{Aliev:2010zk,Kant:2014oha}  \\
$s/t$-channel & + \PYTHIA[8.230]  & & & \\
\\
{\normalfont \bfseries Single top} & \POWHEGBOX[v2] & {\textsc A14} & {\NNPDF[3.0]} & approx. NNLO~\cite{Kidonakis:2010ux,Kidonakis:2011wy}  \\
$Wt$ & + \PYTHIA[8.230]  & & & \\
\\
{\normalfont \bfseries ${\ttbar}W$/${\ttbar}Z$} & \MGNLO[2.3.3]
& {\textsc A14} & {\NNPDF[3.0]} & NLO~\cite{Alwall:2014hca} \\
& + \PYTHIA[8.230] & & & \\
\\
{\normalfont \bfseries ${\ttbar}{\ttbar}$} & \MADGRAPH[2.3.3]
& {\textsc A14} & {\NNPDF[2.3]} & NLO~\cite{Alwall:2014hca} \\
& + \PYTHIA[8.230] & & & \\
\\
 
{\normalfont \bfseries ${\ttbar}H$} & \POWHEGBOX[v2] & {\textsc A14} & {\NNPDF[2.3]} & NLO~\cite{Alwall:2014hca} \\
& + \PYTHIA[8.230] & & & \\
 
\\
{\normalfont \bfseries Diboson}  & \SHERPA[2.2.1]~\cite{Gleisberg:2008ta} & Default & {\NNPDF[3.0]} & NLO~\cite{Gleisberg:2008ta,Hoeche:2012yf} \\
$WW$, $WZ$, $ZZ$ & & & & \\
\\
{\normalfont \bfseries $W$/$Z$+jets} & \SHERPA[2.2.1] & Default & {\NNPDF[3.0]} & NNLO~\cite{Catani:2009sm}  \\
\bottomrule
\end{tabular}
}
 
\end{table}
 
The only significant change with respect to Ref.~\cite{SUSY-2016-10} concerns the modelling of the dominant background in the signal regions -- namely the production of \ttbar\ pairs with additional high transverse momentum (\pt) jets. It was simulated using the \POWHEGBOX~\cite{Alioli:2010xd}~v2 event generator and the \NNPDF[3.0]~\cite{Ball:2014uwa} PDF set with $\alphas(m_Z^2)=0.118$. The parton shower, fragmentation, and the underlying event were simulated using \PYTHIA[8.230]~\cite{Sjostrand:2007gs}. The \hdamp parameter in \POWHEG, which controls the \pt of the first additional emission beyond the Born level and thus regulates the \pt of the recoil emission against the \ttbar\ system, was set to 1.5~times the mass of the top quark (assumed to be $m_\mathrm{top} =  172.5$~\gev)~\cite{ATL-PHYS-PUB-2016-020}.
 
Other changes in generator settings for the modelling of sources of minor backgrounds were found not to affect the sensitivity of this analysis significantly. All simulated background processes are normalised using the best available theoretical calculation for their respective cross-sections.
 
The SUSY signal samples are normalised using the cross-section calculations at next-to-leading order (NLO) in the strong coupling constant, adding the resummation
of soft gluon emission at next-to-leading-logarithm (NLL) accuracy~\cite{Beenakker:1996ch,Kulesza:2008jb,Kulesza:2009kq,Beenakker:2009ha,Beenakker:2011fu}. The masses of the top and bottom squarks are assumed to be much greater than that of the gluino, such that gluinos decay according to 3-body phase-space. For the Gtb benchmark models, the mass difference between the $\chinoonepm$ and the $\ninoone$ is fixed to 2~\GeV. For each signal model, the nominal cross-section and its uncertainty are taken from an envelope of cross-section predictions using different parton shower models and factorisation and renormalisation scales, as described in Ref.~\cite{Borschensky:2014epjc}.

The \EVTGEN[1.6.0] program~\cite{Lange:2001uf} was used to describe the properties of the $b$- and $c$-hadron decays in the signal samples and in the background samples, except those produced with \SHERPA. For all samples the response of the detector to particles was modelled with the full ATLAS detector simulation~\cite{:2010wqa} based on \GEANT~\cite{Agostinelli:2002hh}.  All simulated events were overlaid with multiple $pp$ collisions simulated with \PYTHIA[8.186] using the A3 tune~\cite{A14tune} and the \NNPDF[2.3lo] PDF set. The MC samples were generated with variable levels of pile-up in the same and neighbouring bunch crossings, and were reweighted to match the distribution of the mean number of interactions observed in data in 2015--2018.


\section{Event reconstruction}
\label{sec:object}

Events are required to have a primary vertex~\cite{PERF-2015-01,ATL-PHYS-PUB-2015-026} reconstructed from at least two tracks~\cite{PERF-2015-08} with $\pT>500~\MeV$. Among the vertices found, the vertex with the largest summed $\pT^2$ of the associated tracks~\cite{ATL-PHYS-PUB-2015-026} is designated as the primary vertex.
 
Jets are built from topological clusters of energy in the calorimeter~\cite{PERF-2014-07}, calibrated to the electromagnetic scale, using the anti-$k_t$ algorithm~\cite{Cacciari:2008gp,Fastjet} with radius parameter $R=0.4$. Jet energy scale corrections, derived from MC simulation and data, are used to calibrate the average energies of jet candidates to the scale of their constituent particles~\cite{JETM-2018-05}. Remaining differences between data and simulated events are evaluated and corrected for using in situ techniques, which exploit the transverse momentum balance between a jet and a reference object such as a photon, $Z$ boson, or multijet system in data. After these calibrations, all jets in the event with $\pt > 30$~\GeV\ and $|\eta| < 2.8$ must satisfy a set of loose jet-quality requirements~\cite{Aad:2013zwa}. These requirements are designed to reject jets originating from sporadic bursts of detector noise, large coherent noise or isolated pathological cells in the calorimeter system, hardware issues, beam-induced background or cosmic-ray muons~\cite{ATLAS-CONF-2015-029}. If these jet requirements are not met, the event is discarded. If the event is retained, only the jets with $\pt > 30$~\GeV\ and $|\eta| < 2.8$ are considered by the analysis.  In addition, the `medium' working point of the track-based jet vertex tagger~\cite{PERF-2014-03,ATLAS-CONF-2014-018} must be satisfied for jets with $\pT<120$~\GeV\ and $|\eta|<2.5$, to reject jets that originate from pile-up interactions.
 
Jets which contain $b$-hadrons and are within the inner-detector acceptance ($|\eta|<2.5$) are identified and `$b$-tagged' using a multivariate algorithm (`MV2c10') that exploits the impact parameters\footnote{The transverse impact parameter, $d_{0}$, is defined as the distance of closest approach of a track to the beam-line, measured in the transverse plane. The longitudinal impact parameter, $z_{0}$, corresponds to the $z$-coordinate distance between the point at which $d_{0}$ is measured and the primary vertex.} of the charged-particle tracks, the presence of secondary vertices, and the reconstructed flight paths of $b$- and $c$-hadrons inside the jet~\cite{Aad:2019aic}. The output of the multivariate algorithm is a single $b$-tagging score, which signifies the likelihood of a jet to contain $b$-hadrons. For the chosen selection working point applied to this score, the average identification efficiency for jets containing $b$-hadrons is $77\%$, determined with simulated \ttbar\ events. Using the same simulated sample, a rejection factor of approximately 110 (5) is obtained for jets initiated by light quarks and gluons (charm quarks). Differences in efficiency and mis-tag rate between data and MC simulation are taken into account with correction factors as described in Ref.~\cite{Aad:2019aic}.
 
Electron candidates are reconstructed from clusters of energy deposits in the electromagnetic calorimeter that are matched to a track in the inner detector~\cite{Aad:2019tso}. They are required to have $|\eta|<2.47$ and $\pT>20$~\GeV, and must meet `Loose' likelihood-based identification criteria~\cite{Aad:2019tso}. The impact parameter along the beam direction is required to satisfy $|\zzsth| < 0.5$~mm. The electromagnetic shower of an electron can also be reconstructed as a jet such that a procedure (`overlap removal') is required to resolve this ambiguity. In the case where the separation $\dRy$ ($\dRy \equiv \sqrt{(\Delta y)^2+(\Delta \phi)^2}$, for rapidity $y$) between an electron candidate and a non-$b$-tagged ($b$-tagged) jet is $\dRy< 0.2$, the candidate is considered to be an electron ($b$-tagged jet). This procedure uses a $b$-tagged jet definition that is looser than that described earlier, to avoid selecting electrons from heavy-flavour hadron decays. If the separation between an electron candidate and any jet satisfies $0.2<\dRy < 0.4$, the candidate is considered to be a jet, and the electron candidate is removed.
 
Muons are reconstructed by matching tracks in the inner detector to tracks in the muon spectrometer and must have $|\eta|<2.5$ and $\pT>20$~\GeV~\cite{MUON-2018-03}. The impact parameter along the beam direction is required to satisfy $|\zzsth| < 0.5$~mm. Events containing muons identified as originating from cosmic rays, with $|d_0| > 0.2$~mm and $|z_0| > 1$~mm, or as being poorly reconstructed, with $\sigma(q/p)/|(q/p)| > 0.2$, are removed. Here, $\sigma(q/p)/|(q/p)|$ is a measure of the momentum uncertainty for a particle with charge $q$. Muons are discarded if they are within $\dR = 0.4$ of jets that survive the electron--jet overlap removal, except when the number of tracks associated with the jet is less than three, where the muon is kept and the jet discarded. This last exception retains muons which have undergone bremsstrahlung.
 
After resolving the overlap with leptons, the candidate $R=0.4$ jets are reclustered~\cite{Nachman:2014kla} into larger-radius jets using the anti-$k_{t}$ algorithm with a radius parameter of 0.8.  The calibration from the input $R=0.4$ jets propagates directly to the reclustered jets.  These reclustered jets are then trimmed~\cite{Krohn:2009th,PERF-2012-02,Nachman:2014kla, ATLAS-CONF-2017-064} by removing subjets with \pt below 10\% of the \pt\ of the original reclustered jet. The resulting larger-radius jets are required to have $\pt>100$~\gev\ and $|\eta|<2.0$. No additional overlap removal procedure is applied to such jets after reclustering. When it is not explicitly stated otherwise, in this paper the term `jets' refers to the smaller-radius $R=0.4$ jets, while the reclustered larger-radius $R=0.8$ jets are called `large-$R$ jets'.
 
The requirements on electrons and muons are tightened for the selection of events in signal regions and background control regions requiring at least one electron or muon (Section~\ref{sec:strategy}) . The electrons and muons passing the tight selection are called `signal' electrons or muons in the following, as opposed to `baseline' electrons and muons, which need only pass the requirements described above. Signal electrons (muons) must satisfy the `Fix (Loose)'~\cite{Aad:2019tso,Aaboud:2019ynx} (`FixedCutTightTrackOnly'~\cite{PERF-2015-10,MUON-2018-03}) $\pt$-dependent track-based and calorimeter-based isolation criteria. The calorimeter-based isolation is determined by taking the ratio of the sum of energy deposits in a cone of $\Delta R=0.2$ around the electron or muon candidate to the sum of energy deposits associated with the electron or muon. The track-based isolation is estimated in a similar way but using a variable cone size with a maximum value of $\Delta R=0.2$ for electrons and $\Delta R=0.3$ for muons. Signal electrons are also required to pass a `Tight' likelihood-based selection~\cite{Aad:2019tso,Aaboud:2019ynx}. The impact parameter of the electron in the transverse plane is required to be less than five times the transverse impact parameter uncertainty ($\sigma_{d_0}$). Further selection criteria are also imposed on signal muons: muon candidates are required to pass a `Medium' quality selection and meet a $|d_0| < 3\sigma_{d_0}$ requirement~\cite{PERF-2015-10,MUON-2018-03}.
 
The missing transverse momentum $\vec{p}_{\mathrm{T}}^{\mathrm{miss}}$, with magnitude $\met$, is defined as the negative vector sum of the \pt of all selected and calibrated electrons, muons and jets in the event, with an extra term added to account for energy in the event that is not associated with any of these objects~\cite{ATLAS-CONF-2018-023}. This last `soft term' contribution is calculated from ID tracks with $\pT>500~\MeV$ which are matched to the primary vertex, thus ensuring that this contribution is robust against pile-up contamination, and which are not associated with selected objects~\cite{PERF-2016-07,ATLAS-CONF-2018-023}.
 
This analysis does not consider the contribution of reconstructed photons or hadronically decaying $\tau$-leptons when considering possible overlaps with other objects. They are also not included explicitly in the calculation of $\vec{p}_{\mathrm{T}}^{\mathrm{miss}}$, but the associated energy deposits contribute to this calculation via the overlapping reconstructed jets.


\section{Analysis strategy}
\label{sec:strategy}

Evidence for the presence of SUSY signal events in the data sample is sought by selecting events populating multiple signal regions (SRs), which are expected to be enriched in such events on the basis of simulation studies. The expected yields of major SM background processes in the SRs are determined with a profile likelihood fit (Section~\ref{sec:results}) using MC samples with normalisations constrained to data in dedicated SR-dependent control regions (CRs). The yields of subdominant backgrounds are estimated directly from MC samples, except in the case of multijet backgrounds, where a data-driven procedure is used. The accuracy of the background estimation procedure is verified by comparing data with background predictions in validation regions (VRs) with low signal contamination, which are located between the CRs and SRs in the multidimensional space of event selection variables.
 
Signal regions are defined using two alternative methodologies. The first methodology, the `cut-and-count' (CC) analysis, defines SRs by applying selections independently to a series of observables sensitive to differences in kinematics between signal and background (`discriminating variables'). These are expected to provide rejection of SM background events while retaining events from a broad range of Gtt, Gbb and Gtb signal models to provide maximum discovery power. This methodology is well suited to subsequent reinterpretation of the results in the context of other theories not considered in this paper. The CC SR event selection criteria fall into two broad categories targeting final states which contain no leptons or at least one lepton (referred to as `0-lepton' and `1-lepton' SRs henceforth). The second event selection methodology, the neural network (NN) analysis, classifies events using a supervised machine-learning technique in which correlations between discriminating variables are further exploited to maximise exclusion power for specific Gtt and Gbb signal models. The event selection for each NN SR optimally selects a mixture of events containing different lepton multiplicities, depending on the values of the other discriminating variables. A set of CRs and VRs is associated with each CC or NN SR. The event selection criteria are discussed in Section~\ref{sec:evsel}.
 
\subsection{Discriminating variables}
 
The following discriminating variables are used in the CC and NN SR, CR and VR event selections, in addition to more widely used variables such as $\met$ and the momenta and multiplicities of jets, $b$-jets and leptons.
 
The `inclusive effective mass' (\meff), is defined by the following scalar sum:
\begin{equation*}
\meff = \sum_{i} \pt^{\mathrm{jet}_{i}}  + \sum_{j} \pt^{\ell_j}  + \met,
\end{equation*}
where the first and second sums run over the selected jets ($\njet$) and signal leptons ($N_\mathrm{lepton}$), respectively. This variable is correlated with the invariant mass scale of the final-state particles in the event, and typically takes a higher value for pair-produced gluino events than for SM background events with lower mass scales.
 
In regions with at least one selected lepton, the transverse mass $\mt$ is calculated from the $\pt$ of the leading selected lepton ($\ell$) and $\met$, and is defined as:
\begin{equation*}
\mt = \sqrt{2\pt^\ell \met\{1-\cos[\Delta\phi(\vec{p}_{\mathrm{T}}^{\mathrm{miss}},\vec{p}_{\mathrm{T}}^{\ell})]\}}.
\end{equation*}
This variable is used to reject $\ttbar$ and $W$+jets background events in which one $W$ boson decays leptonically. The $\mt$ distribution for these backgrounds, assuming on-mass-shell $W$ bosons, has an upper bound corresponding to the $W$ boson mass, leading to a Jacobian edge in the observed distribution, and typically has higher values for Gtt signal model events. In addition, the minimum transverse mass formed by $\met$ and any of the three highest-\pt $b$-tagged jets in the event, $\mtb$, is used in regions with any lepton multiplicity:
\begin{equation*}
\mtb =  \mathrm{min}_{i\leq 3}  \left ( \sqrt{2\pt^{b\textrm{-jet}_i}
\met\{1-\cos[\Delta\phi(\vec{p}_{\mathrm{T}}^{\mathrm{miss}},\vec{p}_{\mathrm{T}}^{b\textrm{-jet}_i} )]\}} \right ).
\end{equation*}
The  $\mtb$ distribution has an upper bound corresponding to $\sqrt{m_{\mathrm{top}}^2 - m_W^2}$ for $\ttbar$ events with a single leptonic $W$ boson decay, while extending to higher values for signal events.
 
Another powerful variable for selecting signal events is the total jet mass variable, defined as:
\begin{equation*}
\mjsum = \sum_{i\leq 4} m_{J,i},
\end{equation*}
where $m_{J,i}$ is the mass of large-$R$ jet $i$ in the event. The decay products of a high-$\pt$ (boosted) hadronically decaying top quark can be reconstructed in a single large-$R$ jet, resulting in a jet with a high mass. This variable typically takes larger values for Gtt signal model events than for background events, because the former can contain as many as four hadronically decaying top quarks while the latter typically contain a maximum of two.
 
The requirement of a selected lepton, with the additional requirements on jets, $\met$ and event variables described above, makes the multijet background negligible for the $\ge$ 1-lepton signal regions. For the 0-lepton signal regions, the distribution of the minimum azimuthal angle $\dphimin$ between $\vec{p}_{\mathrm{T}}^{\mathrm{miss}}$ and the $\vec{p}_{\mathrm{T}}$ of the four leading $R=0.4$ jets in the event, defined as:
\begin{equation*}
\dphimin = \textrm{min}_{i\leq 4} \left(|\phi_{\mathrm{jet}_i} - \phi_{\vec{p}_{\mathrm{T}}^{\mathrm{miss}}}| \right),
\end{equation*}
peaks near zero for multijet background events in which large values of $\met$ have been generated by poorly measured jets or by neutrinos emitted close to the axis of a jet. The distributions of $\dphimin$ for signal events and other backgrounds are more uniformly distributed, reflecting the reduced correlation between the missing transverse momentum direction and the transverse momenta of the leading jets.
 
\subsection{Kinematic reweighting of MC samples}
\label{sec:reweight}
 
\begin{table}[htbp]
\caption{Definitions of the reweighting regions used to derive the $m_\mathrm{eff}$ reweighting factors applied to the MC samples. The $N_\mathrm{lepton}$ requirements apply to signal leptons. The $Z$ and $VV$ RR uses a definition of $E_\mathrm{T}^\mathrm{miss}$ ($\hat{E}_\mathrm{T}^\mathrm{miss}$) that adds the lepton pair's transverse momentum to the missing transverse momentum, to simulate $Z\to\nu\nu$ events.}
\label{tab:rwcr}
\vspace{2mm}
\centering
\begin{tabular}{c c c c c}
\toprule
\multicolumn{5}{c}{Criteria common to all regions: $\njet \geq 4$, $E_\mathrm{T}^\mathrm{miss}$ or $\hat{E}_\mathrm{T}^\mathrm{miss}$ $\geq 200$~\GeV} \\ \midrule
Reweighting Region        & $N_\mathrm{lepton}$ & $\nbjet$ & $\mtb$ [\GeV] & $m_{\ell\ell}$ [\GeV]  \\ \midrule
\ttbar     & $=1$                & $=2$   & $\leq 350$ & -                \\
Single top, $\ttbar+W/Z/H$, ${\ttbar}{\ttbar}$   & $=1$                & $=2$   & $> 350$    & -                \\
$W$+jets            & $=1$                & $=0$   & -            & -                \\
$Z$+jets, $VV$             & $=2$ opposite charge          & $=0$   & -            & $\in$ [60,120] \\
\bottomrule
\end{tabular}
\end{table}
 
In signal-depleted regions requiring the presence of exactly one lepton with loose event selections, discrepancies are observed in the shapes of $\pt$-related observables, such as \meff, \mjsum and \met, between data and the MC background expectations. Similar discrepancies are also observed in other similar analyses, e.g. Ref.~\cite{ATL-PHYS-PUB-2016-020,HIGG-2020-23}. No similar discrepancies are visible in the 0-lepton regions, or for events with $\geq 2$ leptons. To reduce these discrepancies, a kinematic reweighting procedure is applied to simulated background events containing at least one signal lepton, prior to their use in the main analysis fits. Dedicated reweighting regions (RRs) with loose event selection criteria are defined for the \ttbar process, the single-top, $\ttbar+W/Z/H$ and ${\ttbar}{\ttbar}$ processes, the $W$+jets process, and the $Z$+jets and electroweak diboson processes, as set out in Table~\ref{tab:rwcr}. Requirements on the number of reconstructed $b$-jets ($\nbjet$) are applied to ensure that the RRs are orthogonal to all analysis signal regions, which include a $\nbjet\geq3$ requirement.  Simulated events for these processes are first normalised to the total number of observed events in the respective RR. The ratio of data events to normalised MC events is then computed as a function of $\meff$, for exclusive bins of $\njet=4,5,6\text{ and\,}\ge 7$ (for the \ttbar and $W$+jets RRs) or for $\njet\ge4$ (for the single-top, $\ttbar+W/Z/H$ and ${\ttbar}{\ttbar}$ RR, and the $Z$+jets and electroweak diboson RR).  This ratio is found to be well-fitted with a decreasing exponential function of $\meff$ for each $\njet$ bin. The resulting fitted functions are then used to reweight MC simulated events with exactly one lepton. The reweighting factors typically take values between $\sim$1.17 and $\sim$0.19 for $\meff$ ranging from threshold to 4200~\GeV\ for the \ttbar and $W$+jets processes, and between $\sim$1.7 and $\sim$0.43 for the same range of values of $\meff$ for the single-top, $\ttbar+W/Z/H$ and ${\ttbar}{\ttbar}$ processes, and the $Z$+jets and electroweak diboson processes. The reweighting procedure reduces the discrepancies between data and MC background expectations in the validation regions used in the analysis (Section~\ref{sec:fitresults}).
 
\subsection{Background estimation}
\label{sec:bckgrndest}
 
The dominant background process in most signal regions is the production of a \ttbar pair in association with heavy- and light-flavour jets. In both the CC and NN SRs the dominant contribution to this background arises from events in which exactly one of the top quarks decays via a $W$ boson to a lepton and a neutrino. In selected background events containing no leptons, the lepton is outside the acceptance of the analysis or is a hadronically decaying $\tau$-lepton.  A normalisation factor for the \ttbar background is extracted for each CC or NN SR from a CR which is defined for a similar but orthogonal region of kinematic phase space. The normalisation factors are derived by dividing the data event yields in the CRs by the equivalent MC predictions, and then applied as factors multiplying the event yields in the SRs predicted by MC simulation. This procedure is equivalent to propagating the CR event yields to the SRs by multiplying with transfer factors derived from MC simulation. Systematic uncertainties associated with the MC simulations used in this procedure are taken into account in the final fit (Section~\ref{sec:syst}).
 
In the CC analysis the CRs are defined with a lepton multiplicity requirement $N_\mathrm{lepton}=1$, for both 0-lepton and 1-lepton SRs, and a requirement on $\mt$ inverted with respect to that used in the 1-lepton SRs~\cite{SUSY-2016-10}. In the NN analysis they are defined with orthogonal selections on the neural network output. In each CR, \ttbar production is the dominant process and the contribution from rare background processes, such as $\ttbar+W/Z/H$, or $\ttbar \ttbar$, is low. At least 20 data events are also required in each $\ttbar$ CR to minimise the data statistical uncertainties of the normalisation factors. The CRs are required to possess a signal-to-background ratio which is expected to be less than 5\% for Gtt, Gbb and Gtb signal models near the expected 95\% CL exclusion contours of the analysis. The CRs are also found to possess less than 10\% potential contamination from signal models beyond the exclusion contours of previous analyses.
 
The normalised $\ttbar$ background estimates extracted from the CRs are cross-checked with VRs that share similar background compositions with the SRs but use an orthogonal event selection. In the CC analysis the VRs incorporate an inverted requirement on one of the SR observables: $\mjsum$, $\meff$ or $\met$. In the NN analysis the VRs apply an orthogonal requirement to the neural network output together with inverted requirements on $\mjsum$ and $\meff$. The signal-to-background ratio in the VRs is expected to be lower than 30\% for benchmark Gtt, Gbb and Gtb signal points near the expected 95\% CL exclusion contours of the analysis. The purity of the CRs and VRs in \ttbar events is expected to be higher than 50\% and 33\%, respectively.
 
The leading non-\ttbar\ backgrounds in this analysis consist of single-top, $W$+jets, $Z$+jets, $\ttbar+W/Z/H$, $t\bar{t}t\bar{t}$ and electroweak diboson ($VV$) events, which are estimated using simulated samples (Table~\ref{t-MCsamples}) normalised to theoretical cross-sections. There is one exception to this procedure -- the $Z$+jets process makes a significant contribution to the total background in the NN Gbb SRs and is therefore normalised, for these SRs only, with dedicated 2-lepton CRs (Section~\ref{subsec:NN}). Due to their relatively low selection efficiency, these CRs contain fewer data events than the \ttbar CRs described above.
 
The remaining multijet background in the 0-lepton CC regions, and in the NN regions, is estimated following the strategy of Ref.~\cite{EXOT-2016-13}. This method estimates the multijet background from a CR with the same requirements as the SR, but with a selection requiring $\dphimin< 0.1$ to enhance the yield of events in which the missing transverse momentum is correlated with the $\pt$ of a leading jet. The yield is extrapolated from the multijet CRs to their corresponding SRs with exponential functions. The decay parameters of these functions are fixed by fits to the $\dphimin$ distribution of events passing the $\met$ trigger and a $\met>200$~\GeV\ requirement. The multijet background prediction is validated by comparing the data with the total prediction in the range $0.1<\dphimin< 0.2$. The contribution of multijet background events to the SRs is found to be $\lesssim 5\%$.


\section{Event selection}
\label{sec:evsel}

\subsection{Preselection}
\label{sec:presel}
Events used in the analysis are required to meet a set of loose preselection criteria. All events (except those used in the $Z$+jets CR described below) are required to possess at least four jets, of which at least three must be $b$-tagged, and $\met > 200$~\gev, which ensures that the efficiency of the $\met$ triggers used in this analysis is close to 100\% for selected events. Events selected in the CC 0-lepton regions are additionally required to contain no baseline leptons and possess $\dphimin>0.4$, while those selected in the CC 1-lepton regions are required to contain at least one signal lepton. No additional requirements on lepton multiplicity are applied at this stage in the NN analysis.
 
Figures~\ref{fig:presel_0l} and \ref{fig:presel_1l} show the multiplicities of selected jets and $b$-tagged jets, and the distributions of \met, \meff, \mtb, and \mjsum\ for events meeting the 0-lepton and 1-lepton preselection criteria, respectively. The reweighting described in Section~\ref{sec:reweight} is applied in the 1-lepton preselection to events with at least one lepton. The uncertainty bands depict the statistical and experimental systematic uncertainties, as described in Section~\ref{sec:syst}, but not the theoretical uncertainties in the background modelling. Distributions for example SUSY signal models are overlaid for comparison.
 
\begin{figure}[htbp]
\centering
\includegraphics[width=0.490\textwidth]{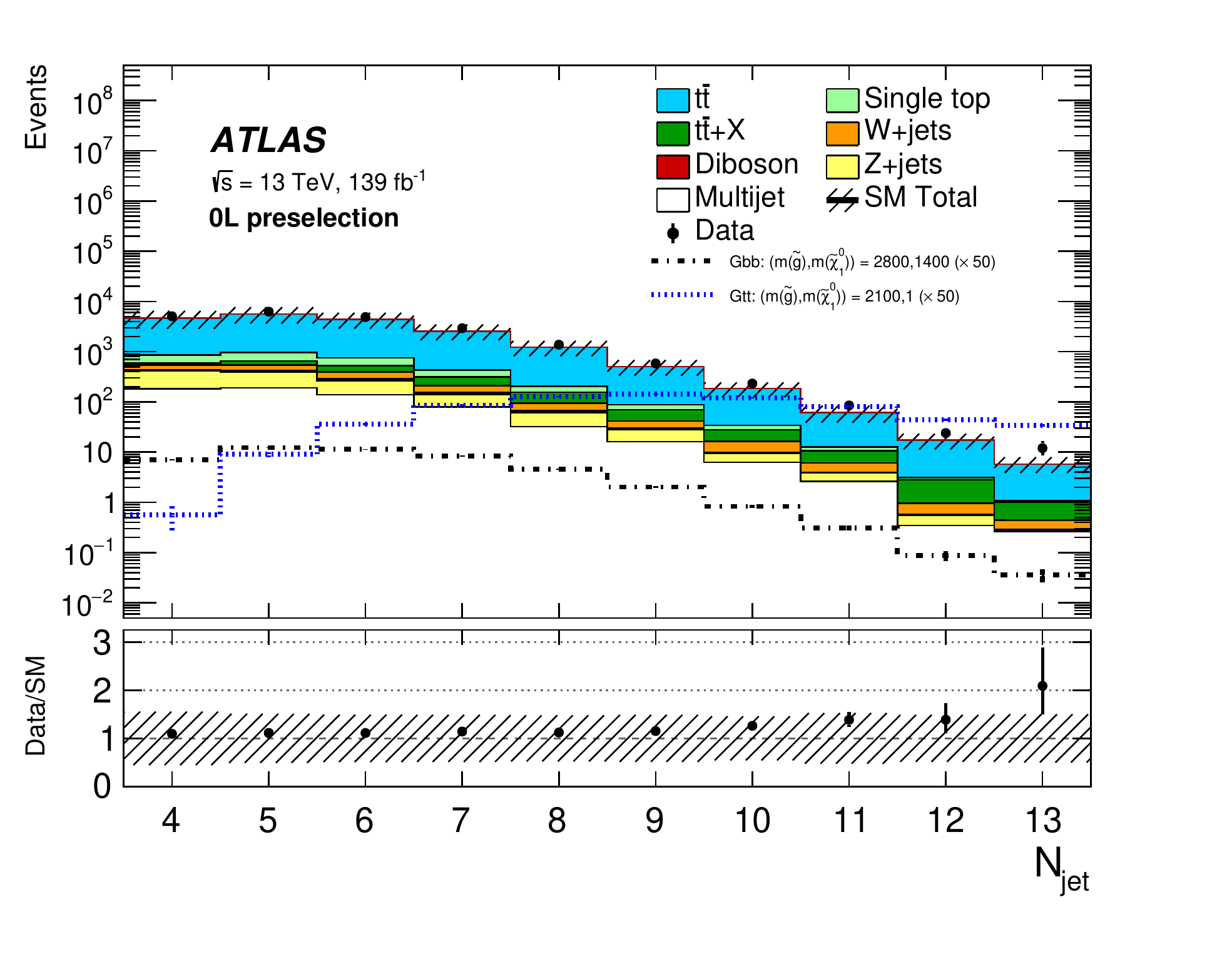}
\includegraphics[width=0.490\textwidth]{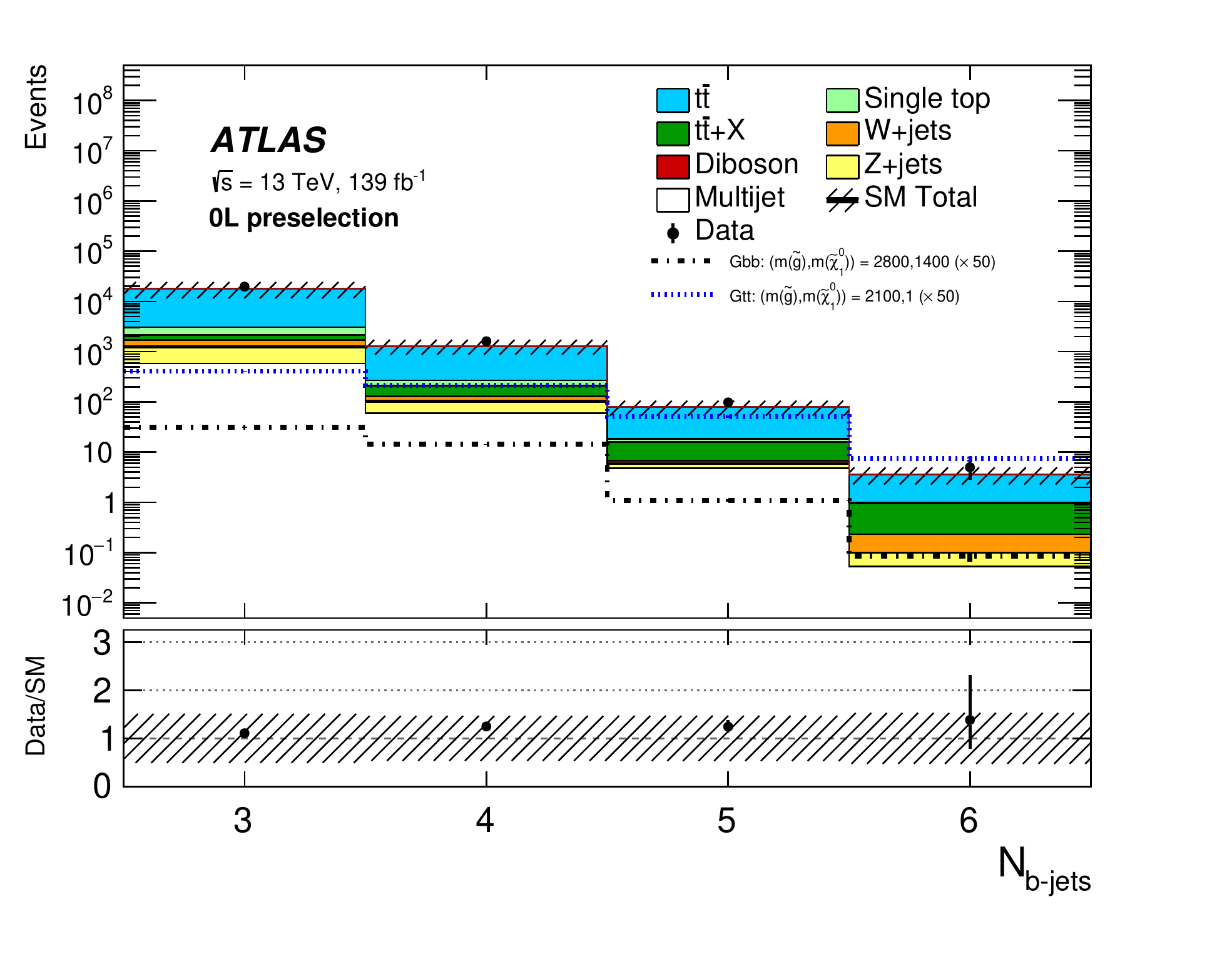}
\includegraphics[width=0.490\textwidth]{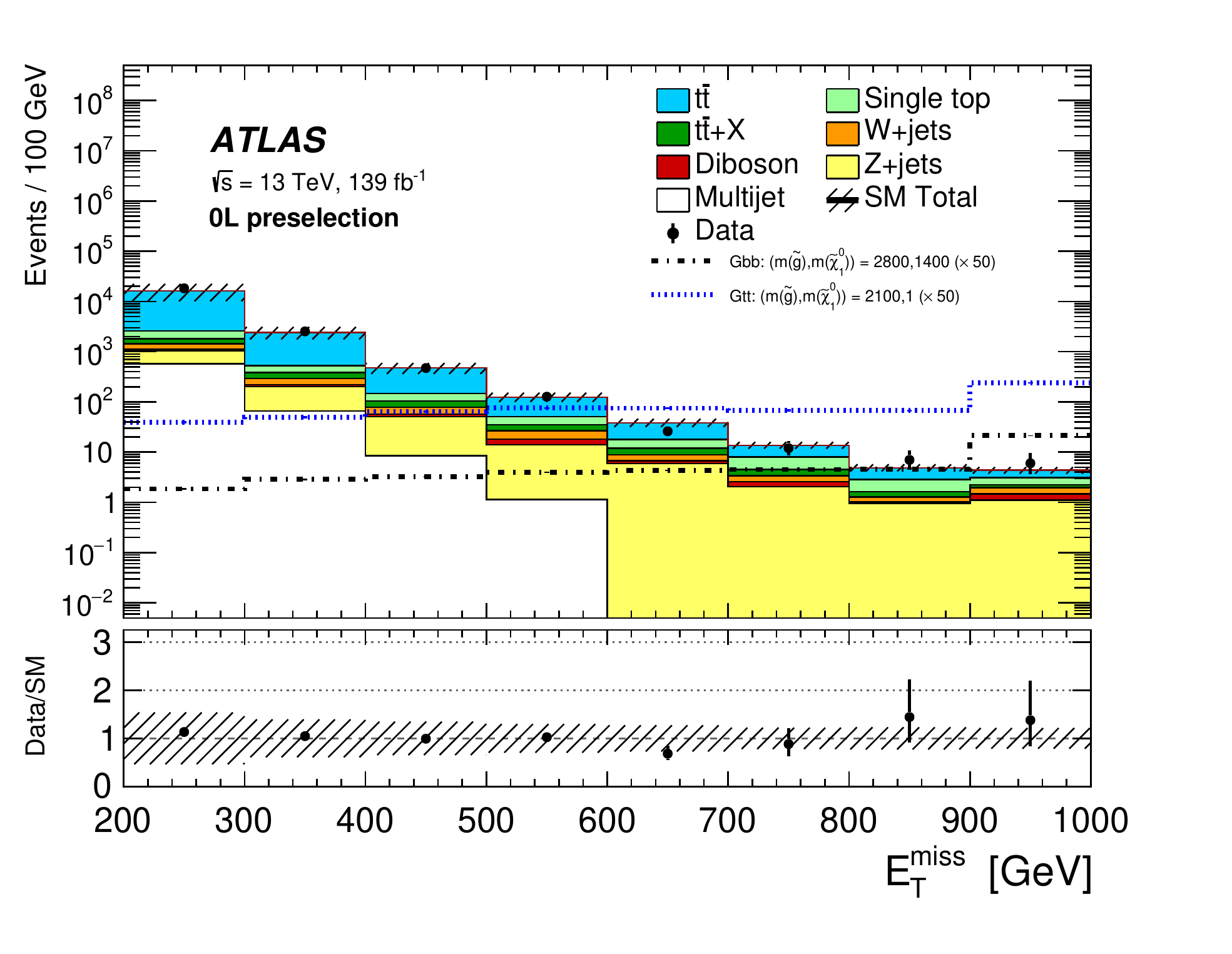}
\includegraphics[width=0.490\textwidth]{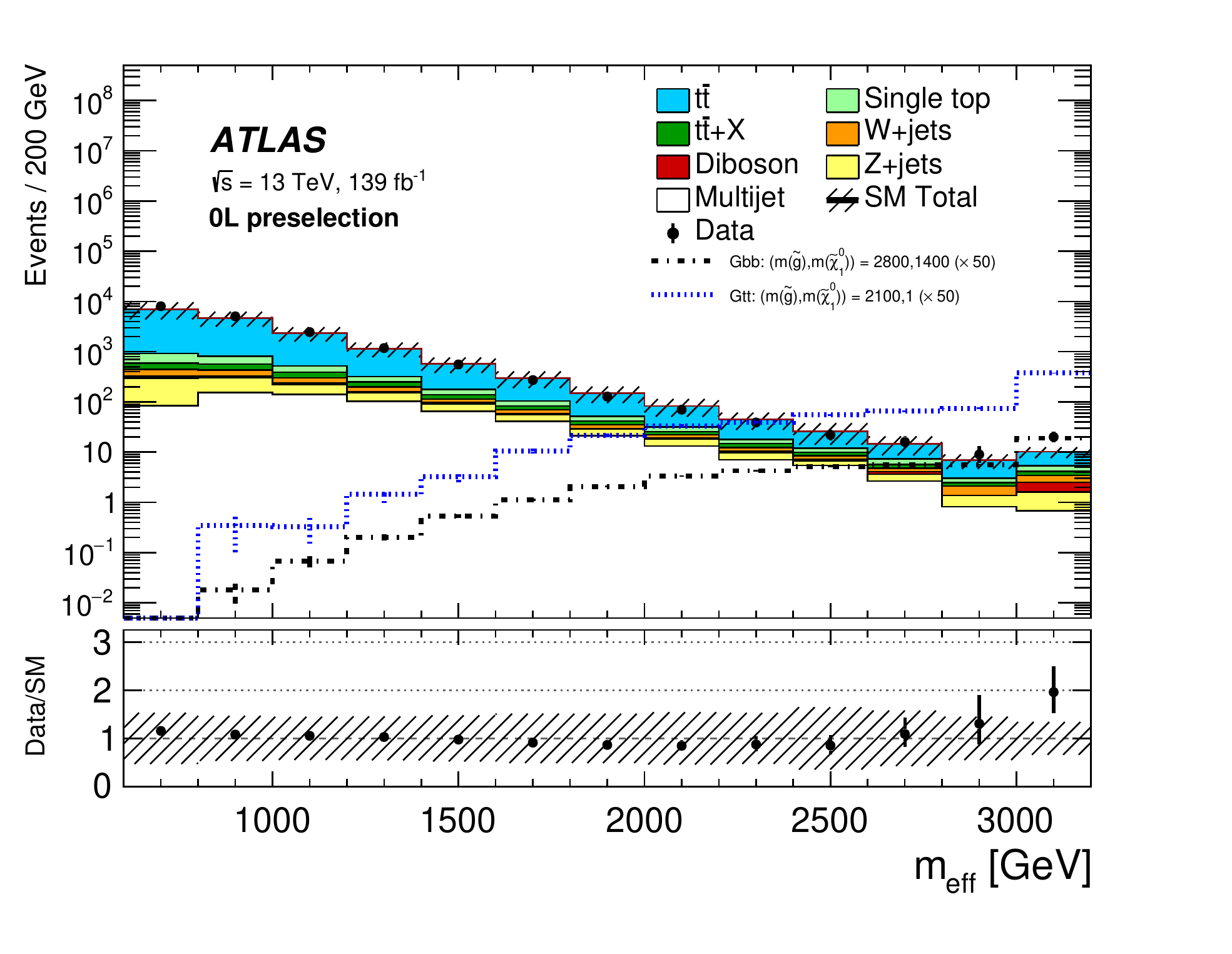}
\includegraphics[width=0.490\textwidth]{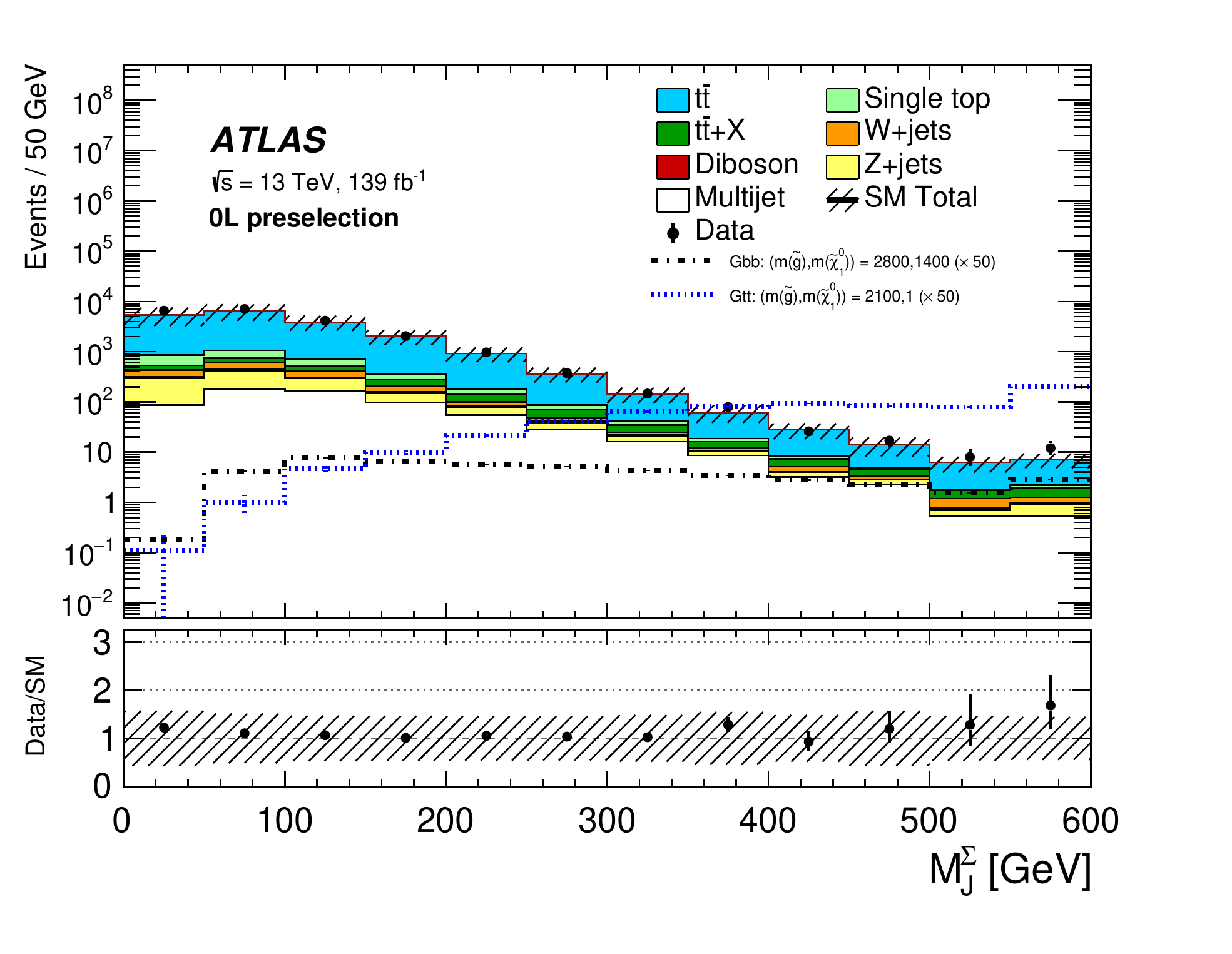}
\includegraphics[width=0.490\textwidth]{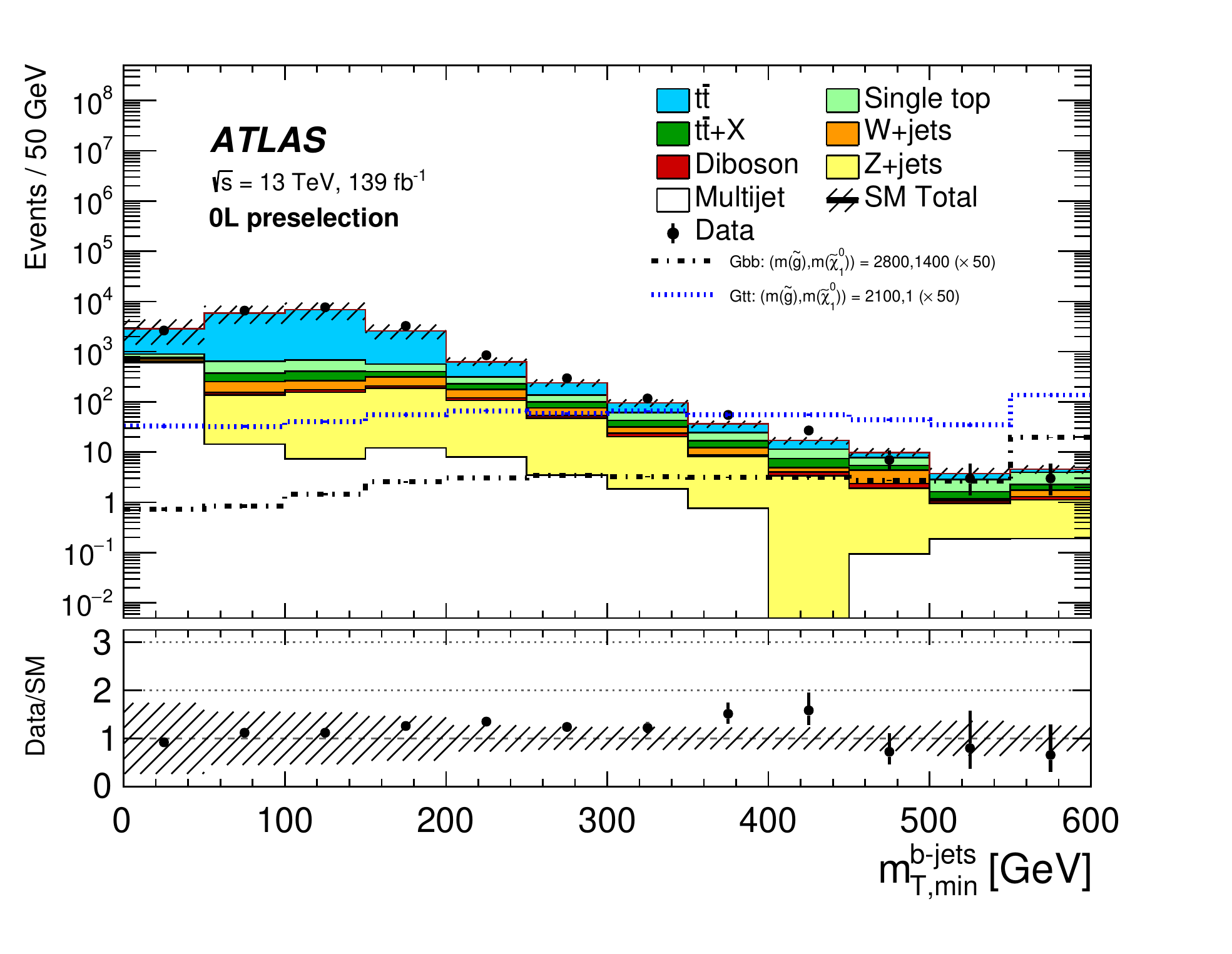}
\caption{Distributions of (top-left) the number of selected jets ($N_\mathrm{jet}$), (top-right) the number of selected $b$-tagged jets ($N_{b-\mathrm{jets}}$), (centre-left) $E_\mathrm{T}^\mathrm{miss}$, (centre-right) $m_\mathrm{eff}$, (bottom-left) $M_J^\Sigma$ and (bottom-right) $m_\mathrm{T, min}^{b-\mathrm{jets}}$ for events meeting the 0-lepton preselection criteria. The statistical and experimental systematic uncertainties (as defined in Section~\ref{sec:syst}) are included in the uncertainty bands. The final bin in each case includes overflow events. The lower panel of each figure shows the ratio of data to the background prediction. All backgrounds (including $t\bar{t}$) are normalised using the theoretical calculations described in Section~\ref{sec:samples}. The background category `$t\bar{t}+X$' includes $t\bar{t} W/Z$, $t\bar{t} H$ and $t\bar{t}t\bar{t}$ events. Distributions for example SUSY signal models, applying a normalisation scaling of 50, are overlaid for comparison.
}
\label{fig:presel_0l}
\end{figure}
 
\begin{figure}[htbp]
\centering
\includegraphics[width=0.490\textwidth]{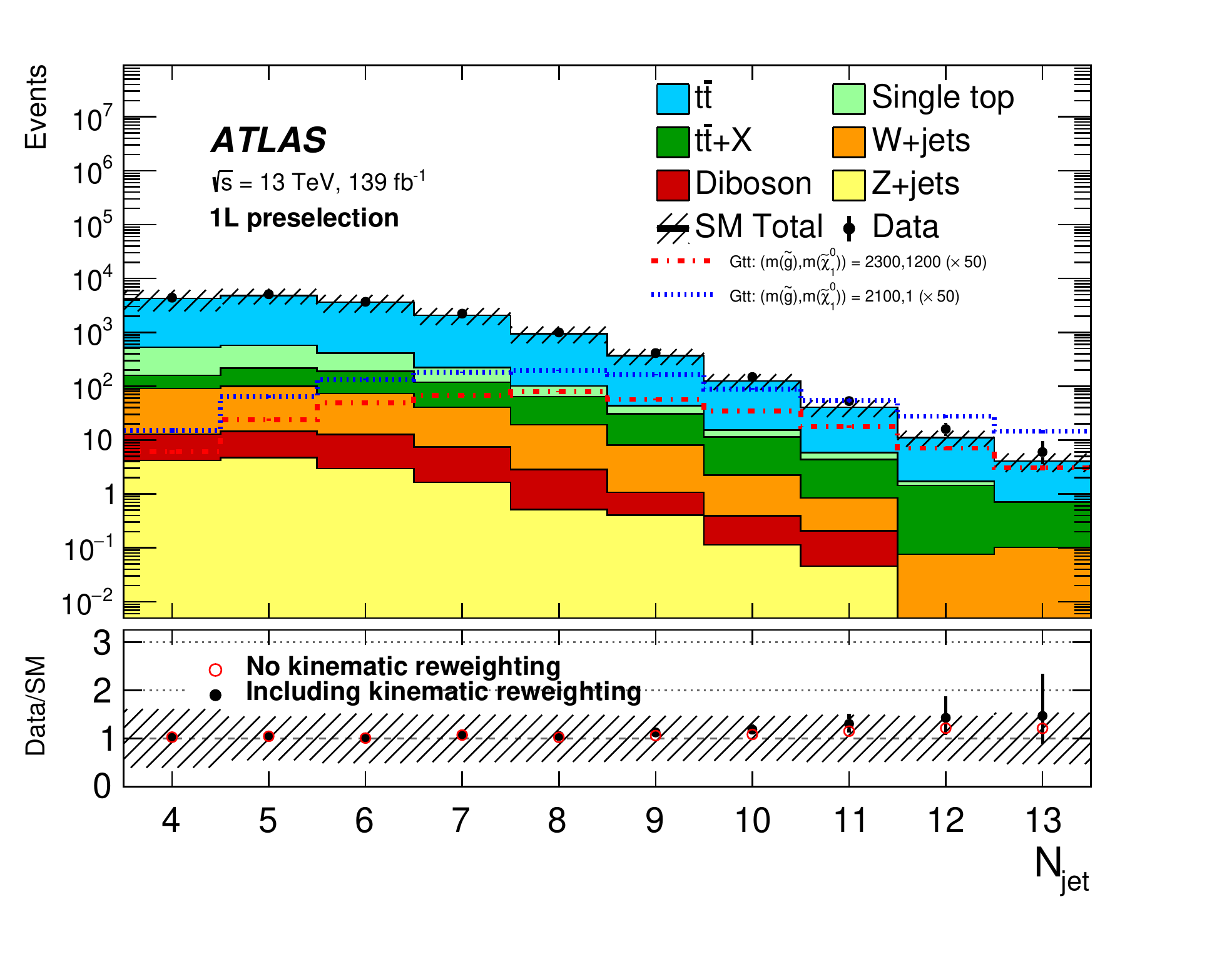}
\includegraphics[width=0.490\textwidth]{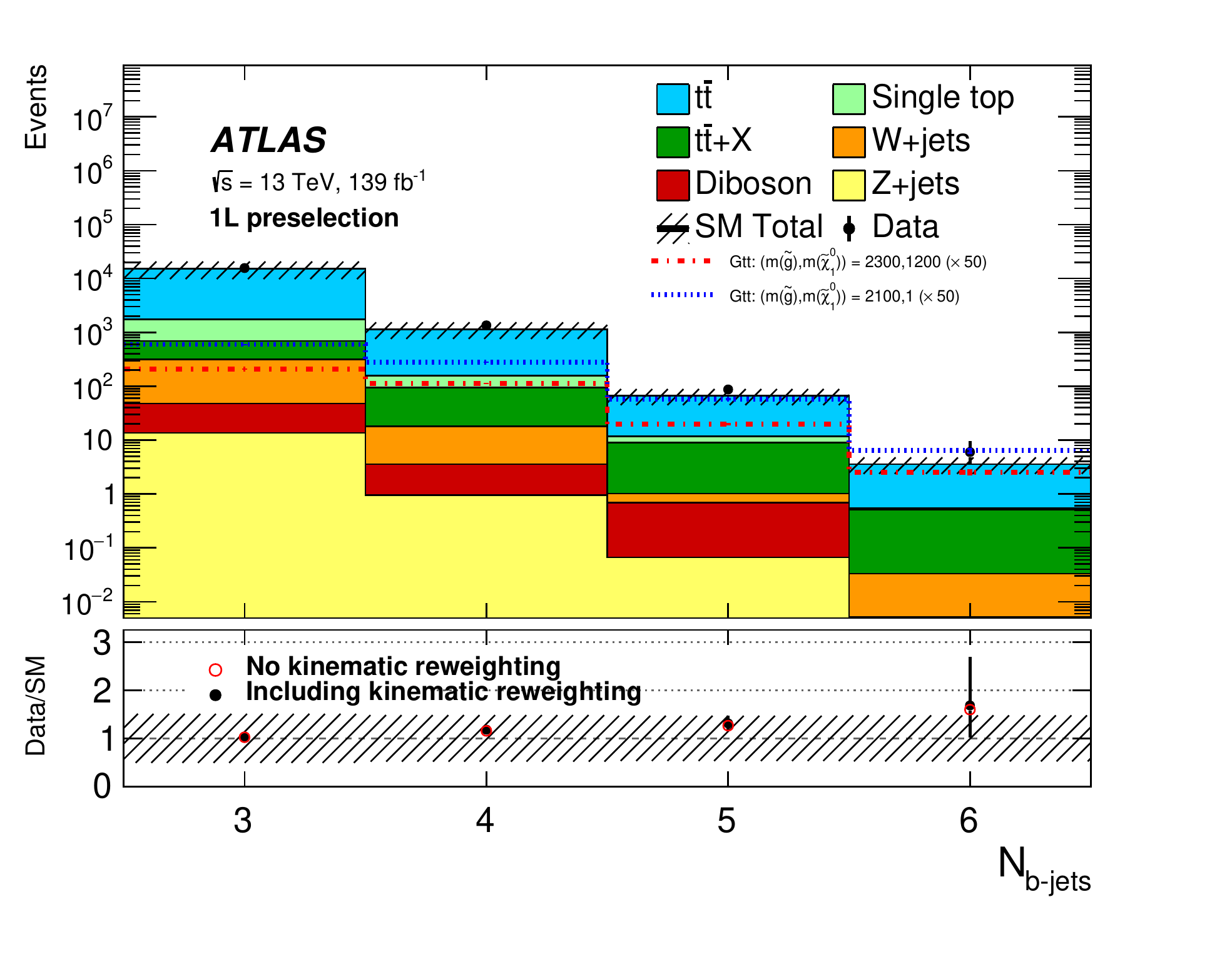}
\includegraphics[width=0.490\textwidth]{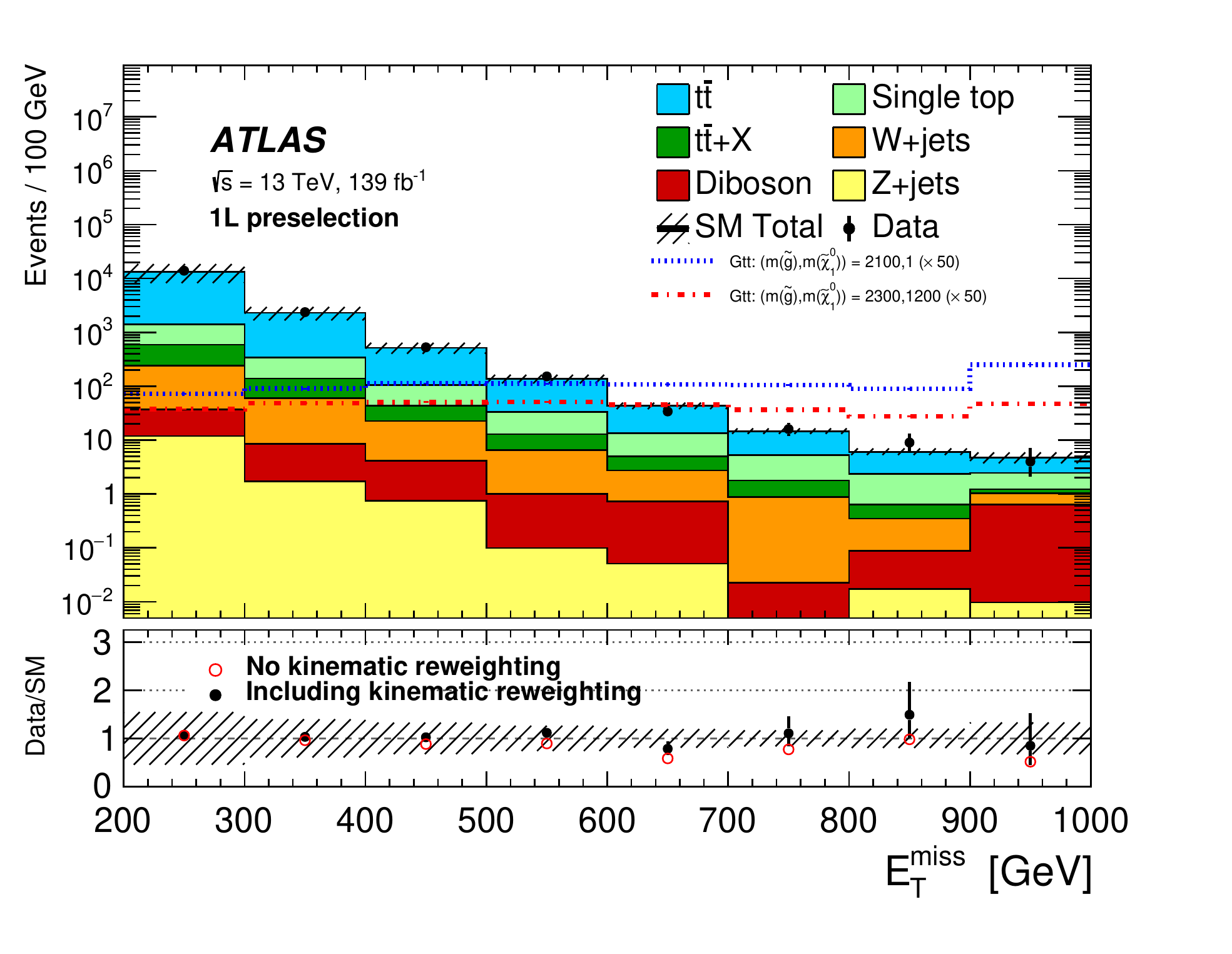}
\includegraphics[width=0.490\textwidth]{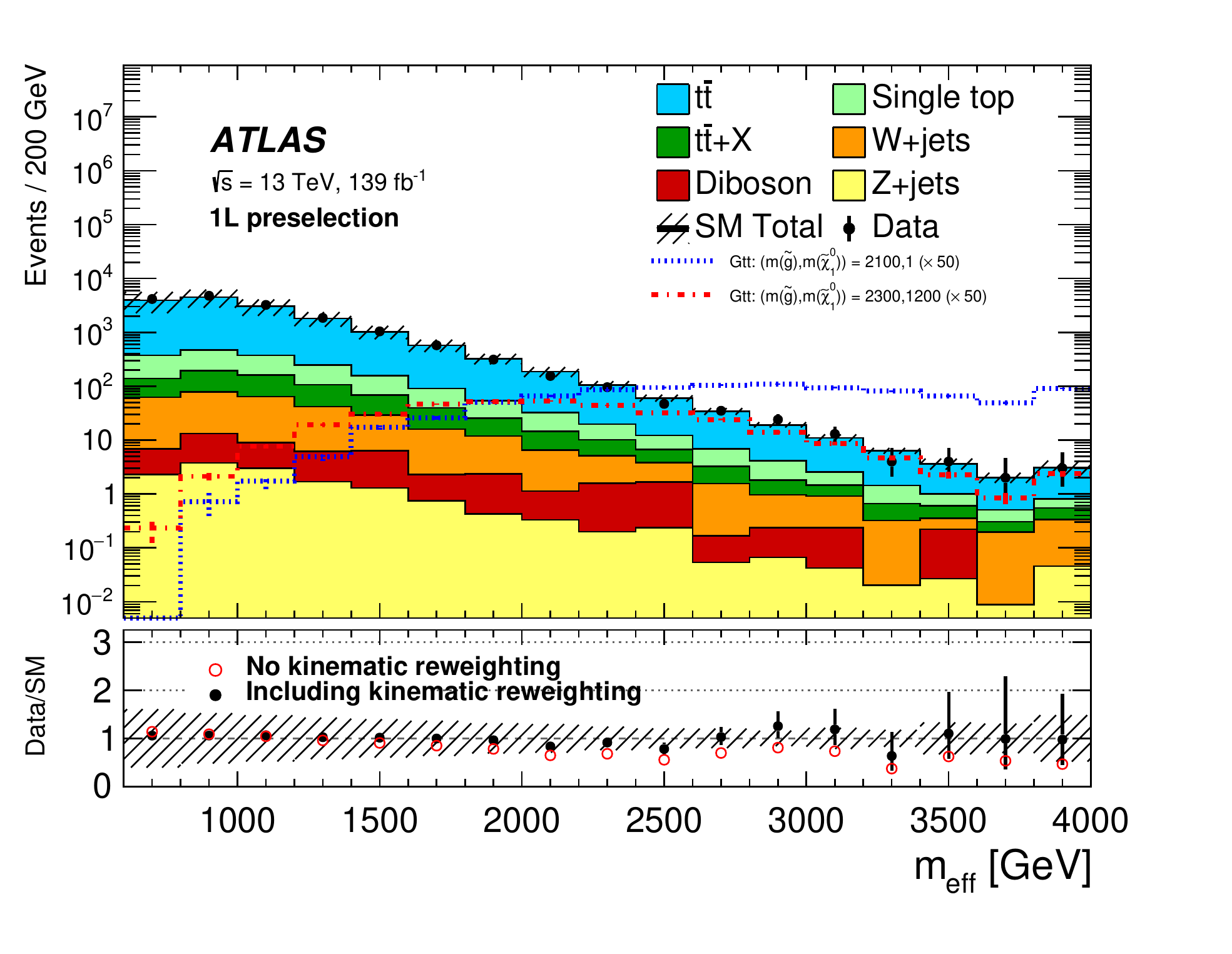}
\includegraphics[width=0.490\textwidth]{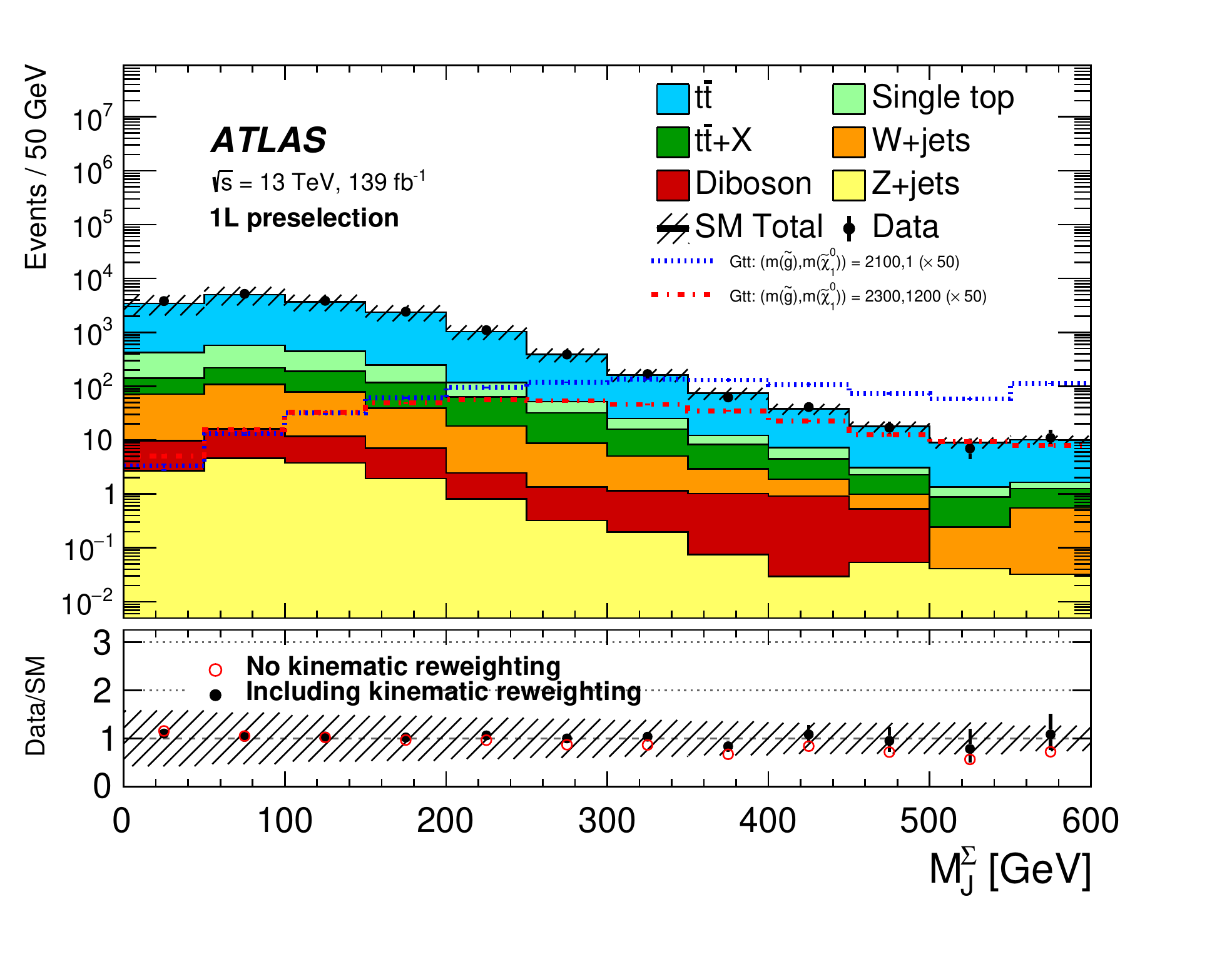}
\includegraphics[width=0.490\textwidth]{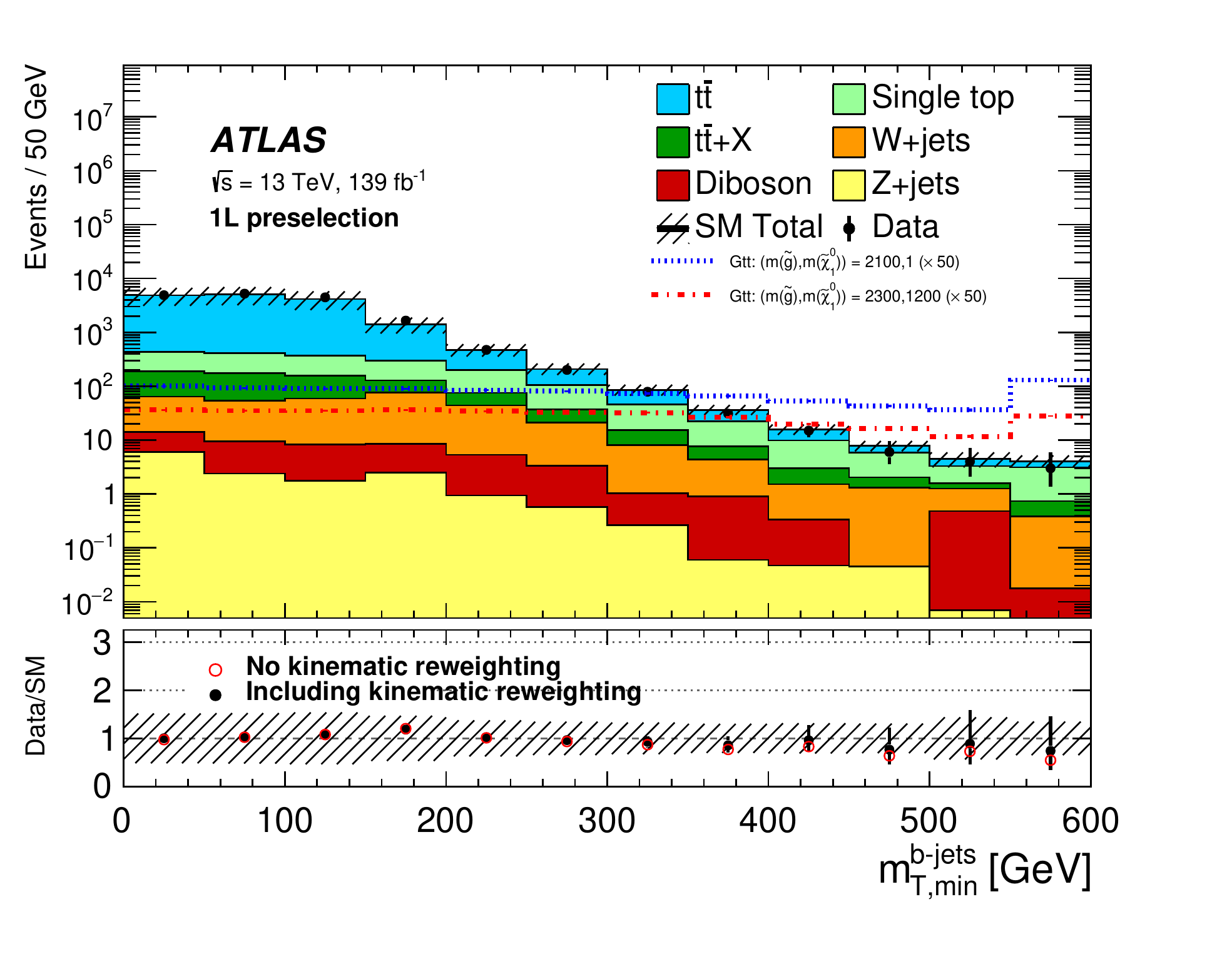}
\caption{Distributions of (top-left) the number of selected
jets ($N_\mathrm{jet}$), (top-right) the number of selected $b$-tagged jets ($N_{b-\mathrm{jets}}$), (centre-left) $E_\mathrm{T}^\mathrm{miss}$, (centre-right) $m_\mathrm{eff}$,
(bottom-left) $M_J^\Sigma$ and (bottom-right) $m_\mathrm{T, min}^{b-\mathrm{jets}}$ for events meeting the 1-lepton preselection criteria, after applying the
kinematic reweighting to the $m_\mathrm{eff}$ distribution as described in the text.
The statistical and experimental systematic uncertainties (as defined in
Section~\ref{sec:syst}) are included in the
uncertainty bands. The final bin in each case includes overflow events.
The lower panel of each figure shows the ratio of data to the background prediction before (red empty circles) and after (black filled circles with error bars) the kinematic reweighting. All backgrounds (including  $t\bar{t}$) are
normalised using the theoretical calculations described in Section~\ref{sec:samples}.
The background category `$t\bar{t}+X$' includes $t\bar{t} W/Z$, $t\bar{t} H$ and $t\bar{t}t\bar{t}$ events.
Distributions for example SUSY signal models, applying a normalisation scaling of 50, are overlaid for comparison.}
\label{fig:presel_1l}
\end{figure}

\subsection{Cut-and-count analysis}
The cut-and-count analysis employs a set of overlapping and not statistically independent single-bin SRs. The event selection criteria for the CC SRs, CRs and VRs are listed in Tables~\ref{tab:CCEvselGtt0lep}--\ref{tab:CCEvselGbbGtb0lep}. The CC SR event selection criteria are optimised to maximise the expected significance of Gtt, Gbb and Gtb models close to the 95\% CL exclusion contours in the $\mglu$--$\mchi$ mass plane set by the previous ATLAS search in this channel using a smaller dataset~\cite{SUSY-2016-10}. Separate SRs are defined for each of these three classes of models. The SRs are further categorised according to whether they target models with large $\gluino$--$\ninoone$ mass splitting ($\msplit = \mglu - \mchi \gtrsim 1.5$~\tev, Regions B), moderate mass splitting (0.3~\tev~$\lsim \msplit \lsim 1.5$~\tev, Regions M1 and M2) or small mass splitting ($\msplit\lsim 0.3$~\tev, Regions C). These regions differ mainly in the selections applied to the $\meff$, $\met$, $\mt$ and $\mjsum$ variables.
 
For each SR, a CR is defined to constrain the \ttbar background (Section~\ref{sec:bckgrndest}). The CRs for the 0-lepton SRs require the presence of exactly one signal lepton. The 1-lepton SRs and the associated CRs require at least one signal lepton, with the latter also implementing an inverted requirement on $\mt$. The CRs also place looser requirements on jet multiplicity and the other discriminating variables.
 
The VRs for the 0-lepton SRs use inverted requirements on $\mjsum$, $\meff$ or $\met$ to remove overlap with the respective SRs. For the 1-lepton SRs, two VRs are defined to validate the background prediction in high-$\mtb$ and high-$\mt$ regions by increasing the threshold for $\njet$ and/or inverting the $\mjsum$ or $\mt$ requirements to remove overlap with both the corresponding SR and the CR.
 
\begin{table}[t]
\centering
\small
\renewcommand{\arraystretch}{1.5}
\caption{Event selection requirements for the CC Gtt 0-lepton SRs together with the associated $t\bar{t}$ CRs and VRs, classified according to the  $\tilde{g}$--$\ninoone$ mass splitting ($\Delta m$) targeted. The thresholds in bold for each control and validation region ensure orthogonality with the corresponding signal region. $N_\mathrm{lepton}=0$ requires zero baseline leptons, while $N_\mathrm{lepton}=1$ requires one signal lepton.
}
\label{tab:CCEvselGtt0lep}
\vspace{2mm}
\resizebox{\textwidth}{!}{
\begin{tabular}{c c c c c c c c c c c}
\toprule
Targeted kinematics & Type & $N_\mathrm{lepton}$ & $\njet$ & $\nbjet$ & $\met$ [\GeV] & $\dphimin$ & $\meff$ [\GeV] & $\mt$ [\GeV] & $\mtb$ [\GeV] & $\mjsum$ [\GeV]  \\ \midrule
\multirow{3}{*}{\begin{minipage}{3cm}\centering Region B\\
(Boosted, Large \msplit) \end{minipage}}
& SR & $=0$ & $\geq5$ & $\geq3$ & $\geq600$ & $\geq0.4$ & $\geq2900$ & $-$ & $\geq120$ & $\geq300$ \\
& CR & \textbf{$=1$} & $\geq4$ & $\geq3$ & $\geq200$ & $-$ & $\geq2000$ & $<150$ & $-$ & $\geq150$ \\ 
& VR & $=0$ & $\geq5$ & $\geq3$ & $\geq250$ & $\geq0.4$ & $\geq2000$ & $-$ & $-$ & \textbf{$<300$} \\ \midrule
\multirow{3}{*}{\begin{minipage}{3cm}\centering Region M1\\
(Moderate \msplit) \end{minipage}}
& SR & $=0$ & $\geq9$ & $\geq3$ & $\geq600$ & $\geq0.4$ & $\geq1700$ & $-$ & $\geq120$ & $\geq300$ \\
& CR & \textbf{$=1$} & $\geq8$ & $\geq3$ & $\geq200$ & $-$ & $\geq1100$ & $<150$ &  $-$ & $\geq150$ \\ 
& VR & $=0$ & $\geq9$ & $\geq3$ & $\geq300$ & $\geq0.4$ & $\geq1400$ & $-$ &  $-$ & \textbf{$<300$} \\ \midrule
\multirow{3}{*}{\begin{minipage}{3cm}\centering Region M2\\
(Moderate \msplit) \end{minipage}}
& SR & $=0$ & $\geq10$~~ & $\geq3$ & $\geq500$ & $\geq0.4$ & $\geq1100$ & $-$ & $\geq120$ & $\geq200$ \\
& CR & \textbf{$=1$} & $\geq9$ & $\geq3$ & $\geq200$ & $-$ & ~~$\geq800$ & $<150$ &  $-$ & $\geq100$ \\ 
& VR & $=0$ & $\geq10$~~ & $\geq3$ & $\geq300$ & $\geq0.4$ & ~~$\geq800$ &  $-$ & $-$ & \textbf{$<200$} \\ \midrule
\multirow{3}{*}{\begin{minipage}{3cm}\centering Region C\\
(Compressed, small \msplit) \end{minipage}}
& SR & $=0$ & $\geq10$~~ & $\geq4$ & $\geq400$ & $\geq0.4$ & ~~$\geq800$ & $-$ & $\geq180$ & $\geq100$ \\
& CR & \textbf{$=1$} & $\geq9$ & $\geq4$ & $\geq200$ & $-$ & ~~$\geq800$ & $<150$ & $-$ & $\geq100$ \\
& VR & $=0$ & $\geq10$~~ & $\geq4$ & $\geq200$ & $\geq0.4$ & ~~$\geq800$ & $-$ & $-$ & \textbf{$<100$} \\ 
\bottomrule
\end{tabular}
}
\end{table}

\begin{table}[t]
\centering
\small
\renewcommand{\arraystretch}{1.5}
\caption{Event selection requirements for the CC Gtt 1-lepton SRs together with the associated $t\bar{t}$ CRs and VRs, classified according to the  $\tilde{g}$--$\ninoone$ mass splitting ($\Delta m$) targeted. The thresholds in bold for each control and validation region ensure orthogonality with the corresponding signal region.
}
\vspace{2mm}
 
\resizebox{\textwidth}{!}{
\begin{tabular}{c c c c c c c c}
\toprule
Targeted kinematics & Type & $\njet$ & $\met$ [\GeV] & $\meff$ [\GeV] & $\mt$ [\GeV] & $\mtb$ [\GeV] & $\mjsum$ [\GeV] \\ \midrule
\multirow{4}{*}{\begin{minipage}{3cm}\centering Region B\\
(Boosted, Large \msplit) \end{minipage}}
& SR & $\geq4$ & $\geq600$ & $\geq2300$ & $\geq150$ & $\geq120$ & $\geq200$ \\
& CR & $=4$ & $\geq200$ & $\geq1500$ & \textbf{$<150$} & $-$ & $-$ \\ 
& VR1 & $\geq4$ & $\geq200$ & $\geq1500$ & $\geq150$ & $-$ & \textbf{$<200$} \\
& VR2 & $\geq 5$ & $\geq200$ & $\geq1200$ & \textbf{$<150$} & $\geq120$ & $\geq200$ \\\midrule
\multirow{4}{*}{\begin{minipage}{3cm}\centering Region M1\\
(Moderate \msplit) \end{minipage}}
& SR & $\geq5$ & $\geq600$ & $\geq2000$ & $\geq200$ & $\geq120$ & $\geq200$ \\
& CR & $=5$ & $\geq200$ & $\geq1200$ & \textbf{$<200$} & $-$ & $-$ \\ 
& VR1 & $\geq5$ & $\geq200$ & $\geq1200$ & $\geq200$ & $-$ & \textbf{$<200$} \\
& VR2 & $\geq 6$ & $\geq200$ & $\geq1000$ & \textbf{$<200$} & $\geq120$ & $\geq100$ \\\midrule
\multirow{4}{*}{\begin{minipage}{3cm}\centering Region M2\\
(Moderate \msplit) \end{minipage}}
& SR & $\geq8$ & $\geq500$ & $\geq1100$ & $\geq200$ & $\geq120$ & $\geq100$ \\
& CR & $=8$ & $\geq200$ & ~~$\geq800$ & \textbf{$<200$} & $-$ & $-$ \\ 
& VR1 & $\geq8$ & $\geq200$ & ~~$\geq800$ & $\geq200$ & $-$ & \textbf{$<100$} \\
& VR2 & $\geq 9$ & $\geq200$ & ~~$\geq800$ & \textbf{$<200$} & $\geq120$ & $\geq100$ \\\midrule
\multirow{4}{*}{\begin{minipage}{3cm}\centering Region C\\
(Compressed, small \msplit) \end{minipage}}
& SR & $\geq9$ & $\geq300$ & ~~$\geq800$ & $\geq150$ & $\geq120$ & $-$ \\
& CR & $=9$ & $\geq200$ & ~~$\geq800$ & \textbf{$<150$} & $-$ & $-$ \\
& VR1 & $\geq9$ & $\geq200$ & ~~$\geq800$ & $\geq150$ & \textbf{$<120$} & $-$ \\
& VR2 & $\geq 10$~~ & $\geq200$ & ~~$\geq800$ & \textbf{$<150$} & $\geq120$ & $-$ \\
\bottomrule
\end{tabular}
}
\end{table}
 
\begin{table}[t]
\centering
\small
\renewcommand{\arraystretch}{1.5}
\caption{Event selection requirements for the CC Gbb 0-lepton SRs together with the associated $t\bar{t}$ CRs and VRs, classified according to the  $\tilde{g}$--$\ninoone$ mass splitting ($\Delta m$) targeted. The thresholds in bold for each control and validation region ensure orthogonality with the corresponding signal region. $N_\mathrm{lepton}=0$ requires zero baseline leptons, while $N_\mathrm{lepton}=1$ requires one signal lepton.
}
\vspace{2mm}
\resizebox{\textwidth}{!}{
\begin{tabular}{c c c c c c c c c c}
\toprule
Targeted kinematics & Type & $N_\mathrm{lepton}$ & ${\pt}^\mathrm{jet}$ [\GeV] & $\meff$ [\GeV] & $\met$ [\GeV] & $\mtb$ [\GeV]& $\mt$ [\GeV] \\ \midrule
\multirow{3}{*}{\begin{minipage}{3cm}\centering Region B\\
(Boosted, Large \msplit) \end{minipage}}
& SR & $=0$& $> 65$ & $>2600$& $>550$& $>130$& &  \\
& CR & \textbf{$=1$}& $> 65$ & $>2600$& $>450$& & $<150$ & \\ 
& VR & $=0$& $> 65$ & \textbf{$<2400$}& $>550$& $>130$& & \\ \midrule
\multirow{3}{*}{\begin{minipage}{3cm}\centering Region M\\
(Moderate \msplit) \end{minipage}}
& SR & $=0$& $> 30$ & $>2000$& $>550$& $>130$& &  \\
& CR & \textbf{$=1$}& $> 30$ & $>2000$& $>550$& & $<150$ & \\ 
& VR & $=0$& $> 30$ & $>1600$& \textbf{$<500$}& ~~$>80$&& \\ \midrule
\multirow{3}{*}{\begin{minipage}{3cm}\centering Region C\\
(Compressed, small \msplit) \end{minipage}}
& SR & $=0$& $> 30$ & $>1600$& $>550$ & $>130$& &  \\
& CR & \textbf{$=1$}& $> 30$ & $>1600$& $>550$ & & $<150$ & \\
& VR & $=0$& $> 30$ & $>1500$& \textbf{$<450$} & $>130$& & \\
\bottomrule
\end{tabular}
}

\end{table}
 
\begin{table}[t]
\centering
\small
\renewcommand{\arraystretch}{1.5}
\caption{Event selection requirements for the CC Gtb 0-lepton SRs together with the associated $t\bar{t}$ CRs and VRs, classified according to the  $\tilde{g}$--$\ninoone$ mass splitting ($\Delta m$) targeted. The thresholds in bold for each control and validation region ensure orthogonality with the corresponding signal region. $N_\mathrm{lepton}=0$ requires zero baseline leptons, while $N_\mathrm{lepton}=1$ requires one signal lepton.
}
\label{tab:CCEvselGbbGtb0lep}
\vspace{2mm}
\resizebox{\textwidth}{!}{
\begin{tabular}{c c c c c c c c c c c}
\toprule
Targeted kinematics & Type & $N_\mathrm{lepton}$ & $\njet$ & $\nbjet$ & $\meff$ [\GeV]  & $\met$ [\GeV] & $\mtb$ [\GeV] &  $\mt$ [\GeV] & $\mjsum$ [\GeV] \\ \midrule
\multirow{3}{*}{\begin{minipage}{3cm}\centering Region B\\
(Boosted, Large \msplit) \end{minipage}}
& SR & $=0$& $\geq 4$& $\geq 3$& $>2500$& $>550$& $>130$& & $> 200$  \\
& CR & \textbf{$=1$}& $\geq 4$& $\geq 3$& $>2200$& $>400$& & $<150$& $> 200$  \\ 
& VR & $=0$& $\geq 4$& $\geq 3$& \textbf{$<2500$}& $>450$& $>130$& & $> 200$ \\ \midrule
\multirow{3}{*}{\begin{minipage}{3cm}\centering Region M\\
(Moderate \msplit) \end{minipage}}
& SR & $=0$& $\geq 6$& $\geq 4$& $>2000$& $>550$& $>130$& & $> 200$  \\
& CR & \textbf{$=1$}& $\geq 6$& $\geq 4$& $>1700$& $>300$& & $<150$& $> 200$  \\ 
& VR & $=0$& $\geq 6$& $\geq 4$& $>1600$& \textbf{$<550$}& $>110$& & $> 200$ \\ \midrule
\multirow{3}{*}{\begin{minipage}{3cm}\centering Region C\\
(Compressed, small \msplit) \end{minipage}}
& SR & $=0$& $\geq 7$& $\geq 4$& $>1300$& $>500$& $>130$& & ~~$> 50$ \\
& CR & \textbf{$=1$}& $\geq 7$& $\geq 4$& $>1300$& $>350$& & $<150$& ~~$> 50$ \\
& VR & $=0$& $\geq 7$& $\geq 4$& $>1300$& \textbf{$<500$}& ~~$>80$& & ~~$> 50$ \\
\bottomrule
\end{tabular}
}
 
\end{table}
 
\clearpage
 
\subsection{Neural network analysis}
\label{subsec:NN}
The NN analysis uses low-level kinematic variables~\cite{Baldi:2014kfa} as inputs to a dense neural network with three hidden layers trained to discriminate Gtt or Gbb model events from SM background events. Each SR is defined by event selection criteria applied to the outputs of the neural network, optimised to maximise the statistical significance of the SUSY model considered. The neural network input variables are:
\begin{itemize}
\item The four-momenta ($p_{\mathrm{T}}$, $\eta$, $\phi$, $m$) of the
10 leading jets, in decreasing order of \pt, and a set of binary variables indicating
which jets are $b$-tagged;
\item The four-momenta of the four leading large-$R$ jets, in decreasing order of \pt;
\item The four-momenta of the four leading leptons ($e$ or $\mu$), in decreasing order of \pt;
\item The two components of the vector $\vec{p}_{\mathrm{T}}^{\mathrm{miss}}$.
\end{itemize}
These four-momenta describe the expected final state of a typical Gtt signal event. If a given event contains fewer jets or leptons than specified above, the remaining inputs are set to zero. This procedure enables the analysis to optimally select a mixture of events with different jet and lepton multiplicities. The neural network generates three output scores measuring the probability of a given event being a signal event ($P$(Gtt) or $P$(Gbb), depending on the signal model targeted), a \ttbar background event ($P$(\ttbar)), or a $Z$+jets background event ($P$($Z$)). The output scores for all processes are normalised such that they sum to unity.
 
The neural network was trained using the Keras~\cite{chollet2015keras} library with the TensorFlow~\cite{tensorflow2015-whitepaper} backend. Training was performed once for each representative model. The neural network hyperparameters were optimised with a random search~\cite{bergstra2012random} followed by a line scan of the learning rate. The number of training epochs was not fixed but instead defined through an early-stopping strategy using the cross-entropy loss as a figure of merit. The training samples consisted of a total of $9.2\times 10^5$ signal events and $5.6\times 10^5$ background events.  Following event preselection, the training samples for all signal and background processes corresponded to the same integrated luminosity as that of the data sample used in the analysis.
 
To ensure optimum discrimination power across the Gtt and Gbb model planes, a parameterised training method~\cite{Baldi:2016fzo} was employed in which the neural network was also given the $(\mglu, \mchi)$ pair of the signal point under consideration as well as a binary variable indicating if discrimination of background versus Gtt or Gbb is required. Background events were assigned random parameter values. After training, the neural network outputs were evaluated by unconditionally setting the parameters to that of the model point under consideration and processing all MC and data events. To reduce the large number of potential SRs that could emerge from such a strategy, i.e.\ one SR per model point, a set-cover algorithm~\cite{10.2307/3689577} was used to iteratively select the SR which excludes the most as-yet non-excluded model points until all such points are exhausted. The result is a minimum set of SRs that are expected to exclude the same number of Gtt or Gbb models as a more extensive set with one SR for each model point. The resulting SRs are optimal for excluding models that are representative of regions of the Gtt or Gbb model parameter space. The resulting representative models are specified by the following $(\mglu, \mchi)$ mass values (in \GeV):
\begin{itemize}
\item \textbf{Gtt}: (2100, 1), (1800, 1), (2300, 1200), (1900, 1400)
\item \textbf{Gbb}: (2800, 1400), (2300, 1000), (2100, 1600), (2000, 1800)
\end{itemize}
There is one SR for each of these eight models. The acceptance times efficiency of the SR selections is typically 1\%--10\% for the representative models that they target. For example, the acceptance times efficiency of the SR-Gtt-2300-1200 selection for the Gtt (2300,~1200) representative model is 6.4\%.
 
The eight NN SRs are defined in Tables~\ref{tab:NN_Gtt} and~\ref{tab:NN_Gbb}, together with the associated CRs and VRs. For each SR, a CR is defined to constrain the \ttbar background (Section~\ref{sec:bckgrndest}). The CRs are defined by placing requirements on the neural network outputs orthogonal to those used in the SRs, and placing additional requirements on \meff and \mjsum to select events with kinematics similar to those in the corresponding SR. For the NN Gbb SRs, the \Zjets\ background process (principally $Z(\to\nu\nu)+$jets) also contributes significantly to the background yield. Dedicated control regions for this process (labelled `CRZ' in Table~\ref{tab:NN_Gbb_ZCR} and thereafter) are therefore defined for the Gbb SRs only. The CRs are used to normalise the \Zjets\ background estimates with the same procedure as is used for the \ttbar background estimates (Section~\ref{sec:bckgrndest}). These CRs (Table~\ref{tab:NN_Gbb_ZCR}) select $\Zboson(\to\ell\ell)+$jets events with a requirement of two opposite-sign same-flavour (OSSF) signal electrons/muons (both with $\pt>30$~\GeV) with an invariant mass in the range $60~\GeV < m_{\ell\ell} < 120$~\GeV\ and combined $\pt > 70$~\GeV. The events must pass the lowest-threshold unprescaled single-lepton triggers used in 2015--2018, which are well modelled by MC simulation and have approximately constant efficiency for leptons with offline $\pt> 27$~\GeV. The momenta of the leptons in selected events are added to $\vec{p}_{\mathrm{T}}^{\mathrm{miss}}$ and the leptons discarded, to mimic $Z(\to\nu\nu)+$jets events, and then the resulting modified value of $\met$ ($\hat{E}_\mathrm{T}^\mathrm{miss}$) is required to exceed 200~\GeV, to replicate the SR event preselection criterion. These events are then passed through the SR neural network and selections are applied to the neural network outputs that are orthogonal to the SR selection criteria.
 
VRs are defined for the NN SRs in a similar way to the CRs, with modified, orthogonal, selections on the neural network outputs and additional selections applied to high-level kinematic variables including $\dphimin$, $\meff$ and $\mjsum$. These VRs are defined in Tables~\ref{tab:NN_Gtt} and~\ref{tab:NN_Gbb} and their relationship to the SRs and \ttbar CRs is illustrated in Figure~\ref{fig:NN_scheme}. Two VRs are defined for each NN Gbb SR to validate both the $\ttbar$ and \Zjets\ background estimates.
 
\begin{figure}[htb]
\centering
\includegraphics[width=0.4\linewidth]{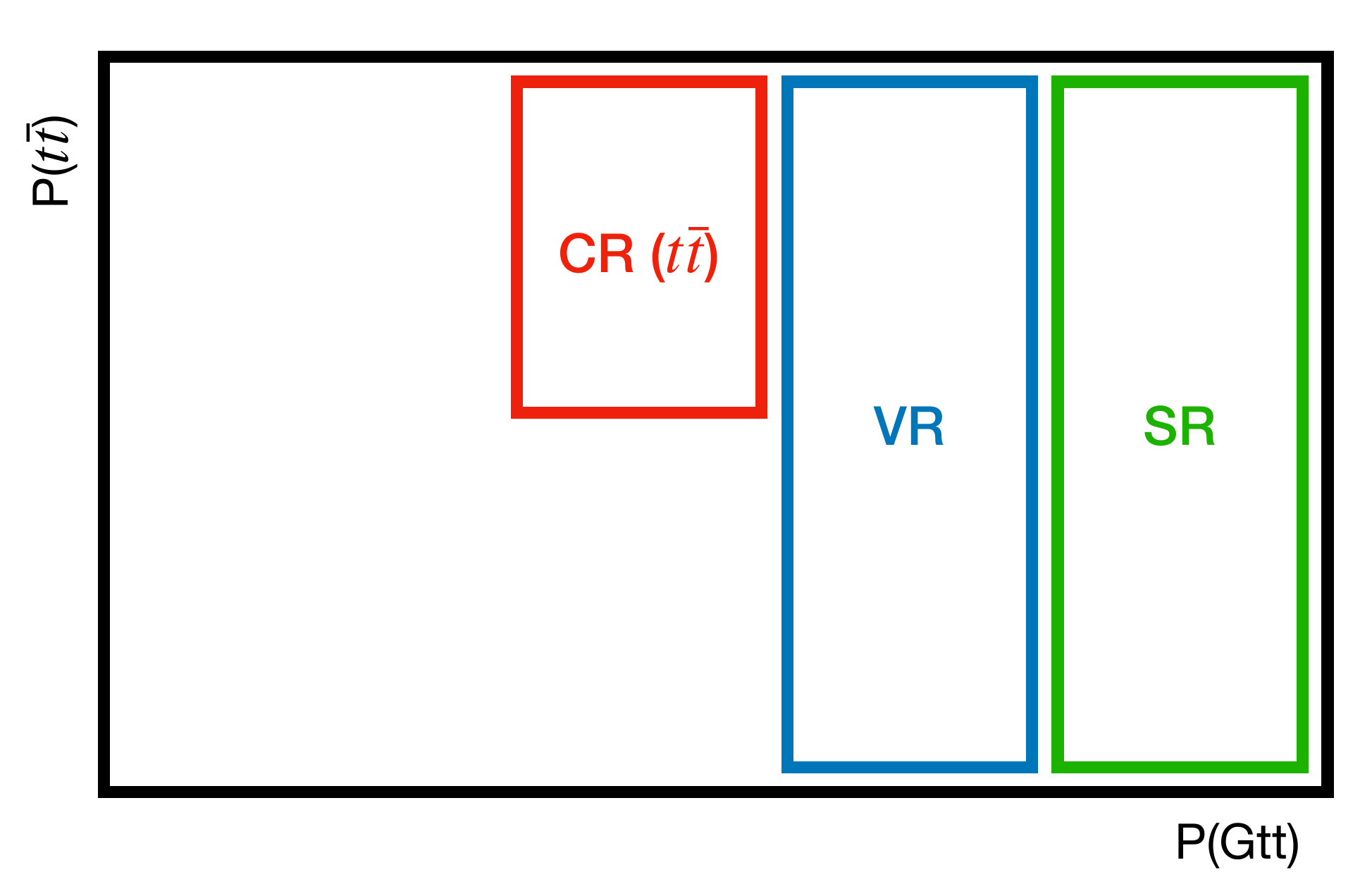}
\includegraphics[width=0.4\linewidth]{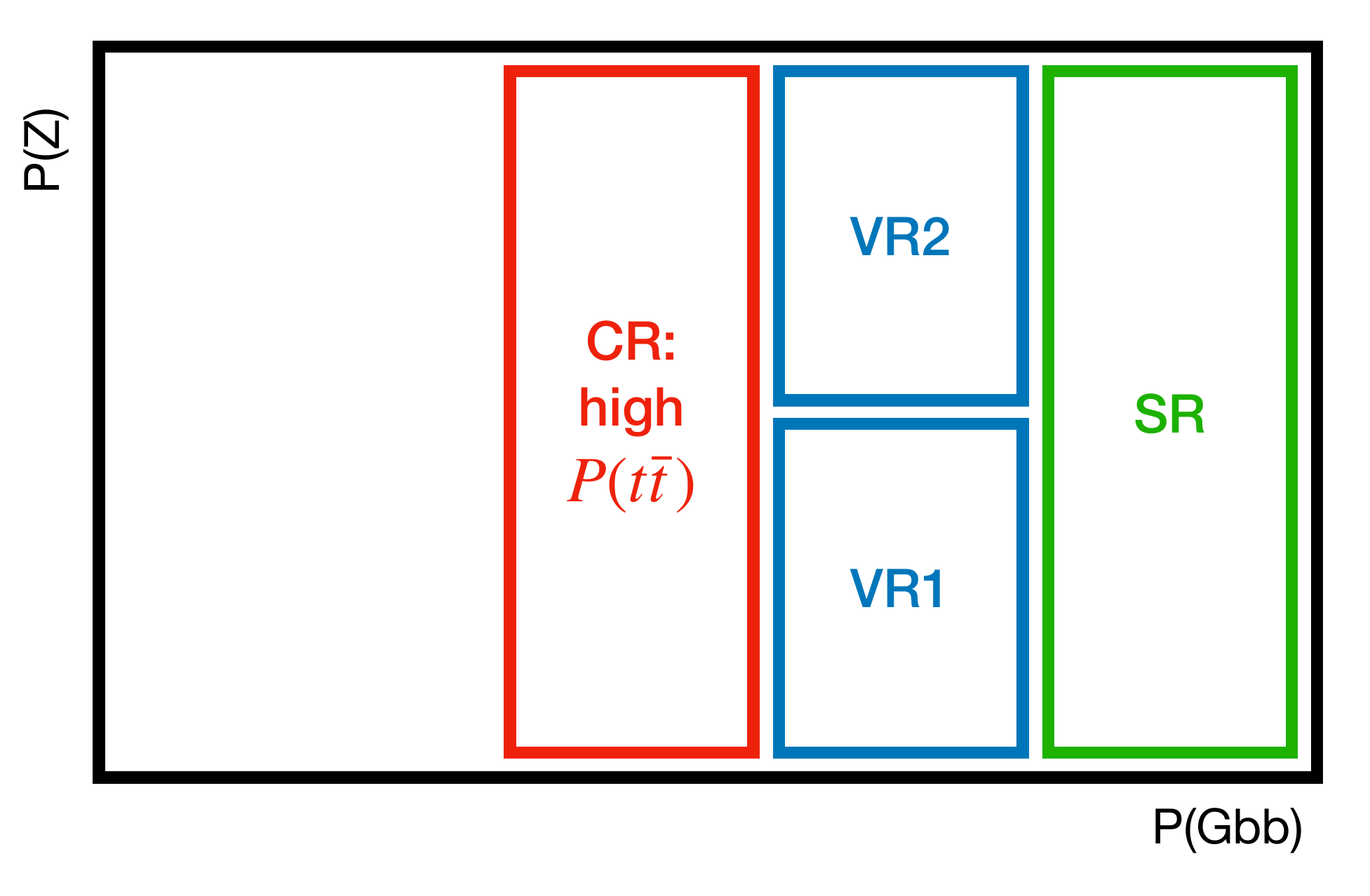}
\caption{Schematic diagram of the inter-relationship of the SRs, $t\bar{t}$ CRs and VRs used in the NN analysis for the Gtt (left) and Gbb (right) SRs.  The $t\bar{t}$ CRs and VRs apply additional selections to $\Delta\phi_\mathrm{min}^\mathrm{4j}$, $m_\mathrm{eff}$ and $M_J^\Sigma$ beyond those used in the SRs.  }
\label{fig:NN_scheme}
\end{figure}

\begin{table}[htb]
\centering
\small
\renewcommand{\arraystretch}{1.5}
\caption{Definitions of the NN Gtt SRs together with the associated $t\bar{t}$ CRs and VRs. In the first column, the two numbers separated by a hyphen specify the values of $m(\tilde{g})$ and $\mchi$ in GeV for the targeted representative Gtt model. The third and fourth columns specify the ranges of probability for the targeted Gtt signal model and $t\bar{t}$ background, respectively, generated by the selection applied to the NN output. The fifth and sixth columns specify the values of additional requirements applied to $m_\mathrm{eff}$ and $M_J^\Sigma$ in the CRs and VRs to select events with kinematics similar to those in the corresponding SRs.}
\label{tab:NN_Gtt}
\vspace{2mm}
\small
\begin{tabular}{c c l l l l l }
\toprule
Representative model &Region & $P(\mathrm{Gtt})$ & $\log_{10}(P(\ttbar))$ & \meff[\GeV] & \mjsum[\GeV] \\ \midrule
 
\multirow{3}{*}{\begin{minipage}{3cm}\centering Gtt-2100-1\end{minipage}}
&SR & $\geq$ 0.9997 & - & - & - \\
&CR & $\in$ (0.68, 0.86) & $\geq -1.8$ & $\geq2000$ & - \\
&VR & $\in$ (0.86, 0.9997) & - & $\geq2000$ & - \\ \midrule
 
\multirow{3}{*}{\begin{minipage}{3cm}\centering Gtt-1800-1\end{minipage}}
&SR & $\geq$ 0.9997 & - & - & - \\
&CR & $\in$ (0.73, 0.89) & $\geq -2.0$ & $\geq2000$ & - \\
&VR & $\in$ (0.89, 0.9997) & - & $\geq2000$ & - \\ \midrule
 
\multirow{3}{*}{\begin{minipage}{3cm}\centering Gtt-2300-1200\end{minipage}}
&SR & $\geq$ 0.9993 & - & - & - \\
&CR& $\in$ (0.78, 0.83) & $\geq -1.6$ & $\geq1400$ & - \\
&VR& $\in$ (0.83, 0.9993) & - & $\geq1800$ & - \\ \midrule
 
\multirow{3}{*}{\begin{minipage}{3cm}\centering Gtt-1900-1400\end{minipage}}
&SR& $\geq$ 0.9987 & - & - & - \\
&CR& $\in$ (0.78, 0.8) & $\geq -1.4$ & ~~$\geq800$ & $<700$ \\
&VR& $\in$ (0.8, 0.9987) & - & ~~$\geq800$ & $<700$ \\

\bottomrule
\end{tabular}
\end{table}

\begin{table}[htb]
\centering
\small
\renewcommand{\arraystretch}{1.5}
\caption{Definitions of the NN Gbb SRs together with the associated $t\bar{t}$ CRs and VRs. In the first column, the two numbers separated by a hyphen specify the values of $m(\tilde{g})$ and $\mchi$ in GeV for the targeted representative Gbb model. The third, fourth and fifth columns specify the ranges of probability for the targeted Gbb signal model and the $t\bar{t}$ or $Z+$jets background, respectively, generated by the selection applied to the NN output. The sixth, seventh and eighth columns specify the values of additional requirements applied to $\Delta\phi_\mathrm{min}^\mathrm{4j}$, $m_\mathrm{eff}$ and $M_J^\Sigma$ in the CRs and VRs to select events with kinematics similar to those in the corresponding SRs.}
\label{tab:NN_Gbb}
\vspace{2mm}
\begin{tabular}{c l l l l l l l}
\toprule
Rep. model &Region & $P(\mathrm{Gbb})$ & $\log_{10}(P(\ttbar))$ &$\log_{10}(P(Z))$ & \dphimin & \meff[\GeV] & \mjsum[\GeV] \\ \hline
 
\multirow{3}{*}{\begin{minipage}{3cm}\centering Gbb-2800-1400\end{minipage}}
&SR & $\geq$ 0.999  & - & - & $\geq0.6$ & - & - \\
&CR & $\in$ (0.43, 0.76) & $\geq -0.7$ & - & $\geq0.5$  & $\geq1400$ & $<800$ \\
&VR1 & $\in$ (0.76, 0.999)  & - & $< -1.7$ & $\geq0.5$ & $\geq2500$ & $<800$ \\
&VR2 & $\in$ (0.76, 0.999)  & - & $\geq -1.7$ & $\geq0.5$ & - & - \\ \hline
 
\multirow{3}{*}{\begin{minipage}{3cm}\centering Gbb-2300-1000\end{minipage}}
&SR & $\geq$ 0.9994  & - & - & $\geq0.6$ & - & - \\
&CR & $\in$ (0.52, 0.77) & $\geq -0.8$ & - & $\geq0.5$  & $\geq1400$ & $<800$ \\
&VR1 & $\in$ (0.77, 0.9994)  & - & $< -1.3$ & $\geq0.5$ & $\geq2400$ & $<800$ \\
&VR2 & $\in$ (0.77, 0.9994)  & - & $\geq -1.3$ & $\geq 0.5$ & - & - \\  \hline
 
\multirow{3}{*}{\begin{minipage}{3cm}\centering Gbb-2100-1600\end{minipage}}
&SR & $\geq$ 0.9993  & - & - & $\geq0.4$ & - & - \\
&CR& $\in$ (0.88, 0.91) & $\geq -1.3$ & - & $\geq0.4$  & $\geq800$ & $<500$ \\
&VR1 & $\in$ (0.91, 0.9993)  & - & $< -1.4$ & $\geq0.4$ & $\geq800$ & $<500$ \\
&VR2 & $\in$ (0.91, 0.9993)  & - & $\geq -1.4$ & $\geq0.4$ & - & - \\  \hline
 
\multirow{3}{*}{\begin{minipage}{3cm}\centering Gbb-2000-1800\end{minipage}}
&SR & $\geq$ 0.997  & - & - & $\geq0.4$ & - & - \\
&CR& $\in$ (0.92, 0.93) & $\geq -1.9$ & - & $\geq0.4$  & $\geq400$ & $<400$ \\
&VR1 & $\in$ (0.93, 0.997)  & - & $< -1.4$ & $\geq0.4$ & $\geq400$ & $<400$ \\
&VR2 & $\in$ (0.93, 0.997)  & - & $\geq-1.4$ & $\geq0.4$ & - & - \\  \hline

\end{tabular}
\end{table}
 
\begin{table}[htb]
\centering
\small
\renewcommand{\arraystretch}{1.5}
\caption{Definitions of the NN Gbb $Z+$jets CRs (CRZ). The first seven rows specify the event selection criteria. The final three rows specify the ranges of probability for the targeted Gbb signal model and the $t\bar{t}$ or $Z+$jets background, respectively, generated by the selection applied to the NN output. $N_\mathrm{lepton}=2$ requires two signal leptons.}
\label{tab:NN_Gbb_ZCR}
\vspace{2mm}
\begin{tabular}{l l l l l l}
\toprule
& Gbb-2800-1400 & Gbb-2300-1000 & Gbb-2100-1600 & Gbb-2000-1800\\
\midrule
$N_{\mathrm{lepton}}$ ($\pt(\ell)>30$ GeV) & \multicolumn{4}{c}{\,$= 2$} \\
$m_{\ell(1),\ell(2)}$ [\GeV] & \multicolumn{4}{c}{$\in[60,120]$}\\
$\pt(\ell(1),\ell(2))$ [\GeV] & \multicolumn{4}{c}{$>70$~~}\\
$\hat{E}_\mathrm{T}^\mathrm{miss}$ [\GeV] & \multicolumn{4}{c}{$>200$~~~~} \\
$p(\ell(1)),p(\ell(2))$ & \multicolumn{4}{c}{Set to zero after adding to $\met$} \\
$N_\mathrm{jet}$ & \multicolumn{4}{c}{$\geq 4$} \\
$N_{b\textrm{-jets}}$ & \multicolumn{4}{c}{$\geq 3$} \\
\midrule
$P(\mathrm{Gbb})$ & ~~\,$\geq 0.6$ & ~~\,$\geq 0.8$ & ~~\,$\geq 0.9$ & ~~\,$\geq0.9$ \\
$\log_{10}(P(\ttbar))$ & $<-1.3$ & $<-1.3$ & $<-1.7$ & $<-2.2$ \\
$\log_{10}(P(Z))$ & $\geq -3.0$ & $\geq -2.7$ &$\geq -4.9$ & $\geq-3.7$ \\
 
\bottomrule
 
\end{tabular}
\end{table}
\clearpage


\section{Systematic uncertainties}
\label{sec:syst}

The magnitudes of the post-fit uncertainties in the background estimates for the various signal regions, obtained following the profile likelihood fit described in Section~\ref{sec:fitresults}, are summarised in Figure~\ref{fig:summary_uncerts}. The uncertainties considered include experimental and theoretical systematic uncertainties, and statistical uncertainties in data CR yields and MC background samples.
\begin{figure}[htbp]
\centering
\includegraphics[width=0.95\textwidth]{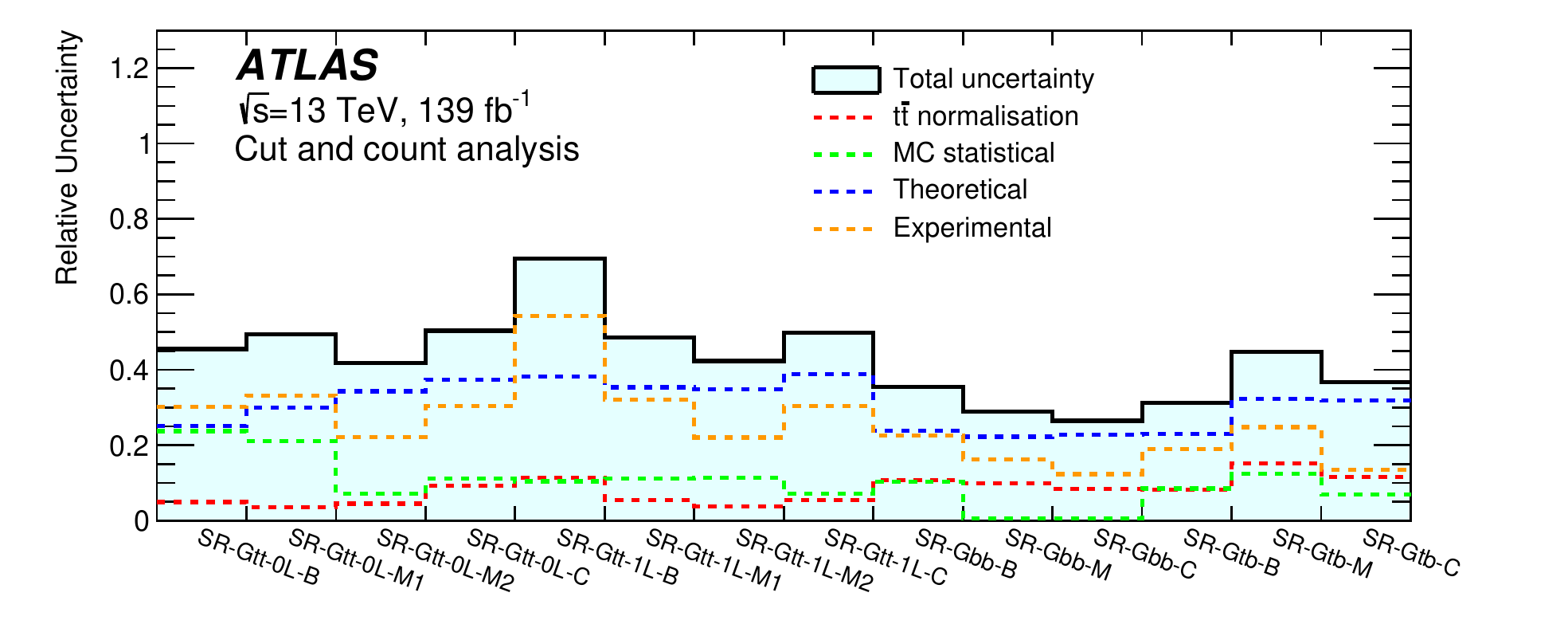}
\includegraphics[width=0.95\textwidth]{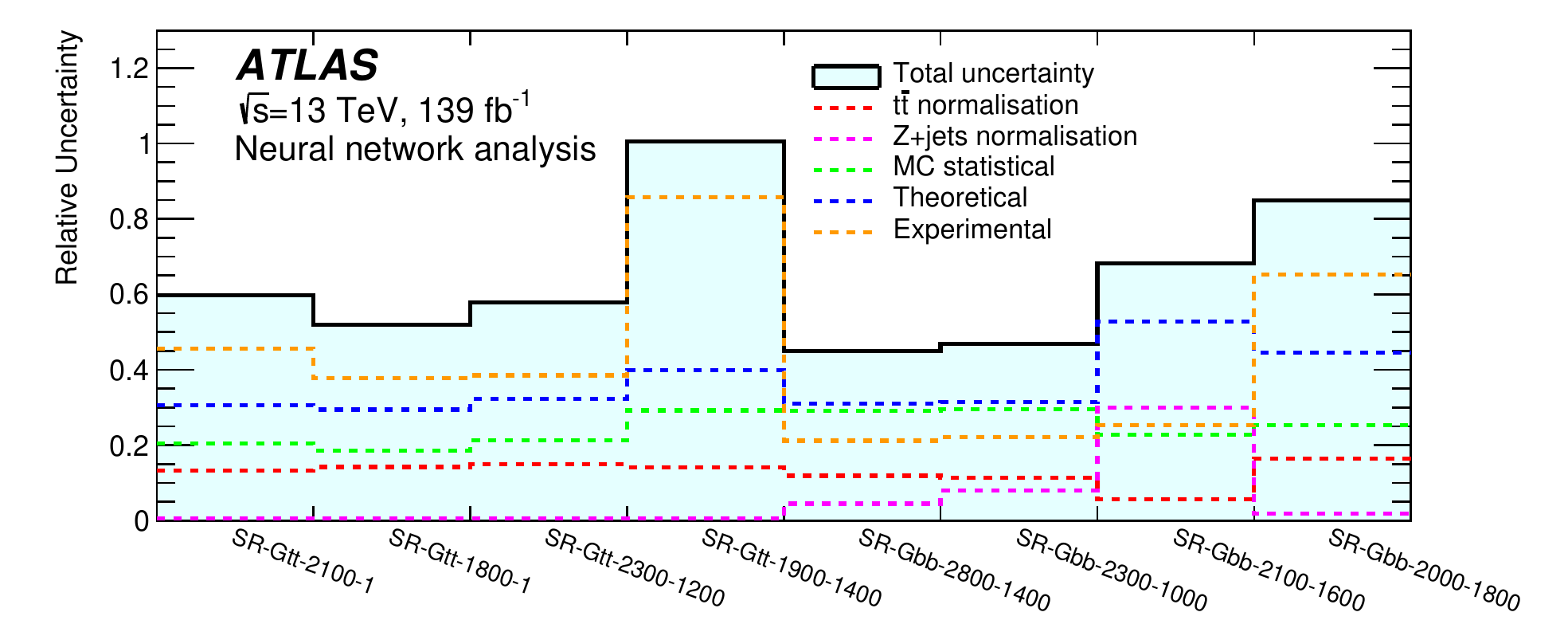}
\caption{A summary of the uncertainties in the background estimates for each of the signal regions of the CC (top) and NN (bottom) analyses. The individual experimental and theoretical uncertainties are assumed to be uncorrelated and are combined by adding in quadrature.}
\label{fig:summary_uncerts}
\end{figure}
 
Detector-related systematic uncertainties affect both the background estimate and the signal yield. The sources of the largest experimental uncertainties are related to the jet energy scale (JES) and resolution (JER)~\cite{PERF-2016-04}, the jet mass scale (JMS), and the $b$-tagging efficiencies and mis-tagging rates~\cite{ATL-PHYS-PUB-2016-012,ATL-PHYS-PUB-2017-013}. The jet energy-related uncertainties are also propagated to the reclustered large-$R$ jets, which use them as inputs. Jet mass scales are evaluated by comparing the masses reconstructed via calorimeter- and track-based measurements~\cite{JETM-2018-02}. The impact of the JES uncertainties on the expected background yields is between 0.6\% and 38\% (with the largest uncertainty observed in SR-Gbb-2000-1800), while JER uncertainties affect the background yields by 1\%--49\% in the various regions (with the largest uncertainty observed in SR-Gtt-1900-1400). This JER variation is the principal uncertainty contributing to the large total uncertainty observed in SR-Gtt-1900-1400 in Figure~\ref{fig:summary_uncerts}. The impact of JMS uncertainties on the expected background yields is 0.3\%--41\%, depending on the region (with the largest uncertainty observed in SR-Gtt-2100-1). Uncertainties in the measured $b$-tagging efficiencies and mis-tagging rates are 0.3\%--17\% (with the largest uncertainty observed in SR-Gbb-B).
 
The experimental uncertainties due to lepton reconstruction, identification and isolation efficiency differences between data and simulation~\cite{ATLAS-CONF-2016-024,PERF-2015-10} are also taken into account, and so are the uncertainties in lepton energy measurements~\cite{PERF-2013-05}. These uncertainties contribute at most 14\% to the total uncertainty (in SR-Gtt-0L-B). All lepton and jet measurement uncertainties are propagated to the calculation of \met, and additional uncertainties in the scale and resolution of the soft term~\cite{PERF-2016-07} are included. The overall impact of the \met soft-term uncertainties is at most 13\% (in SR-Gtt-0L-B).
 
Considering theoretical uncertainties, hadronisation and parton showering model uncertainties of the \ttbar\ background are evaluated by comparing two samples generated with \POWHEGBOX and showered by either \HERWIG[7.04] or \PYTHIA[8.230]~\cite{SUSY-2016-10}. In addition, systematic uncertainties in the modelling of initial- and final-state radiation are explored with \POWHEGBOX samples showered with two alternative settings~\cite{TOPQ-2017-19} of \PYTHIA[8.230]~\cite{Skands:2010ak}. The uncertainty due to the choice of matrix-element event generator is estimated by comparing the expected yields obtained using \ttbar samples generated with either \MGNLO or \POWHEGBOX. The total theoretical uncertainty in the \ttbar background estimation is taken as the sum in quadrature of these individual components, corresponding to an impact of 6\%--42\% (with the largest uncertainty observed in SR-Gbb-2000-1800). Moreover, an additional uncertainty of 30\% is assigned to the fraction of \ttbar events produced in association with additional heavy-flavour jets~\cite{SUSY-2016-10,TOPQ-2014-10} (i.e.\ $\ttbar~+ \geq \mathrm{1}b$ and $\ttbar~+ \geq \mathrm{1}c$), which has an impact of at most 10\%.
 
Modelling uncertainties affecting single-top processes arise especially from the interference between the \ttbar\ and $Wt$ processes. This uncertainty is estimated using inclusive $WWbb$ events, generated using \MGMCatNLO, which are compared with the sum of \ttbar and $Wt$ events. Furthermore, as in the \ttbar\ modelling uncertainties, variations of \PYTHIA[8.230] settings increasing or decreasing the amount of radiation are also used. An additional uncertainty is included in the cross-section of single-top processes~\cite{Kant:2014oha}, and is at most 5\%. Overall, the modelling uncertainties affecting single-top processes lead to changes of 4\%--19\%  in total yields in the various regions (with the largest uncertainty observed in SR-Gbb-2800-1400).
 
Uncertainties related to factorisation and renormalisation scales and affecting the matching procedure between the matrix element and parton shower in the $W/Z$+jets backgrounds are also taken into account~\cite{SUSY-2016-10}. The resulting uncertainties in the total yield range up to 53\% in the various regions (with the largest uncertainty in SR-Gbb-2100-1600). A constant 30\% uncertainty in the heavy-flavour content of $W/Z$+jets is assumed, motivated by Ref.~\cite{STDM-2017-38}, which contributes at most 8\% (with the largest uncertainty observed in SR-Gtt-0L-B). Furthermore, a 50\% uncertainty is assigned to $\ttbar+W/Z/H$, $\ttbar\ttbar$ and diboson backgrounds, and is assumed to be uncorrelated across all regions. It is found to have no significant impact on the sensitivity of this analysis, and contributes at most 15\% (with the largest uncertainty observed in SR-Gtt-2100-1) to the total background uncertainty. The effect of the uncertainties related to the parton distribution functions affect the background yields by less than 2\%, and therefore are neglected here. Uncertainties due to the limited number of events in the MC background samples are included and can reach 30\% in regions targeting moderate/large mass-splittings.
 
Systematic uncertainties are also assigned to the kinematic reweighting procedure, by propagating the statistical uncertainties in the parameters of the exponential fits (Section~\ref{sec:reweight}). In addition, the changes in estimated background yield obtained by omitting the reweighting procedure are added in quadrature to conservatively assess the impact of the procedure on the final results, which was observed to contribute to total yield variations of 0\%--49\% across all regions (with the largest observed in SR-Gtt-1L-B). These uncertainties are applied to all simulated background events containing at least one signal lepton. The uncertainties are treated as nuisance parameters in the fits, in the same way as the other uncertainties.
 
The uncertainties in the cross-sections of signal processes are determined from different cross-section predictions extracted using alternative event generators and parton shower variations, as described in Section~\ref{sec:samples}. These are at most 30\% in all SRs.


\section{Results and interpretation}
\label{sec:results}

\subsection{Results}
\label{sec:fitresults}
The SM background yields are estimated separately for each SR with a profile likelihood fit~\cite{Cowan:2010js} implemented in the HistFitter framework~\cite{HFpaper}, referred to as a background-only fit. The fit (one for each SR) uses the observed event yield in the associated CR(s) to adjust the $\ttbar$ normalisation (and $Z+$jets normalisation for the NN Gbb regions), assuming that no signal contributes to this yield, and applies that normalisation factor to the number of background events predicted by simulation for the equivalent process in the SR. The normalisation factors are allowed to vary freely, without application of constraints derived from uncertainties in theoretical cross-sections. The mean yields predicted by simulation in the SR and CR(s) are used for all background processes. The numbers of events observed in each region are described by Poisson probability density functions. The systematic uncertainties in the expected values are included in the fit as nuisance parameters. They are constrained by Gaussian distributions with widths corresponding to the sizes of the uncertainties and are treated as correlated, when appropriate, between the various regions. The product of the various probability density functions forms the likelihood, which the fit maximises by adjusting the $\ttbar$ (and, in the NN Gbb regions, the $Z+$jets) normalisation and the nuisance parameters. The values of the normalisation factors range from 0.8 to 1.9 depending on the CRs (Figure~\ref{fig:pullCR}). The normalisation factors are broadly consistent with unity within uncertainties, with the largest pull across the 26 CRs being 2.3$\sigma$ (for the $\ttbar$ normalisation factor in CC SR Gtt-0L-M2).
 
\begin{figure}[htbp]
\centering
\includegraphics[width=0.9\textwidth]{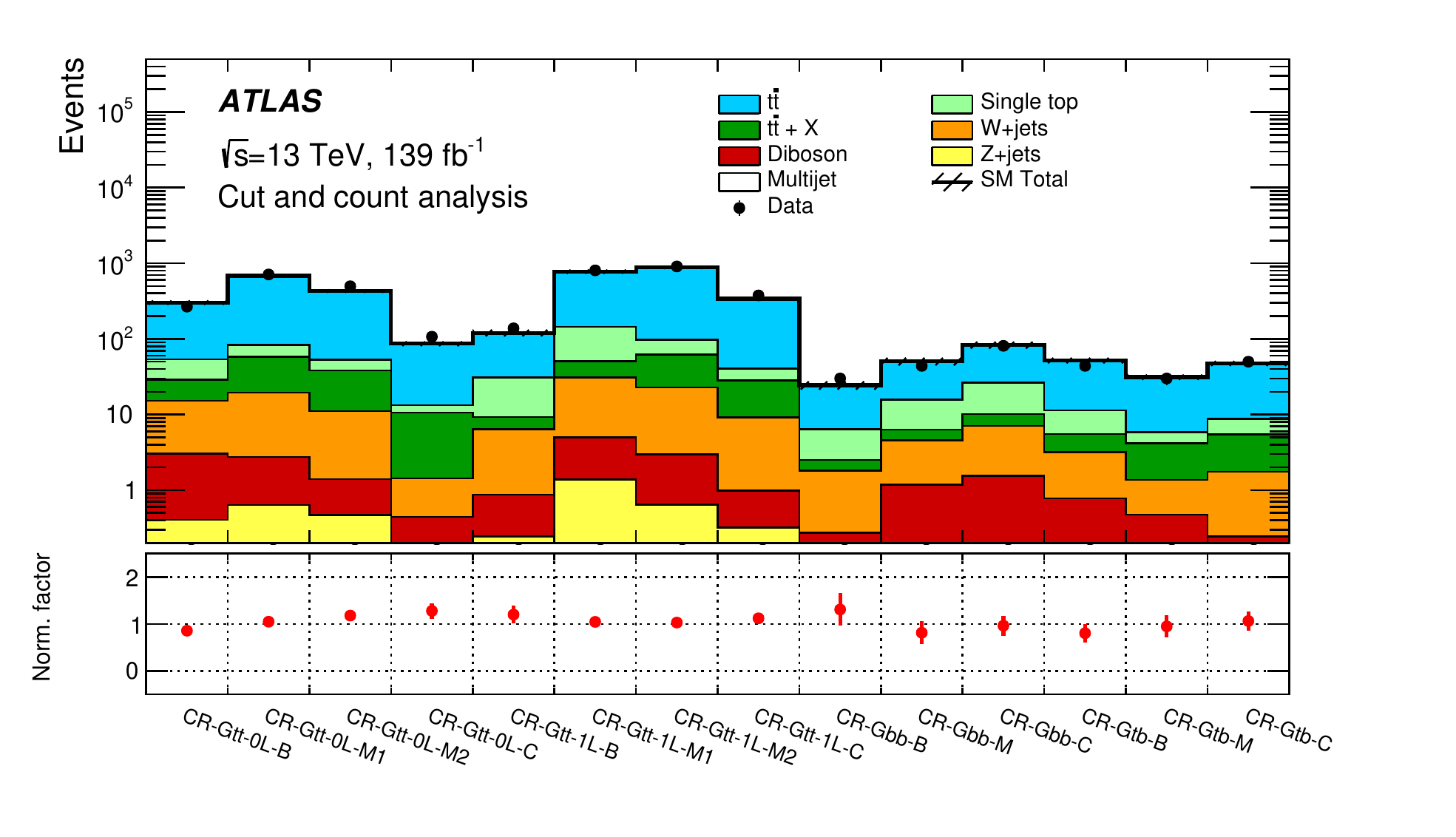} \\
\includegraphics[width=0.9\textwidth]{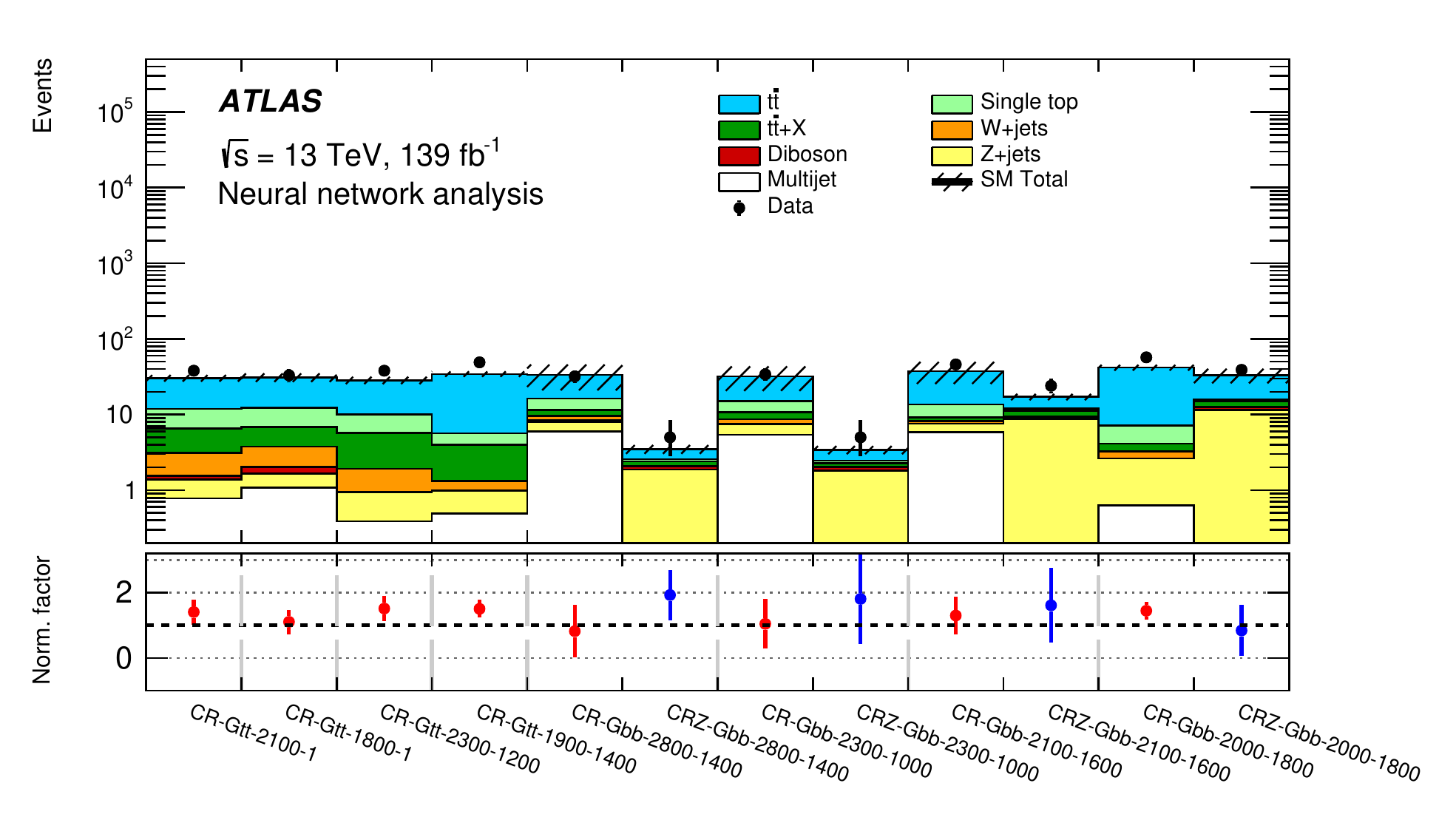}
\caption{Pre-fit event yield in control regions and related background normalisation factors after the background-only fit for
the CC (top) and NN (bottom) analyses. The upper panel shows the observed number of events and the MC background yield before the fit. The uncertainty bands include both the experimental and theoretical systematic uncertainties. The background category `$t\bar{t}+X$' includes $t\bar{t} W/Z$, $t\bar{t} H$ and $t\bar{t}t\bar{t}$ events. The background normalisation factors obtained from the fit are displayed in the bottom panels, with the error bars representing the combined statistical+systematic uncertainty of each normalisation factor, obtained from the background-only fit. Red (blue) points correspond to the $t\bar{t}$ ($Z$+jets) CRs.
}
\label{fig:pullCR}
\end{figure}
 
The results of the background-only fits are extrapolated to the VRs, following the definitions of Section~\ref{sec:evsel}. A comparison of the observed and expected yields in the VRs after the fit is shown in Figure~\ref{fig:pullVR} for the CC and NN analyses. The largest yield differences in the CC and NN VRs are observed for VR2-Gtt-1L-C (1.3$\sigma$ across 18 VRs) and VR2-Gbb-2000-1800 (1.8$\sigma$ across 12 VRs) respectively. The MC reweighting gives better agreement between the observed and expected yields in the VRs; in the CC analysis the change in the predicted yields due to the MC reweighting, which is applied to events containing exactly one signal lepton, is on average $5\%$ and $10\%$ for the 0-lepton and 1-lepton channels, respectively. This reweighting affects the CC 0-lepton channels, which require exactly zero signal leptons, via the $\ttbar$ normalisation factors obtained from CRs requiring exactly one signal lepton.
\begin{figure}[!htbp]
\centering
\includegraphics[width=0.9\textwidth]{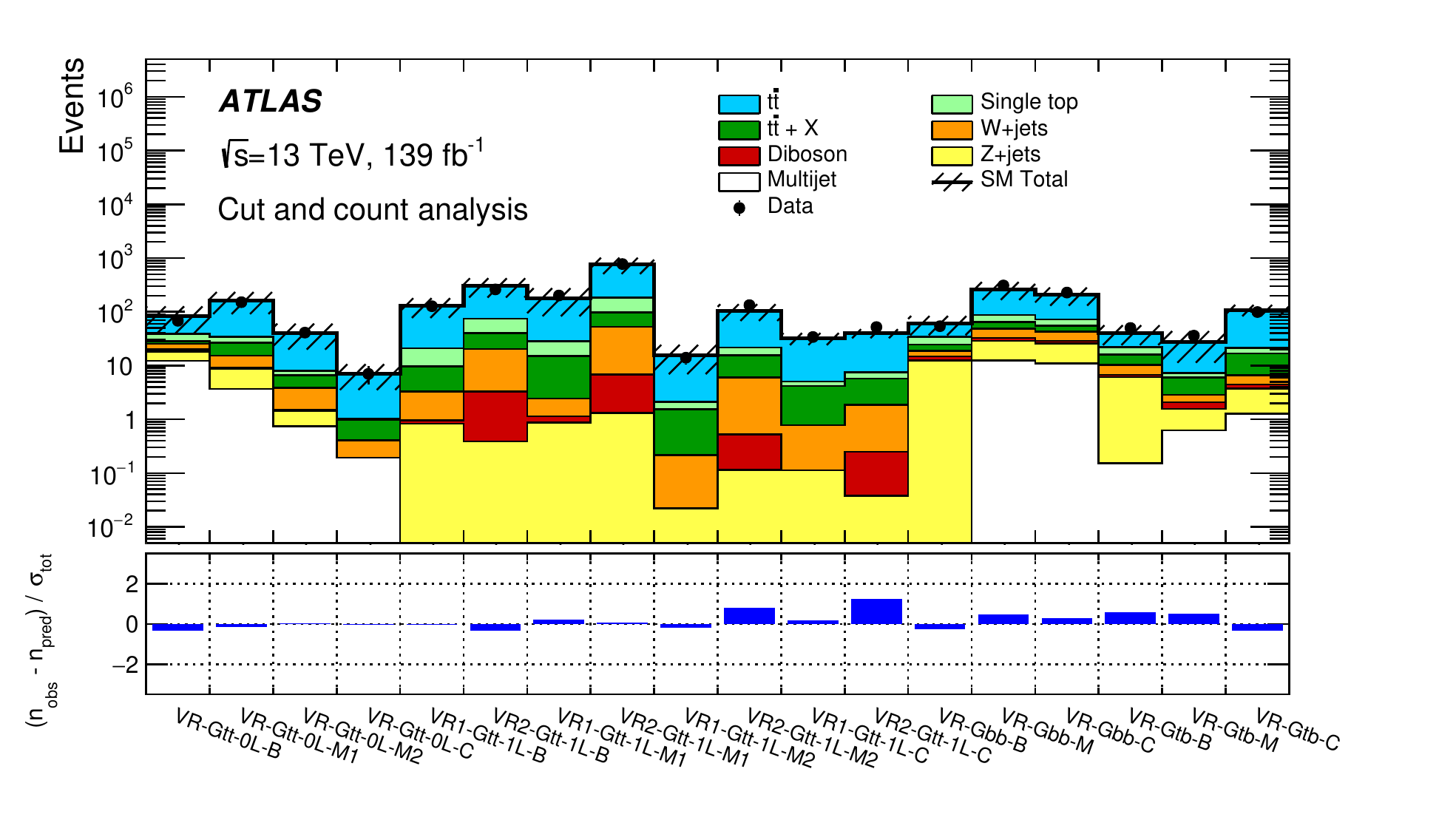}
\clearpage
\includegraphics[width=0.9\textwidth]{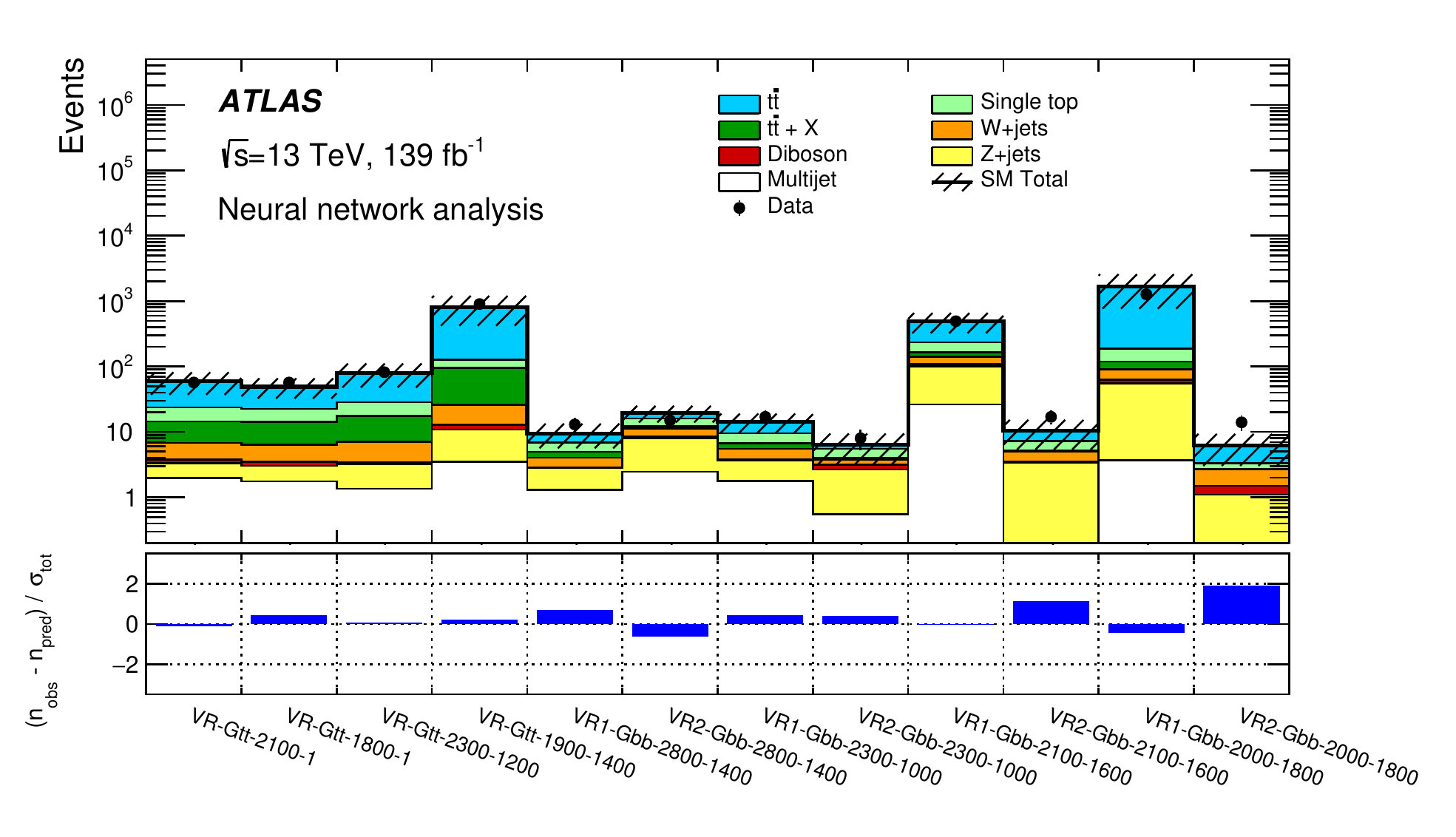}
\caption{Data and fitted background yields in the VRs for
the CC (top) and NN (bottom) analyses.  The upper panels show the observed numbers of events and the expected background
yields obtained from the background-only fits. The uncertainty bands include both the experimental and theoretical systematic uncertainties. The background
category `$t\bar{t}+X$' includes $t\bar{t} W/Z$, $t\bar{t} H$ and $t\bar{t}t\bar{t}$ events. The bottom panels show the Gaussian significance $(n_\textrm{obs} - n_\textrm{pred}) / \sigma_\textrm{tot}$ of the deviation of the
observed yield ($n_\textrm{obs}$) from the background expectation ($n_\textrm{pred}$) in each VR, obtained using the combined statistical+systematic background uncertainty ($\sigma_{\textrm{tot}}$). }
\label{fig:pullVR}
\end{figure}
 
The observed and expected event yields in the SRs for the CC and NN analyses are shown in Figure~\ref{fig:pullSR} and Tables~\ref{tab:yield_discovery_Gtt0L1L}--\ref{tab:yieldsNN}. The significances of the deviations of the observed data from the background expectations, evaluated using a model-independent fit (described in Section~\ref{sec:interpretation}) with pseudo-experiments, are presented in the lower panels of Figure~\ref{fig:pullSR}, and in Table~\ref{mod-ind-lim-CCA}. No statistically significant deviation from the background expectation is found for any of the presented SRs or analysis methodologies. The largest deviation across the 22 SRs is observed in SR-Gtb-B, with a significance of 2.3$\sigma$. In some cases the SR event selections are not orthogonal and hence the significances can be correlated between regions.
 
\begin{figure}[!htbp]
\centering
\includegraphics[width=0.9\textwidth]{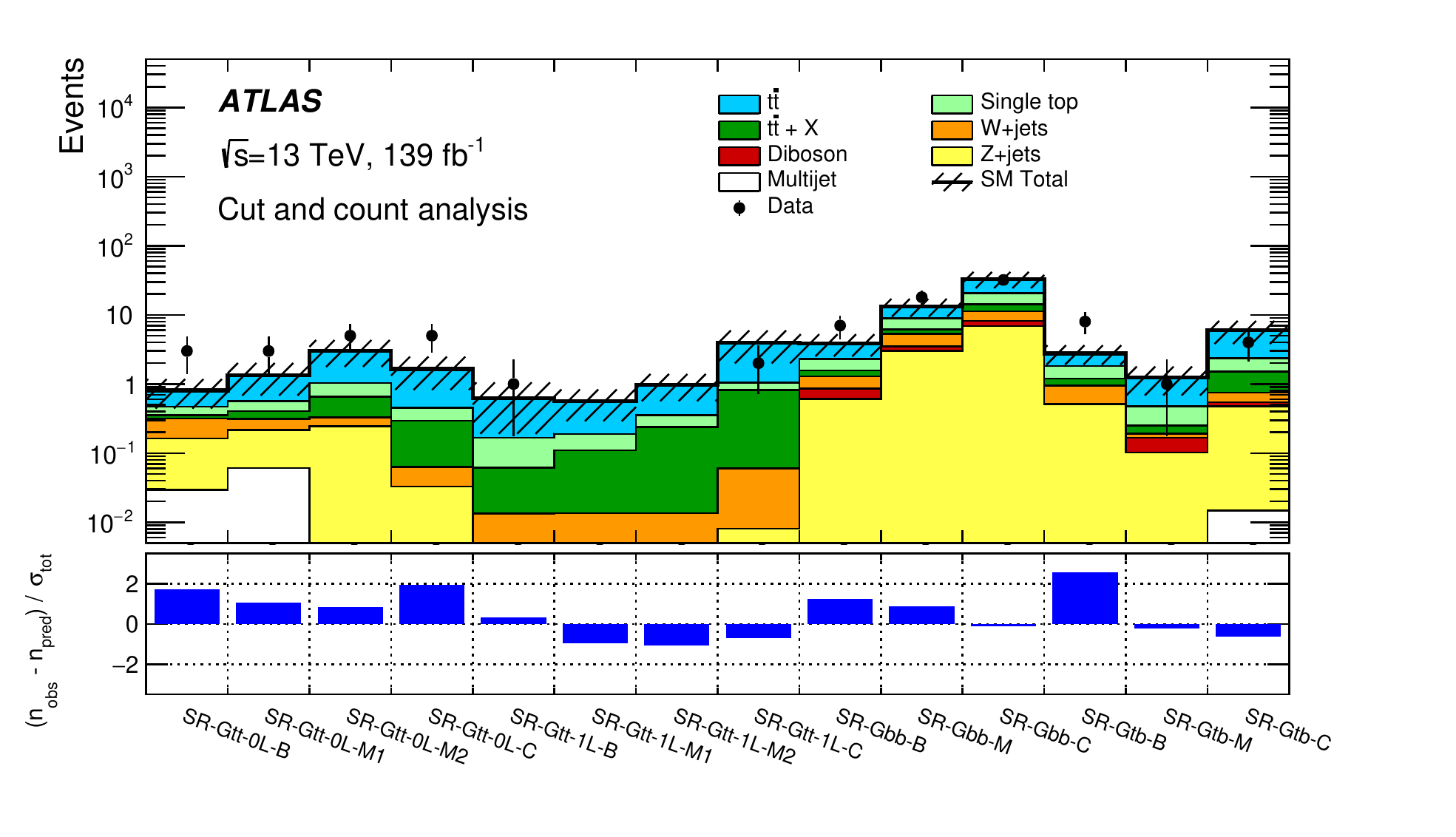}\\
\includegraphics[width=0.9\textwidth]{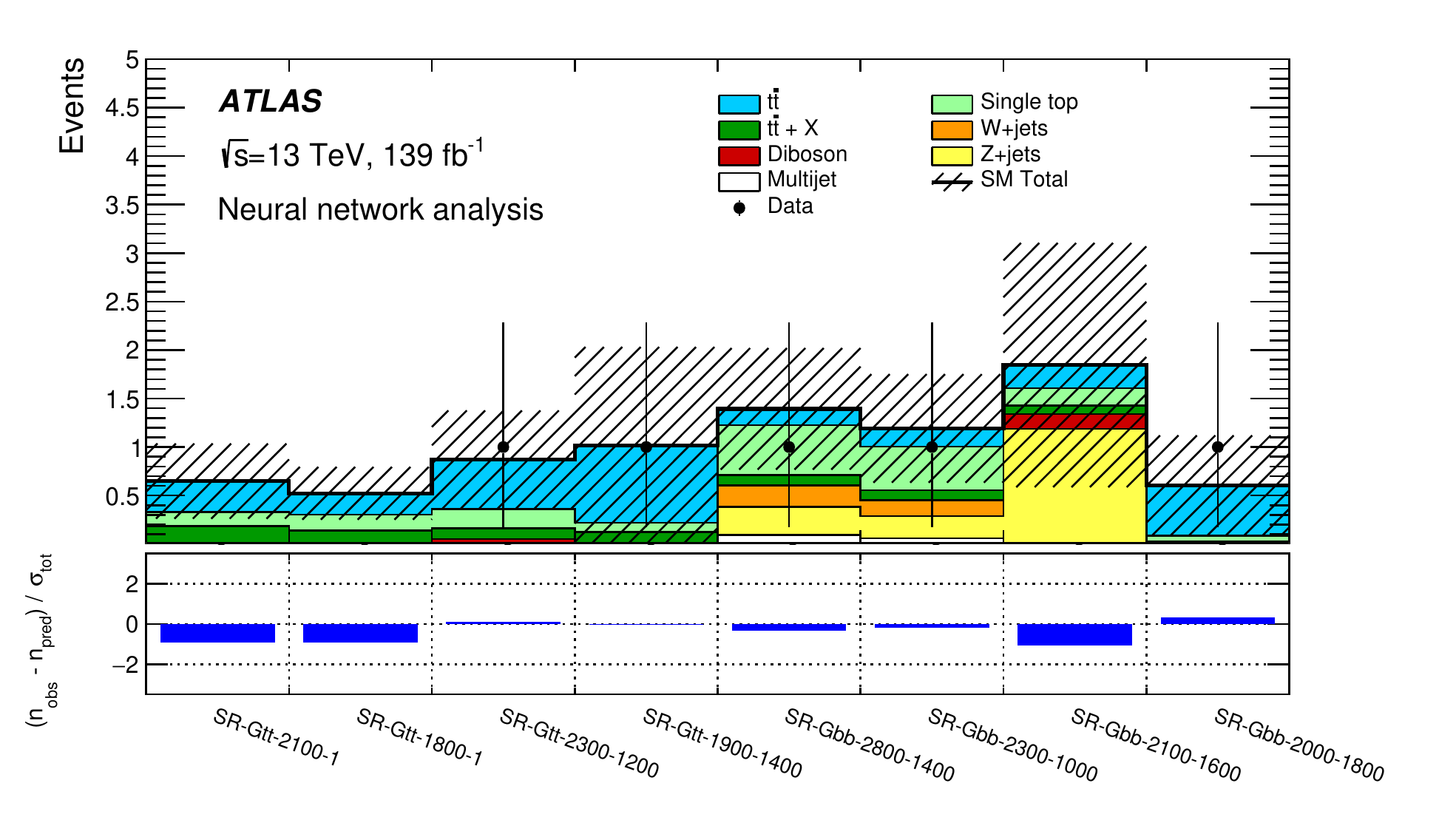}\\
\caption{Data and fitted background yields in the SRs for the CC (top) and NN (bottom) analyses.
The upper panels show the observed numbers of events and the expected background yields
obtained from the background-only fits. Bins without data-points correspond to zero observed events.
The uncertainty bands include both the experimental and theoretical systematic uncertainties.
The background category `$t\bar{t}+X$' includes $t\bar{t} W/Z$, $t\bar{t} H$ and $t\bar{t}t\bar{t}$ events.
The bottom panels show the Gaussian significance $(n_\textrm{obs} - n_\textrm{pred}) / \sigma_\textrm{tot}$ of the deviation of the observed yield ($n_\textrm{obs}$) from the background expectation ($n_\textrm{pred}$) in each SR, obtained using the combined statistical+systematic background uncertainty ($\sigma_{\textrm{tot}}$). }
\label{fig:pullSR}
\end{figure}
 
\begin{table*}[!htbp]
\centering
\caption{Post-fit results of the background-only fit extrapolated to the Gtt 0-lepton and Gtt 1-lepton SRs in
the CC analysis, for both the total expected background yields and the main contributing background processes.
The quoted uncertainties include both the experimental and theoretical systematic uncertainties. The data in the SRs are not included in the fit.
The background category `$t\bar{t}+X$' includes $t\bar{t} W/Z$, $t\bar{t} H$ and $t\bar{t}t\bar{t}$ events.
The row `Pre-fit background' provides the total background prediction when the
$t\bar{t}$ normalisation is obtained from a theoretical calculation~\cite{Czakon:2011xx}, taking into account the kinematic weights described in Section~\ref{sec:reweight}. Yields are obtained for the large $\Delta m$ (`B'), moderate $\Delta m$ (`M1' and `M2') and small $\Delta m$ (`C') scenarios.
}
\vspace{5mm}
 
\label{tab:yield_discovery_Gtt0L1L}

\begin{tabular}{lcccc}
\toprule
\multirow{2}{*}{Region} & \multicolumn{4}{c}{SR-Gtt-0L} \\
 
& B            		&   	M1   		&   M2   			&   C                \\[-0.05cm]
\midrule
Observed events & $ 3   $& $ 3   $& $ 5 $& $ 5 $ \\
\midrule
Fitted background & $0.81 \pm 0.32 $& $1.3 \pm 0.6 $& $3.0 \pm 1.3 $& $1.7 \pm 0.8 $ \\
\midrule
\ttbar & $0.34 \pm 0.21 $& $0.8 \pm 0.5 $& $2.0 \pm 1.2 $& $1.2 \pm 0.8 $ \\
Single top & $0.11 \pm 0.10 $& $0.16 \pm 0.10 $& $0.38 \pm 0.23 $& $0.16 \pm 0.11 $ \\
$t\bar{t}+X$ & $0.04^{+0.05}_{-0.04} $~\, & $0.09^{+0.13}_{-0.09} $~ & $0.32 \pm 0.20 $& $0.23 \pm 0.16 $ \\
\Zjets & $0.13 \pm 0.08 $& $0.16 \pm 0.08 $& $0.25 \pm 0.14 $& $0.03^{+0.08}_{-0.03}$~\, \\
\Wjets & $0.15 \pm 0.11 $& $0.10 \pm 0.09 $& $0.08 \pm 0.05 $& $0.03 \pm 0.02 $ \\
Diboson & $<0.01 $& $<0.01$ & $<0.01$& $<0.01 $ \\
Multijet & $0.03^{+0.20}_{-0.03} $~\, & $0.06^{+0.25}_{-0.06} $~\, & $<0.01 $& $<0.01 $ \\
\midrule
Pre-fit background & $0.87$ & $1.3$ & $2.7$ & $1.4$ \\
\bottomrule
 
\end{tabular}

\vspace{0.5cm}

\begin{tabular}{lcccc}
\toprule
\multirow{2}{*}{Region} & \multicolumn{4}{c}{SR-Gtt-1L} \\
 
& B            		&   	M1   		&   M2   			&   C                \\[-0.05cm]
 
\midrule
Observed events & $ 1 $& $ 0 $& $ 0 $& $ 2 $ \\
\midrule
Fitted background & $0.6 \pm 0.4 $& $0.56 \pm 0.27 $& $1.0 \pm 0.4 $& $4.0 \pm 2.0 $ \\
\midrule
\ttbar & $0.5 \pm 0.4 $& $0.37 \pm 0.23 $& $0.63 \pm 0.34 $& $2.9 \pm 1.8 $ \\
Single top & $0.11 \pm 0.08 $& $0.08 \pm 0.07 $& $0.12 \pm 0.08 $& $0.23 \pm 0.15 $ \\
$t\bar{t}+X$ & $0.05 \pm 0.04 $& $0.10 \pm 0.06 $& $0.23 \pm 0.15 $& $0.8 \pm 0.5 $ \\
\Zjets & $<0.01 $& $<0.01$ & $<0.01$& $<0.01 $ \\
\Wjets & $0.013^{+0.021}_{-0.013} $~\, & $<0.01$ & $<0.01$ & $0.052 \pm 0.034 $\\
Diboson & $<0.01 $& $<0.01 $& $<0.01 $& $<0.01 $ \\
Multijet & $<0.01 $& $<0.01 $& $<0.01 $& $<0.01 $ \\
\midrule
Pre-fit background & $0.55$ & $0.55$ & $1.0$ & $3.7$ \\
\bottomrule
\end{tabular}

\end{table*}
 
\begin{table*}[!htbp]
\centering
\caption{Post-fit results of the background-only fit extrapolated to the Gbb and Gtb SRs in
the CC analysis, for both the total expected background yields and the main contributing background processes.
The quoted uncertainties include both the experimental and theoretical systematic uncertainties. The data in the SRs are not included in the fit.
The background category `$t\bar{t}+X$' includes $t\bar{t} W/Z$, $t\bar{t} H$ and $t\bar{t}t\bar{t}$ events.
The row `Pre-fit background' provides the total background prediction when the
$t\bar{t}$ normalisation is obtained from a theoretical calculation~\cite{Czakon:2011xx}, taking into account the kinematic weights described in Section~\ref{sec:reweight}. Yields are obtained for the large $\Delta m$ (`B'), moderate $\Delta m$ (`M') and small $\Delta m$ (`C') scenarios.
}
\vspace{5mm}
 
\begin{tabular}{lccc}
\toprule
\multirow{2}{*}{Region}& \multicolumn{3}{c}{SR-Gbb} \\
 
& B            		& M         			&   C               \\[-0.05cm]
\midrule
Observed events & $ 7   $& $ 18 $& $ 32$ \\
\midrule
Fitted background & $3.9 \pm 1.4 $& $13 \pm 4 $& $33 \pm 9 $ \\
\midrule
\ttbar & $1.6 \pm 1.0 $& $4.3 \pm 3.0 $& $12 \pm 7 $ \\
Single top & $0.7 \pm 0.4 $& $2.7 \pm 1.4 $& $6.3 \pm 3.4 $ \\
$t\bar{t}+X$ & $0.27 \pm 0.16 $& $0.9 \pm 0.5 $& $3.0 \pm 1.5 $ \\
\Zjets & $0.61 \pm 0.26 $& $3.0 \pm 1.0 $& $6.9 \pm 2.3 $ \\
\Wjets & $0.43 \pm 0.21 $& $1.8 \pm 1.2 $& $3.1 \pm 1.6 $ \\
Diboson & $0.3 ^{+0.4}_{-0.3} $\, & $0.5 \pm 0.4 $& $1.3 \pm 0.8 $ \\
Multijet & $<0.01 $& $<0.01 $& $<0.01 $ \\
\midrule
Pre-fit background & $3.5$ & $14$ & $33$ \\
\bottomrule
 
\end{tabular}
 
\vspace{0.5cm}
 
\begin{tabular}{lccc}
\toprule
\multirow{2}{*}{Region}& \multicolumn{3}{c}{SR-Gtb} \\
 
& B            		& M         			&   C               \\[-0.05cm]
\midrule
Observed events & $ 8 $& $ 1 $& $ 4 $ \\
\midrule
Fitted background & $2.8 \pm 0.9 $& $1.2 \pm 0.6 $& $6.0 \pm 2.2 $ \\
\midrule
\ttbar & $0.9 \pm 0.6 $& $0.8 \pm 0.5 $& $3.6 \pm 2.1 $ \\
Single top & $0.62 \pm 0.35 $& $0.23 \pm 0.12 $& $0.9 \pm 0.5 $ \\
$t\bar{t}+X$ & $0.25 \pm 0.15 $& $0.06 \pm 0.04 $& $0.8 \pm 0.4 $ \\
\Zjets & $0.52 \pm 0.26 $& $0.10 \pm 0.10 $& $0.46 \pm 0.17 $ \\
\Wjets & $0.43 \pm 0.23 $& $0.02^{+0.07}_{-0.02} $~\, & $0.21 \pm 0.11 $ \\
Diboson & $<0.01 $& $0.06^{+0.16}_{-0.06} $~\, & $0.06^{+0.16}_{-0.06} $~\,  \\
Multijet & $<0.01 $& $<0.01 $& $0.01^{+0.64}_{-0.01} $~\,  \\
\midrule
Pre-fit background & $3.0$ & $1.3$ & $5.8$ \\
\bottomrule
\end{tabular}
\end{table*}

\begin{table*}[htbp]
\centering
\caption{Post-fit results of the background-only fit extrapolated to the SRs in
the NN analysis, for both the total expected background yields and the main contributing background processes.
The quoted uncertainties include both the experimental and theoretical systematic uncertainties. The data in the SRs are not included in the fit.
The background category `$t\bar{t}+X$' includes $t\bar{t} W/Z$, $t\bar{t} H$ and $t\bar{t}t\bar{t}$ events.
The row `Pre-fit background' provides the total background prediction when the
$t\bar{t}$ and $Z+$jets normalisations are obtained from theoretical calculation~\cite{Czakon:2011xx,Catani:2009sm}, taking into account the kinematic weights described in Section~\ref{sec:reweight}.
}
\vspace{5mm}
 
\label{tab:yieldsNN}
 
\vspace{0.5 cm}

\begin{tabular}{lcccc}
\toprule
\multirow{2}{*}{Region} & \multicolumn{4}{c}{SR-Gtt} \\
& 2100-1 & 1800-1 & 2300-1200 & 1900-1400\\
 
\midrule
Observed events & $ 0 $& $ 0 $& $ 1 $& $ 1 $ \\
\midrule
Fitted background & $0.6 \pm 0.4 $& $0.52 \pm 0.27 $& $0.9 \pm 0.5 $& $1.0 \pm 1.0 $ \\
\midrule
\ttbar & $0.32 \pm 0.24 $& $0.22 \pm 0.15 $& $0.51 \pm 0.34 $& $0.8 \pm 0.8 $ \\
Single top & $0.14 \pm 0.12 $& $0.17 \pm 0.13 $& $0.20 \pm 0.14 $& $0.10^{+0.23}_{-0.10} $~\, \\
$t\bar{t}+X$ & $0.19^{+0.24}_{-0.19} $~\, & $0.14 \pm 0.14 $& $0.11^{+0.21}_{-0.11} $~\, & $0.13^{+0.22}_{-0.13} $~\, \\
\Zjets & $<0.01$ & $<0.01$ &  $0.01^{+0.07}_{-0.01} $~\, & $<0.01 $ \\
\Wjets & $<0.01 $& $<0.01 $& $<0.01 $& $<0.01 $ \\
Diboson & $<0.01 $& $<0.01 $& $0.041 \pm 0.023 $& $<0.01 $ \\
Multijet & $<0.01 $& $<0.01 $& $<0.01 $& $<0.01 $ \\
\midrule
Pre-fit background & $0.56$ & $0.50$ & $0.7$ & $0.7$ \\
\midrule
\end{tabular}
 
\vspace{0.5 cm}
 
\begin{tabular}{lcccc}
\toprule
\multirow{2}{*}{Region} & \multicolumn{4}{c}{SR-Gbb} \\
& 2800-1400 & 2300-1000 & 2100-1600 & 2000-1800\\

\midrule
Observed events & $ 1 $& $ 1 $& $ 0 $& $ 1 $ \\
\midrule
Fitted background & $1.4 \pm 0.6 $& $1.2 \pm 0.6 $& $1.8 \pm 1.3 $& $0.6 \pm 0.5 $ \\
\midrule
\ttbar & $0.17^{+0.21}_{-0.17} $~\, & $0.19^{+0.21}_{-0.19} $~\, & $0.24 \pm 0.21 $& $0.5 \pm 0.5 $ \\
Single top & $0.51 \pm 0.31 $& $0.45 \pm 0.28 $& $0.18^{+0.20}_{-0.18} $~\, & $0.06 \pm 0.05 $ \\
$t\bar{t}+X$ & $0.11^{+0.12}_{-0.11} $~\, & $0.10 \pm 0.10 $& $0.09^{+0.11}_{-0.09} $~\, & $0.01^{+0.04}_{-0.01} $~\, \\
\Zjets & $0.29 \pm 0.19 $& $0.23 \pm 0.17 $& $1.2 \pm 1.2 $& $0.017^{+0.035}_{-0.017} $~\, \\
\Wjets & $0.22^{+0.27}_{-0.22} $~\, & $0.17^{+0.20}_{-0.17} $~\, & $<0.01 $& $<0.01 $ \\
Diboson & $<0.01 $& $<0.01 $& $0.15 \pm 0.31 $& $<0.01 $ \\
Multijet & $0.09^{+0.19}_{-0.09} $~\, & $0.06^{+0.12}_{-0.06} $~\, & $<0.01 $& $<0.01 $ \\
\midrule
Pre-fit background & $1.3$ & $1.1$ & $1.3$ & $0.4$ \\
\bottomrule

\end{tabular}

\end{table*}
 
\begin{table}[b]
\centering
\caption{Summary of compatibility of SR observations with background expectations in the CC and NN SRs. The second column shows the $p_0$-values of the background-only hypothesis together with the equivalent Gaussian significances $Z$. Also shown are the resulting 95$\%$ CL upper limits on the visible cross-section ($\sigma^{95}_\mathrm{vis}$), and the observed and expected 95$\%$ CL upper limits on the number of beyond the Standard Model signal events ($S_{\textrm{obs}}^{95}$ and $S_{\textrm{exp}}^{95}$). The $p_0$-values are capped at 0.5.
}
\vspace{5mm}
\label{mod-ind-lim-CCA}
\small
\begin{tabular}{lcccc}
\noalign{\smallskip}\toprule\noalign{\smallskip}
Signal Region & $p_0$ ($Z$)  & $\sigma^{95}_\mathrm{vis}$ [fb]  &  $S_{\textrm{obs}}^{95}$  & $S_{\textrm{exp}}^{95}$   \\
\noalign{\smallskip}\midrule \noalign{\smallskip}
SR-Gtt-0L-B    & $ 0.03$~$(1.82)$ & $0.05$ &  $6.4$ & $ { 3.7 }^{ +1.2 }_{ -0.4 }$ \\[1mm]
SR-Gtt-0L-M1    & $ 0.13$~$(1.13)$ & $0.04$ &  $6.1$ & $ { 4.3 }^{ +1.6 }_{ -1.0 }$ \\[1mm]
SR-Gtt-0L-M2    & $ 0.18$~$(0.91)$ & $0.06$ &  $7.7$ & $ { 5.7 }^{ +2.2 }_{ -1.2 }$ \\[1mm]
SR-Gtt-0L-C    & $ 0.03$~$(1.83)$ & $0.06$ &  $8.5$ & $ { 4.9 }^{ +2.0 }_{ -1.0 }$ \\[1mm]
\noalign{\smallskip}\midrule \noalign{\smallskip}
SR-Gtt-1L-B    & $ 0.29$~$(0.56)$ & $0.03$ &  $3.9$ & $ { 3.3 }^{ +1.2 }_{ -0.2 }$ \\[1mm]
SR-Gtt-1L-M1    & $ 0.5$~~~$(0.0)$~~ & $0.02$ &  $3.0$ & $ { 3.1 }^{ +1.2 }_{ -0.1 }$ \\[1mm]
SR-Gtt-1L-M2    & $ 0.5$~~~$(0.0)$~~ & $0.02$ &  $3.0$ & $ { 3.4 }^{ +1.3 }_{ -0.4 }$ \\[1mm]
SR-Gtt-1L-C    & $ 0.5$~~~$(0.0)$~~ & $0.03$ &  $4.6$ & $ { 5.3 }^{ +2.2 }_{ -1.5 }$ \\[1mm]
\noalign{\smallskip}\midrule \noalign{\smallskip}
SR-Gbb-B    & $ 0.11$~$(1.22)$ & $0.07$ &  $9.5$ & $ { 6.2 }^{ +2.6 }_{ -1.4 }$ \\[1mm]
SR-Gbb-M    & $ 0.18$~$(0.93)$ & $0.11$ &  $16.0$~~ & $ { 11.4 }^{ +5.0 }_{ -2.7 }$~\,\, \\[1mm]
SR-Gbb-C    & $ 0.5$~~~$(0.0)$~~ & $0.14$ &  $19.4$~~ & $ { 19.5 }^{ +5.5 }_{ -4.6 }$~\,\, \\[1mm]
\noalign{\smallskip}\midrule \noalign{\smallskip}
SR-Gtb-B    & $ 0.01$~$(2.30)$ & $0.08$ &  $11.3$~~ & $ { 5.4 }^{ +2.2 }_{ -1.3 }$ \\[1mm]
SR-Gtb-M    & $ 0.5$~~~$(0.0)$~~ & $0.03$ &  $3.7$ & $ { 3.8 }^{ +1.5 }_{ -0.5 }$ \\[1mm]
SR-Gtb-C    & $ 0.5$~~~$(0.0)$~~ & $0.04$ &  $5.7$ & $ { 6.7 }^{ +2.6 }_{ -1.8 }$ \\[1mm]
\noalign{\smallskip}\midrule\noalign{\smallskip}
SR-Gtt-2100-1     &  $ 0.5$~~~$(0.0)$~~  &  $0.02$  &   $3.0$  &  $ { 3.1 }^{ +1.1 }_{ -0.2 }$ \\
SR-Gtt-1800-1     &  $ 0.5$~~~$(0.0)$~~  &  $0.02$  &   $3.0$  &  $ { 3.0 }^{ +1.1 }_{ -0.1 }$ \\
SR-Gtt-2300-1200     &  $ 0.40$~$(0.26)$  &  $0.03$  &   $3.8$  &  $ { 3.5 }^{ +1.4 }_{ -0.3 }$ \\
SR-Gtt-1900-1400     &  $ 0.5$~~~$(0.0)$~~  &  $0.03$  &   $4.2$  &  $ { 4.1 }^{ +1.5 }_{ -1.1 }$ \\
\noalign{\smallskip}\midrule\noalign{\smallskip}
SR-Gbb-2800-1400     &  $ 0.5$~~~$(0.0)$~~  &  $0.03$  &   $3.7$  &  $ { 3.9 }^{ +1.4 }_{ -0.8 }$ \\
SR-Gbb-2300-1000     &  $ 0.5$~~~$(0.0)$~~  &  $0.03$  &   $3.8$  &  $ { 3.8 }^{ +1.3 }_{ -0.7 }$ \\
SR-Gbb-2100-1600     &  $ 0.36$~$(0.35)$  &  $0.02$  &   $3.0$  &  $ { 3.2 }^{ +1.3 }_{ -0.1 }$ \\
SR-Gbb-2000-1800     &  $ 0.29$~$(0.55)$  &  $0.03$  &   $4.0$  &  $ { 3.4 }^{ +1.2 }_{ -0.6 }$ \\
 
\noalign{\smallskip}\bottomrule\noalign{\smallskip}
 
\end{tabular}
\end{table}


\clearpage
 
\subsection{Interpretation}
\label{sec:interpretation}

In the absence of any significant excess over the expected background from SM processes, the data are used to derive one-sided upper limits at 95\% confidence level (CL). Two levels of interpretation are provided in this paper: model-independent exclusion limits and model-dependent exclusion limits set on the Gtt, Gbb and Gtb models. Exclusion limits are evaluated using pseudo-experiments.
 
To set upper limits on the number of `beyond the Standard Model' (BSM) signal events in each SR, a model-independent fit is used~\cite{HFpaper}. This fit proceeds in the same way as the background-only fit, where yields in the CRs are used to constrain the predictions of backgrounds in each SR, while the SR yield is also used in the likelihood function with an additional parameter-of-interest describing potential signal contributions. The observed
and expected upper limits at 95\%~CL on the number of events from BSM phenomena for each signal region ($S_{\textrm{obs}}^{95}$ and $S_{\textrm{exp}}^{95}$) are derived using the CL$_\mathrm{s}$ prescription~\cite{Read:2002hq}, neglecting any possible signal contamination in the CRs. These limits, when normalised by the integrated luminosity of the data sample, may be interpreted as upper limits on the visible cross-section of BSM physics ($\sigma^{95}_\mathrm{vis}$), where the visible cross-section is defined as the product of production cross-section, acceptance and efficiency. All SRs from both the CC and NN analyses are considered, to aid in the reinterpretation of the results of this analysis. The results of the model-independent fit are presented in Table~\ref{mod-ind-lim-CCA}.
 
Model-dependent fits~\cite{HFpaper} in all the SRs are used to set limits in the mass or branching ratio planes of the Gtt, Gbb and Gtb models. Such a fit proceeds in the same way as the model-independent fit, except that both the signal yield in the SR and the signal contamination in the CR(s) are taken into account. Correlations between signal and background systematic uncertainties are taken into account where appropriate. Systematic uncertainties in the assumed signal yields due to detector effects and the theoretical uncertainties in the signal acceptance are included in the fit. The NN analysis SRs provide the higher expected exclusion sensitivity for the Gtt and Gbb models and hence are used to obtain the exclusion limits for these models. For the Gtb models, for which the NN analysis SRs were not optimised, the CC analysis SRs are used. In all cases, at each model point the result obtained from the SR with the best expected CL$_\mathrm{s}$ value is used. For the CC analysis applied to the Gtb models, all CC SRs are considered in this optimisation. The Gtt or Gbb SRs are found to be optimal for Gtb models in which the gluino branching ratio is dominated respectively by $\gluino \to t\bar{t}\ninoone$ or $\gluino \to b\bar{b}\ninoone$, while the Gtb SRs are found to be optimal otherwise. All the fits for the various model points and parameter spaces considered give fitted SUSY signal cross-sections consistent with zero within uncertainties.
 
The 95\%~CL observed and expected exclusion limits for the Gtt and Gbb models, obtained from the NN analysis, are shown in the $\gluino$--$\ninoone$ mass plane in Figure~\ref{fig:limits_NN}. The $\pm1 \sigma^{\textrm{SUSY}}_{\textrm{theory}}$ lines around the observed limits are obtained by changing the SUSY production cross-section by one standard deviation ($\pm1\sigma$), as described in Section~\ref{sec:samples}. The yellow band around the expected limit shows the $\pm1\sigma$ uncertainty, including all statistical and systematic uncertainties except the theoretical uncertainties in the SUSY cross-section. The expected limits on the gluino mass, assuming a massless neutralino LSP, obtained from the CC analysis are ${\sim}150$~\GeV\ and ${\sim}50$~\GeV\ lower for the Gtt and Gbb models, respectively. Compared to the previous results~\cite{SUSY-2016-10}, the expected limits on the gluino mass from the current search (assuming a massless $\ninoone$) have improved by 280~\gev\ and 330~\gev\ for the Gtt and Gbb models, respectively. Gluinos with masses below 2.44~\TeV\ (2.35~\TeV) are excluded at 95\%~CL for massless neutralinos in the Gtt (Gbb) models. The most stringent exclusion limits on the neutralino mass are approximately 1.35~\TeV\ and 1.65~\TeV\ for the Gtt and Gbb models, obtained for a gluino mass of approximately 2.20~\TeV\ and 2.10~\TeV, respectively.

\begin{figure}[htbp]
\centering
\includegraphics[width=0.8\textwidth]{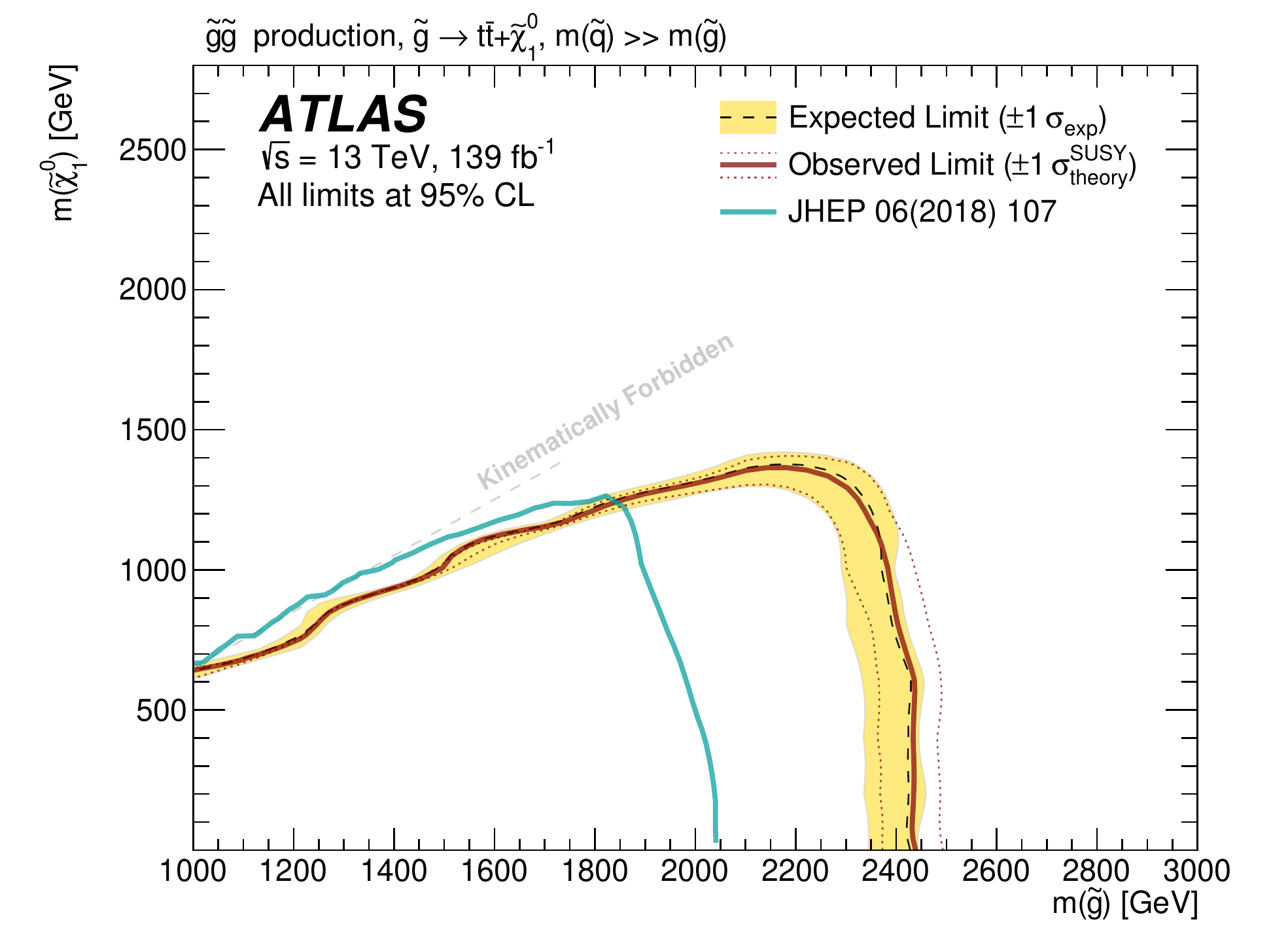}
\includegraphics[width=0.8\textwidth]{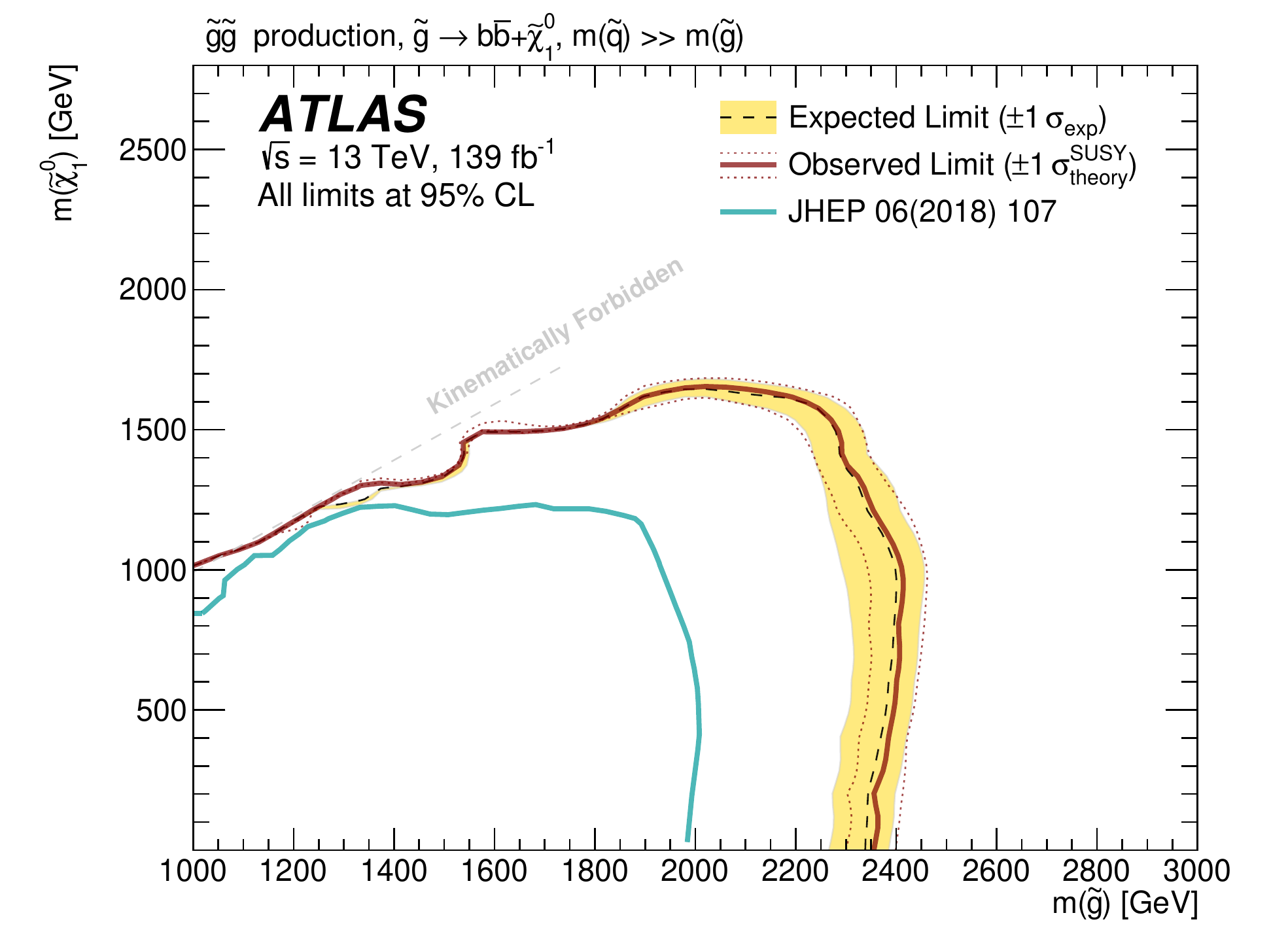}
\caption{Exclusion limits in the $\tilde{g}$--$\ninoone$ mass plane
for the Gtt (top) and Gbb (bottom) models obtained
from the NN analysis. The dashed and solid bold lines
show the 95\% CL expected and observed limits, respectively. The
shaded bands around the expected limits show the
impact of the
experimental and background uncertainties. The dotted
lines show the impact on the observed limit of the variation of the
nominal signal cross-section by $\pm 1 \sigma$ of its theoretical
uncertainty.
The blue contours indicate the exclusion limits from the previous published result~\cite{SUSY-2016-10}.
\label{fig:limits_NN}}
\end{figure}
 
The 95\%~CL observed and expected exclusion limits for the Gtb model, with $\mchi$ set to 1~\GeV, are shown in
Figure~\ref{fig:limits_triangle_1}. The results, which are obtained from the CC analysis, are presented as limits on the mass of the gluino as a function of the branching ratios for $\gluino \to t\bar{t}\ninoone$ and $\gluino \to b\bar{b}\ninoone$. The sum of these branching ratios with that for the decays $\gluino \to t \bar{b} \chinoonem$ and $\gluino \to \bar{t} b \chinoonep$ (which are set to be equal in the simulation) is set to unity. The exclusion limits are strongest when $\mathcal{B}(\gluino \to b\bar{b}\ninoone$) or $\mathcal{B}(\gluino \to t\bar{t}\ninoone$) saturate the total branching ratio (top-left and bottom-right corners of the figures), and weakest when $\mathcal{B}(\gluino \to t \bar{b} \chinoonem$/$\gluino \to \bar{t} b \chinoonep$) saturates. Similar results are obtained for $\mchi = 600$~\GeV\ (Figure~\ref{fig:limits_triangle_600}) and $\mchi = 1000$~\GeV\ (Figure~\ref{fig:limits_triangle_1000}).
 
\begin{figure}[htbp]
\centering
\includegraphics[width=0.49\textwidth]{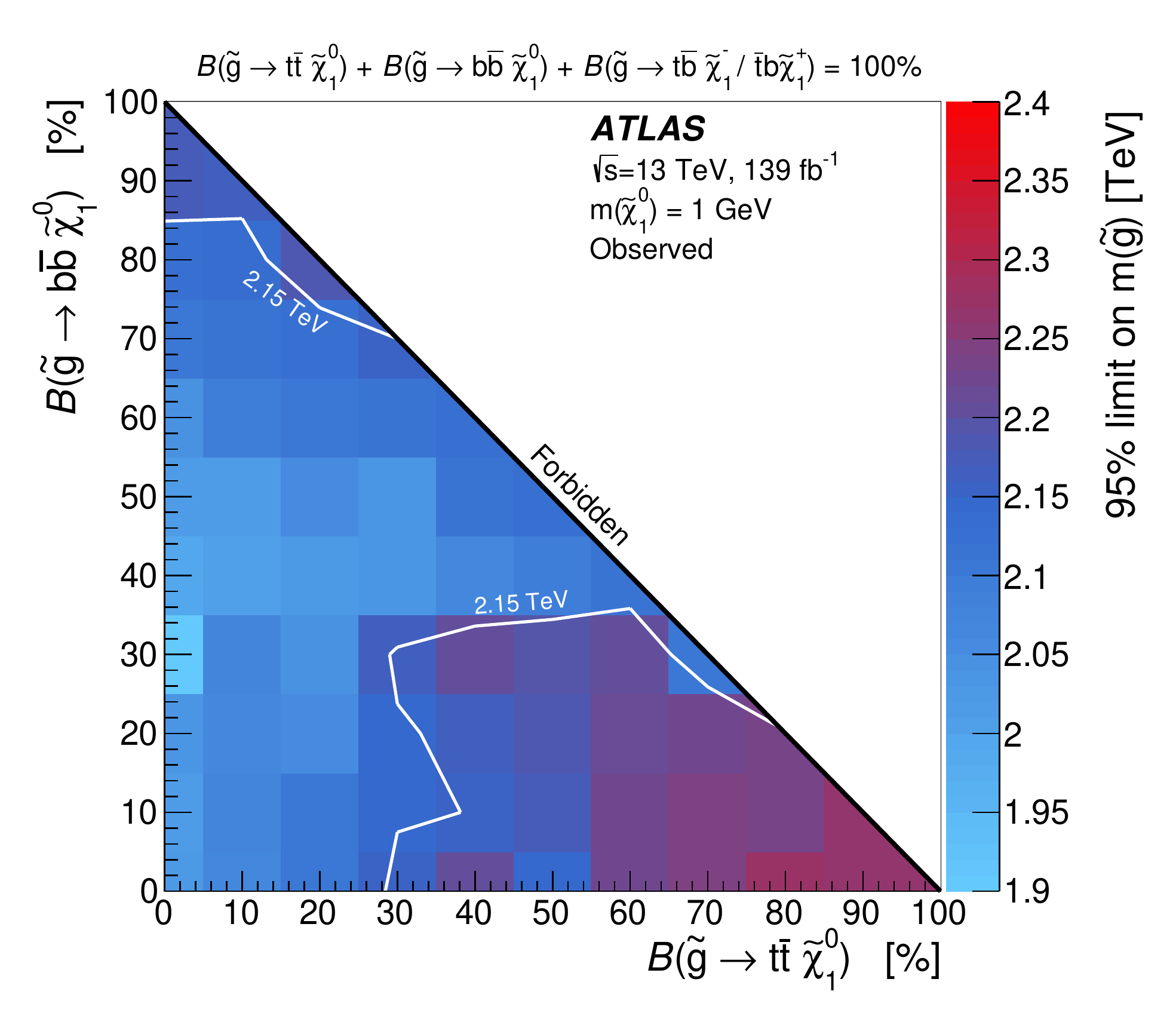}
\includegraphics[width=0.49\textwidth]{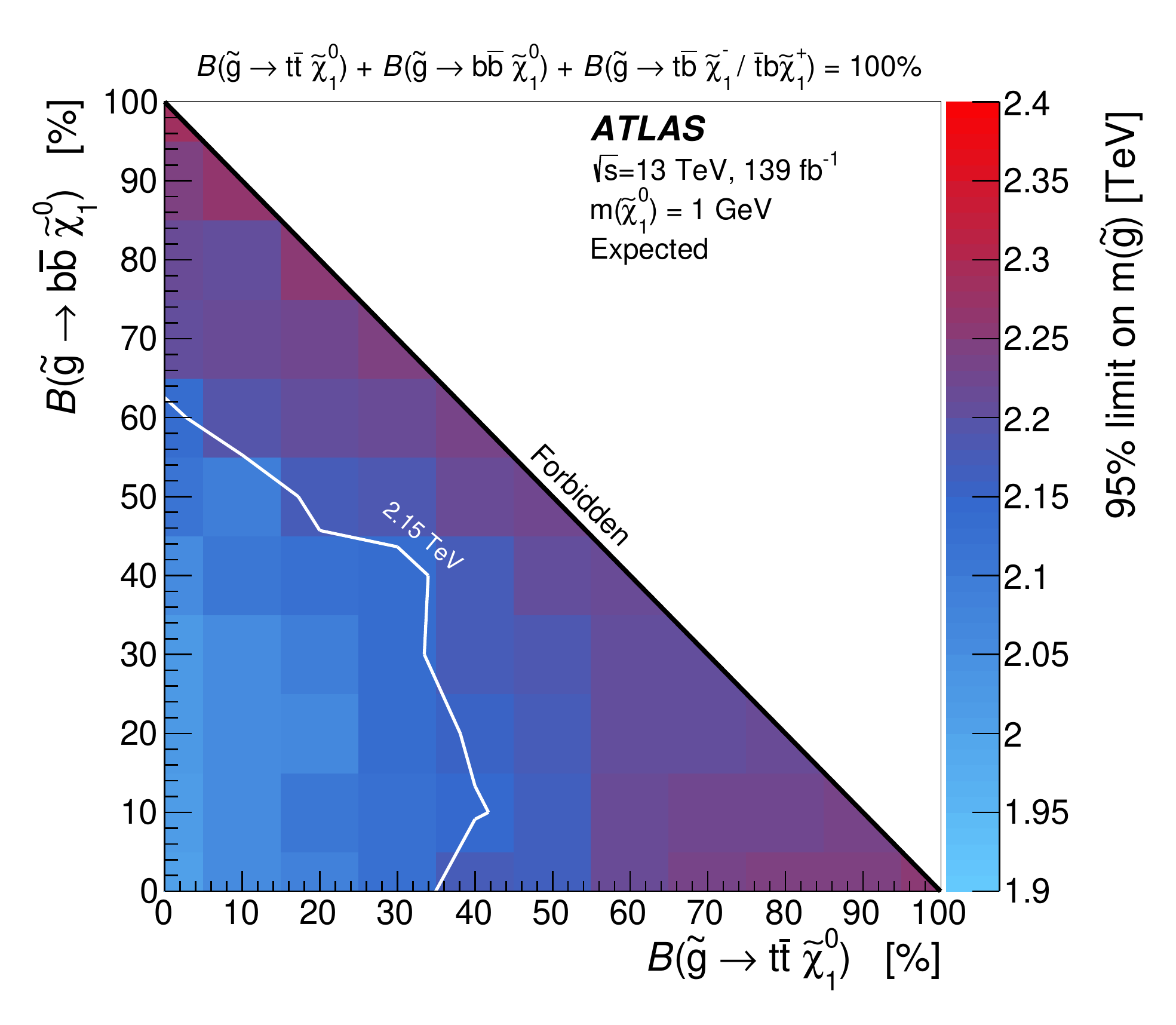}
\caption{The observed (left) and expected (right) 95\%~CL exclusion limits on the gluino mass as a function of $\mathcal{B}(\tilde{g} \to b\bar{b}\ninoone$) (vertical) and $\mathcal{B}(\tilde{g} \to t\bar{t}\ninoone$) (horizontal) for Gtb models with $\mchi = 1$~\GeV, obtained from the CC analysis. The shading indicates the highest excluded gluino mass for each point in the plane. The white line indicates the contour of fixed $m(\tilde{g}) = 2150$~\GeV.}
\label{fig:limits_triangle_1}
\end{figure}
 
\begin{figure}[b!]
\centering
\includegraphics[width=0.49\textwidth]{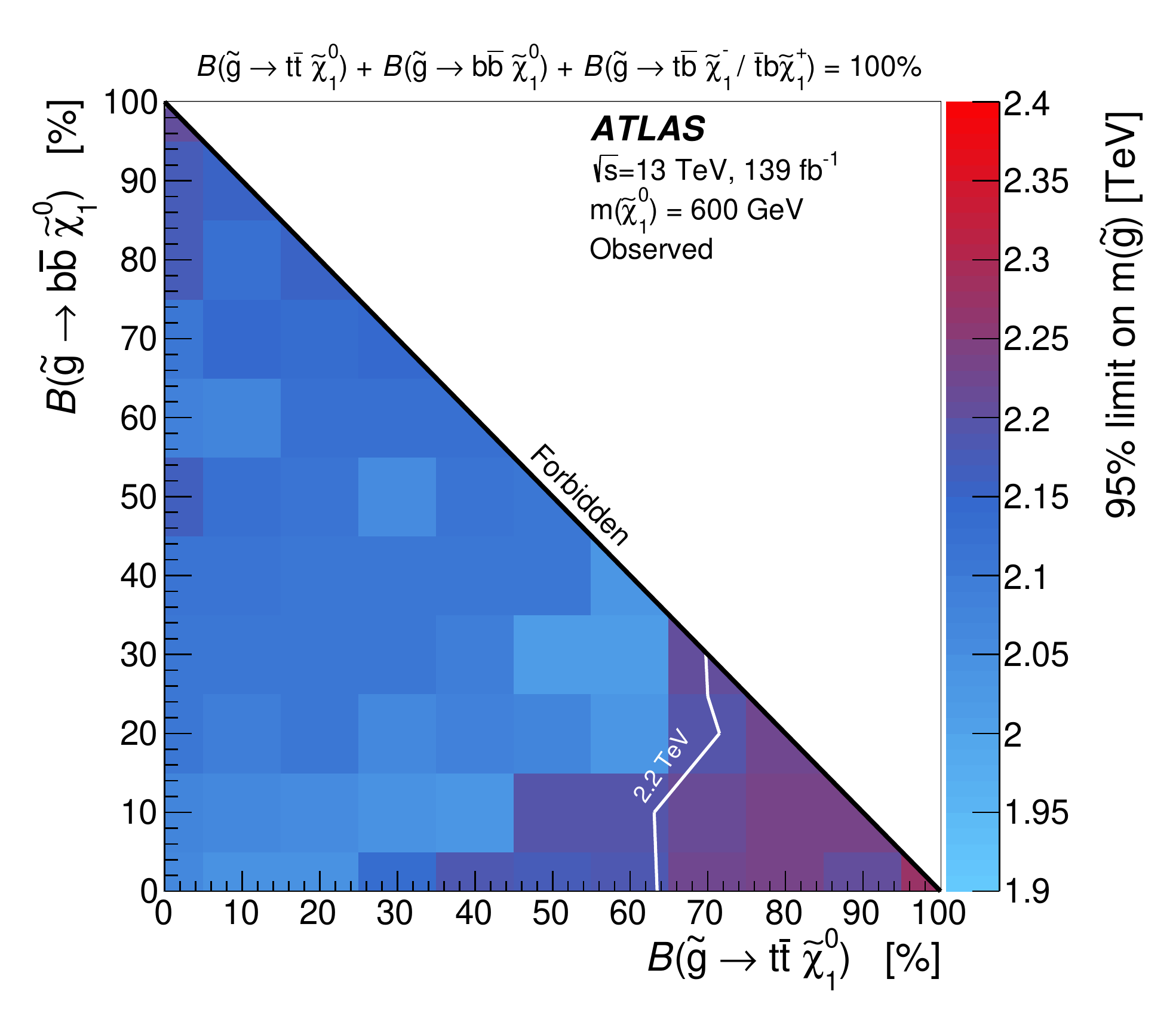}
\includegraphics[width=0.49\textwidth]{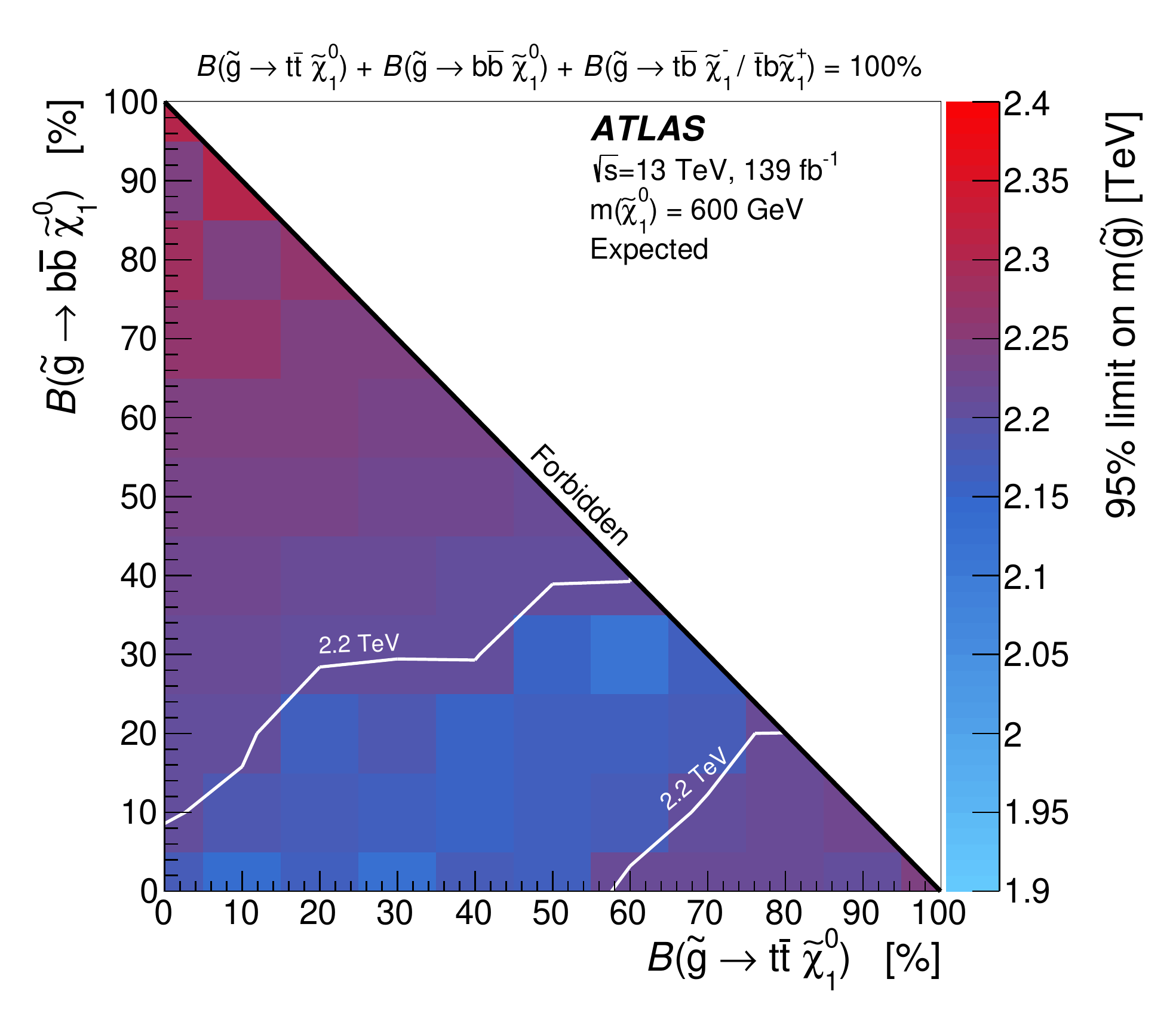}
\caption{The observed (left) and expected (right) 95\%~CL exclusion limits on the gluino mass as a function of $\mathcal{B}(\tilde{g} \to b\bar{b}\ninoone$) (vertical) and $\mathcal{B}(\tilde{g} \to t\bar{t}\ninoone$) (horizontal) for Gtb models with $\mchi = 600$~\GeV, obtained from the CC analysis. The shading indicates the highest excluded gluino mass for each point in the plane. The white line indicates the contour of fixed $m(\tilde{g}) = 2200$~\GeV.}
\label{fig:limits_triangle_600}
\end{figure}
 
\begin{figure}[htbp]
\centering
\includegraphics[width=0.49\textwidth]{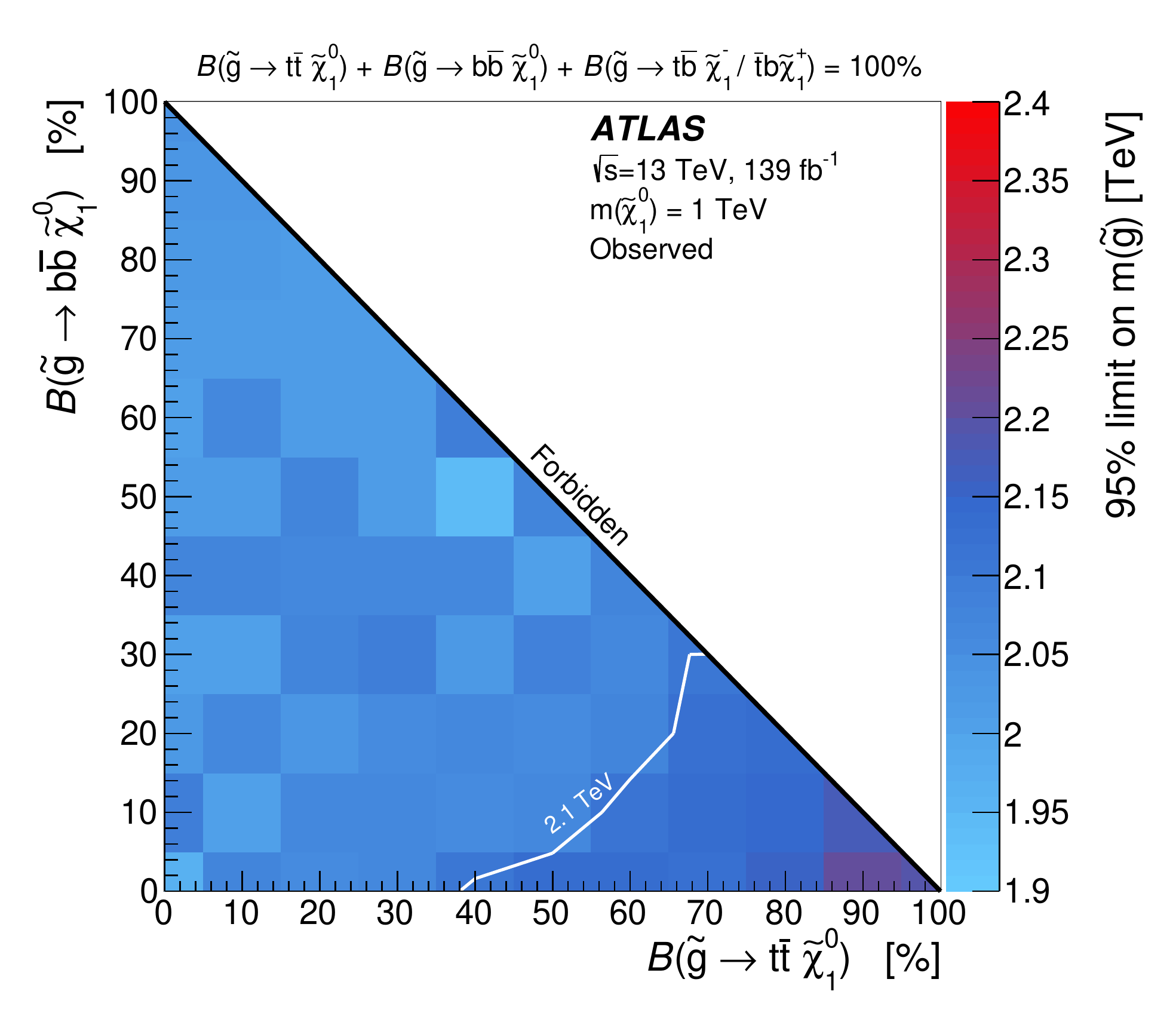}
\includegraphics[width=0.49\textwidth]{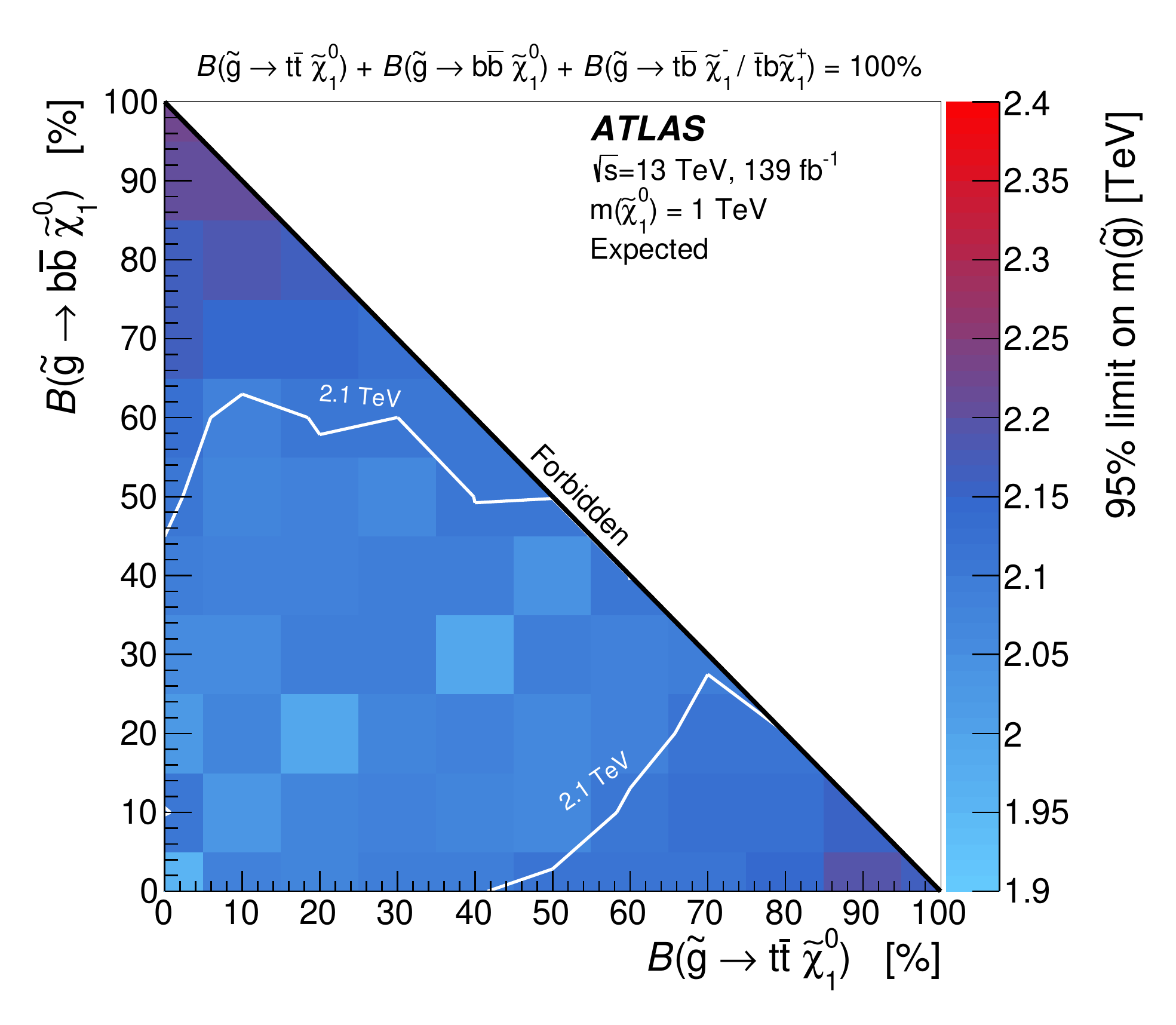}
\caption{The observed (left) and expected (right) 95\%~CL exclusion limits on the gluino mass as a function of $\mathcal{B}(\tilde{g} \to b\bar{b}\ninoone$) (vertical) and $\mathcal{B}(\tilde{g} \to t\bar{t}\ninoone$) (horizontal) for Gtb models with $\mchi = 1000$~\GeV, obtained from the CC analysis. The shading indicates the highest excluded gluino mass for each point in the plane. The white line indicates the contour of fixed $m(\tilde{g}) = 2100$~\GeV.}
\label{fig:limits_triangle_1000}
\end{figure}


\FloatBarrier
\section{Conclusion}
\label{sec:conclusion}

This paper has discussed a search for supersymmetry involving the pair production of gluinos decaying via off-shell third-generation squarks into a lightest neutralino LSP. The analysis exploits proton--proton collision data at a centre-of-mass energy $\sqrt{s} = 13$~\TeV\ with an integrated luminosity of 139~fb$^{-1}$ collected with the ATLAS detector at the LHC from 2015 to 2018. The search uses events containing large missing transverse momentum, zero or one electrons or muons, and several energetic jets, at least three of which must be identified as containing $b$-hadrons. Two analysis methodologies are used: one applying selections independently to a series of kinematic observables in multiple signal regions sensitive to a broad range of SUSY models, and one targeting specific signal models using a deep neural-network to further exploit correlations between observables constructed from the four-vectors of the reconstructed final-state particles. The latter methodology, which also combines events with differing numbers of final-state electrons or muons, provides enhanced discovery and exclusion power for the specific signals targeted. No significant excess is found with respect to the predicted background. Model-independent limits are set on the visible cross-section for new physics processes. Exclusion limits are set on the gluino and neutralino LSP masses in two simplified models where the gluino decays exclusively as $\gluino \to t\bar{t}\ninoone$ or $\gluino \to b\bar{b}\ninoone$. For a massless neutralino, gluino masses of less than 2.44~\TeV\ or 2.35~\TeV\ are excluded at 95\% CL in these two scenarios. The results are also interpreted in a model with variable gluino branching ratios to $b\bar{b}\tilde\chi^0_1$, $t\bar{t}\tilde\chi^0_1$ and $t\bar{b}\tilde\chi^-_1$ / $\bar{t}b\tilde\chi^+_1$. These limits represent a substantial increase in performance over previous ATLAS analyses exploiting smaller datasets collected at the LHC.


\section*{Acknowledgements}


We thank CERN for the very successful operation of the LHC, as well as the
support staff from our institutions without whom ATLAS could not be
operated efficiently.
 
We acknowledge the support of
ANPCyT, Argentina;
YerPhI, Armenia;
ARC, Australia;
BMWFW and FWF, Austria;
ANAS, Azerbaijan;
CNPq and FAPESP, Brazil;
NSERC, NRC and CFI, Canada;
CERN;
ANID, Chile;
CAS, MOST and NSFC, China;
Minciencias, Colombia;
MEYS CR, Czech Republic;
DNRF and DNSRC, Denmark;
IN2P3-CNRS and CEA-DRF/IRFU, France;
SRNSFG, Georgia;
BMBF, HGF and MPG, Germany;
GSRI, Greece;
RGC and Hong Kong SAR, China;
ISF and Benoziyo Center, Israel;
INFN, Italy;
MEXT and JSPS, Japan;
CNRST, Morocco;
NWO, Netherlands;
RCN, Norway;
MEiN, Poland;
FCT, Portugal;
MNE/IFA, Romania;
MESTD, Serbia;
MSSR, Slovakia;
ARRS and MIZ\v{S}, Slovenia;
DSI/NRF, South Africa;
MICINN, Spain;
SRC and Wallenberg Foundation, Sweden;
SERI, SNSF and Cantons of Bern and Geneva, Switzerland;
MOST, Taiwan;
TENMAK, T\"urkiye;
STFC, United Kingdom;
DOE and NSF, United States of America.
In addition, individual groups and members have received support from
BCKDF, CANARIE, Compute Canada and CRC, Canada;
PRIMUS 21/SCI/017 and UNCE SCI/013, Czech Republic;
COST, ERC, ERDF, Horizon 2020 and Marie Sk{\l}odowska-Curie Actions, European Union;
Investissements d'Avenir Labex, Investissements d'Avenir Idex and ANR, France;
DFG and AvH Foundation, Germany;
Herakleitos, Thales and Aristeia programmes co-financed by EU-ESF and the Greek NSRF, Greece;
BSF-NSF and MINERVA, Israel;
Norwegian Financial Mechanism 2014-2021, Norway;
NCN and NAWA, Poland;
La Caixa Banking Foundation, CERCA Programme Generalitat de Catalunya and PROMETEO and GenT Programmes Generalitat Valenciana, Spain;
G\"{o}ran Gustafssons Stiftelse, Sweden;
The Royal Society and Leverhulme Trust, United Kingdom.
 
The crucial computing support from all WLCG partners is acknowledged gratefully, in particular from CERN, the ATLAS Tier-1 facilities at TRIUMF (Canada), NDGF (Denmark, Norway, Sweden), CC-IN2P3 (France), KIT/GridKA (Germany), INFN-CNAF (Italy), NL-T1 (Netherlands), PIC (Spain), ASGC (Taiwan), RAL (UK) and BNL (USA), the Tier-2 facilities worldwide and large non-WLCG resource providers. Major contributors of computing resources are listed in Ref.~\cite{ATL-SOFT-PUB-2021-003}.


\printbibliography
 
\clearpage
\input{atlas_authlist}

\end{document}

%% file: atlas_authlist.tex
 
\begin{flushleft}
\hypersetup{urlcolor=black}
{\Large The ATLAS Collaboration}

\bigskip

\AtlasOrcid[0000-0002-6665-4934]{G.~Aad}$^\textrm{\scriptsize 101}$,
\AtlasOrcid[0000-0002-5888-2734]{B.~Abbott}$^\textrm{\scriptsize 119}$,
\AtlasOrcid[0000-0002-7248-3203]{D.C.~Abbott}$^\textrm{\scriptsize 102}$,
\AtlasOrcid[0000-0002-1002-1652]{K.~Abeling}$^\textrm{\scriptsize 55}$,
\AtlasOrcid[0000-0002-8496-9294]{S.H.~Abidi}$^\textrm{\scriptsize 29}$,
\AtlasOrcid[0000-0002-9987-2292]{A.~Aboulhorma}$^\textrm{\scriptsize 35e}$,
\AtlasOrcid[0000-0001-5329-6640]{H.~Abramowicz}$^\textrm{\scriptsize 150}$,
\AtlasOrcid[0000-0002-1599-2896]{H.~Abreu}$^\textrm{\scriptsize 149}$,
\AtlasOrcid[0000-0003-0403-3697]{Y.~Abulaiti}$^\textrm{\scriptsize 116}$,
\AtlasOrcid[0000-0003-0762-7204]{A.C.~Abusleme~Hoffman}$^\textrm{\scriptsize 136a}$
% ...
\bigskip
\\

% ...
$^{29}$Physics Department, Brookhaven National Laboratory, Upton NY; United States of America.\\
% ...
$^{35}$$^{(a)}$Facult\'e des Sciences Ain Chock, R\'eseau Universitaire de Physique des Hautes Energies - Universit\'e Hassan II, Casablanca;$^{(b)}$Facult\'{e} des Sciences, Universit\'{e} Ibn-Tofail, K\'{e}nitra;$^{(c)}$Facult\'e des Sciences Semlalia, Universit\'e Cadi Ayyad, LPHEA-Marrakech;$^{(d)}$LPMR, Facult\'e des Sciences, Universit\'e Mohamed Premier, Oujda;$^{(e)}$Facult\'e des sciences, Universit\'e Mohammed V, Rabat;$^{(f)}$Institute of Applied Physics, Mohammed VI Polytechnic University, Ben Guerir; Morocco.\\
% ...
$^{55}$II. Physikalisches Institut, Georg-August-Universit\"{a}t G\"ottingen, G\"ottingen; Germany.\\
% ...
$^{101}$CPPM, Aix-Marseille Universit\'e, CNRS/IN2P3, Marseille; France.\\
$^{102}$Department of Physics, University of Massachusetts, Amherst MA; United States of America.\\
% ...
$^{116}$Department of Physics, New York University, New York NY; United States of America.\\
% ...
$^{119}$Homer L. Dodge Department of Physics and Astronomy, University of Oklahoma, Norman OK; United States of America.\\
% ...
$^{136}$$^{(a)}$Departamento de F\'isica, Pontificia Universidad Cat\'olica de Chile, Santiago;$^{(b)}$Millennium Institute for Subatomic physics at high energy frontier (SAPHIR), Santiago;$^{(c)}$Instituto de Investigaci\'on Multidisciplinario en Ciencia y Tecnolog\'ia, y Departamento de F\'isica, Universidad de La Serena;$^{(d)}$Universidad Andres Bello, Department of Physics, Santiago;$^{(e)}$Instituto de Alta Investigaci\'on, Universidad de Tarapac\'a, Arica;$^{(f)}$Departamento de F\'isica, Universidad T\'ecnica Federico Santa Mar\'ia, Valpara\'iso; Chile.\\
% ...
$^{149}$Department of Physics, Technion, Israel Institute of Technology, Haifa; Israel.\\
$^{150}$Raymond and Beverly Sackler School of Physics and Astronomy, Tel Aviv University, Tel Aviv; Israel.\\
% ...
$^{a}$ Also Affiliated with an institute covered by a cooperation agreement with CERN.\\
$^{b}$ Also at An-Najah National University, Nablus; Palestine.\\
$^{c}$ Also at Borough of Manhattan Community College, City University of New York, New York NY; United States of America.\\
$^{d}$ Also at Bruno Kessler Foundation, Trento; Italy.\\
$^{e}$ Also at Center for High Energy Physics, Peking University; China.\\
$^{f}$ Also at Center for Interdisciplinary Research and Innovation (CIRI-AUTH), Thessaloniki ; Greece.\\
% ...

\end{flushleft}
